\documentstyle[12pt]{report}
\input{epsf}

\def\sect{\section}
\def\a{\alpha} \def\b{\beta} \def\g{\gamma} \def\G{\Gamma}
\def\d{\delta}  

\def\k{\kappa} \def\l{\lambda} \def\L{\Lambda} \def\m{\mu} \def\n{\nu}
\def\cs{\xi}    \def\r{\rho}
\def\s{\sigma} \def\S{\Sigma} \def\t{\tau} 
 \def\f{\phi} \def\F{\Phi}  
\def\O{\Omega} \def\na{\nabla}
\def\pa{\partial}   \def\half{{1\over
2}} 
\def\oa{${\cal O}(\a ')$}\def\oaa{${\cal O}(\a '^2)$}
\def\op{{\cal O}}
\def\cd{\cdot}
\def\beq{\begin{equation}}
\def\eeq{\end{equation}}
\def\bea{\begin{eqnarray}}
\def\ena{\end{eqnarray}}
\def\EQ{\begin{equation}}
\def\EN{\end{equation}}
\newcommand{\be}{\begin{equation}}
\newcommand{\ee}{\end{equation}}
\newcommand{\eea}{\end{eqnarray}}

\begin{document}

\begin{titlepage}
\vspace{7cm}
\begin{center}
{\Huge\bf Aspects of Duality}\\[4cm]

{\Large\bf Kasper Olsen \\
Niels Bohr Institute\\University of Copenhagen}\\[2cm]

\begin{figure}[h]
\begin{center}
\mbox{
\epsfysize4cm
\epsffile{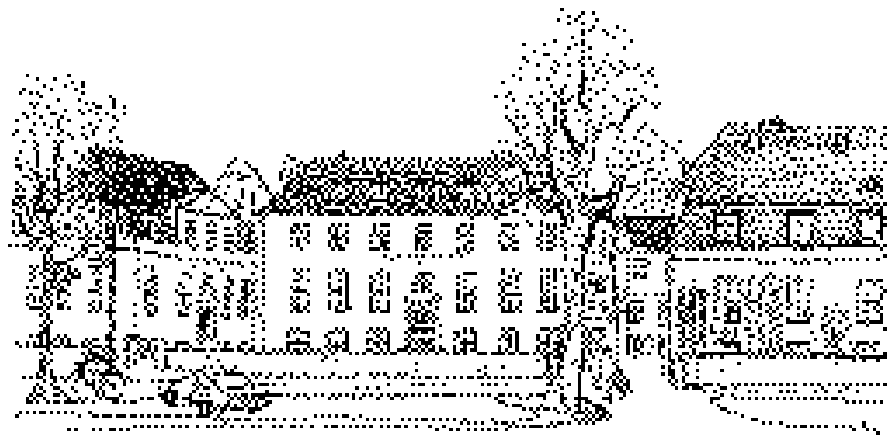}
}
\end{center}
\end{figure}
Thesis accepted for the Ph.D. degree in physics \\
at the Faculty of Science, University of Copenhagen.\\
{\bf March 1999}
\end{center}
\end{titlepage}

\renewcommand{\thefootnote}{\arabic{footnote}}
\setcounter{footnote}{0}
\addcontentsline{toc}{chapter}{Acknowledgements}

\chapter*{Acknowledgements}

I would like to thank my advisor Poul Henrik Damgaard for his guidance and
inspiring discussions throughout the past three years of study.

I am grateful to Peter E. Haagensen and Ricardo Schiappa for very interesting
discussions and collaborations in which a major part of this work was done.

In that connection I would like to thank M.I.T., Department of Theoretical Physics, for
kind hospitality during my four-months stay - especially my officemates Victor
Martinez, Karl Millar and Oliver de Wolfe. 

I would also like to thank
Harvard University, Lyman Laboratory of Physics for hospitality during a four-months
visit  - and Michael Spalinski and John Barrett for creating a nice atmosphere.  

At the Niels Bohr Institute, Department of High Energy Physics, I would like to thank all
those who created a pleasant environment, especially: J\o rgen Rasmussen, Morten Weis,
Jakob L. Nielsen, Lars Jensen, Martin G. Harris and Juri Rolf.

It is a pleasure to acknowledge discussions with V. Balasubramanian, M. Blau, H. Dorn, F.
Larsen, J.M. Maldacena, H.B. Nielsen, M. Spalinski, C. Vafa,  E. Witten and  D. Zanon.

Finally, I would like to thank Don Bennett, Matthias Blau, Jakob
L. Nielsen, J\o rgen Rasmussen and Paolo Di Vecchia 
for a careful reading of various parts of the manuscript.

\tableofcontents

\chapter{Introduction}

Understanding the restrictions imposed by duality in quantum field theory and
string theory is an important goal. One motivation for studying
duality is to understand the degrees of freedom in terms of which a
certain theory 
should be formulated. Another related
motivation is, given the right degrees of freedom, to describe the dynamics of the theory in
the different regimes. 

Just understanding the appropriate degrees of freedom and formulation
in terms of which quantum field theory and string theory should be described -- at
least at strong coupling -- has been a recurring theme. The question "what is quantum field
theory?" has basically only been answered in terms of perturbation theory and certain non-perturbative
methods, such as lattice models. To give a more complete answer we
need to understand what happens at strong coupling.
This is where duality comes in. In general terms, duality means an
equivalence between two or more descriptions of a physical system (or model).
The duality typically connects descriptions in terms of different fields, but this mapping is not in
any way guaranteed to be simple: for example one description could be in terms of a gauge theory in
$d$ dimensions while the other (dual) description could be in terms of
a string theory in $d+1$ dimensions. An example is the $AdS/CFT$ duality of Maldacena \cite{malda}. 

How can duality now be used? Typically,
duality exchanges weak coupling with strong coupling through $\l\rightarrow \l_{D}= 1/\l$, where
$\l$ is a coupling constant so that, in principle, a computation at strong coupling (where $\l$ is
large) can be worked out by looking at the dual weakly coupled formulation of the theory (where $\l_{D}$
is small). This kind of duality is called $S$-duality.
The moduli space of the theory has accordingly at least two "cusps" with $\l$ being small near one
of them and $\l_{D}$ small near the other. So duality relates the
physics at these two cusps despite the fact that the perturbative description of the theory in
terms of actions, fields and symmetries is generally different at the
various cusps (see fig. \ref{mspace} in Chapter 4 for an illustration -- with six
cusps -- in the context of string theory). 

It would of course be very nice if we could for example understand QCD both in the
asymptotically free and the confining regime -- and also "prove" confinement!
However, such a hope has not as yet been realized. 
At least in four and higher dimensions, it
has hereunto seemed that supersymmetry is necessary in order 
to have any duality relations and that only such
supersymmetric versions of QCD could be understood at strong coupling
(the recent work of Maldacena \cite{malda} seems to be an exception). 
One of the most
important works in this direction is the solution of ${\cal N}=2$ supersymmetric
Yang-Mills theory by Seiberg and Witten \cite{witten9407,witten9408} from 1994. It turns
out that the relevant degrees of freedom at strong coupling are not the basic fields in
the theory but rather monopoles and dyons. 
Also certain theories with
${\cal N}=1$ supersymmetry have been shown by Seiberg 
\cite{seiberg9411} and others to exhibit a form of duality.

It is remarkable that the use of duality is not confined to physics. In a major
breakthrough, Witten \cite{witten9411} has demonstrated how the 
above-mentioned ${\cal N}=2$ duality can be
applied to the study of four-dimensional manifolds and their (smooth) invariants: instead of computing
the so-called Donaldson invariants from
$SU(2)$ instanton solutions, Witten has shown that one can obtain the same invariants from
the solutions of the dual equations which include Abelian gauge fields and
monopoles and are therefore much simpler to analyze. These results have made a profound impact on
the mathematics community.  

The similar question for string theory "what is string theory?" seems much more difficult to
answer. Before 1994 string theory was understood only as a theory of interacting one-dimensional
objects. It has turned out that there is not just one perturbative string theory but rather five of them
which can be consistently formulated at weak coupling (they are the Type IIA and Type IIB theories
which have
${\cal N}=2$ supersymmetry, and the three theories with ${\cal N}=1$ 
supersymmetry: the heterotic $SO(32)$,
heterotic $E_8\times E_8$ theory and a Type I theory). With the help of duality it has been
conjectured that these five superstring theories are non-perturbatively
equivalent. As an example the strong coupling limit of Type I open string
theory can be described as a weakly coupled closed
$SO(32)$ heterotic string theory. A tool central to this understanding has been the interpretation of
certain string solitons as the source of a R-R field
\cite{pol9510a}: these are the so-called D-branes which are
$p$-dimensional extended objects (with
$p=0,2,4,6,8$ in the Type IIA string theory for example) with tension that varies as $1/\l$. So, string
theory does not appear to be a theory of strings only. 

There is another surprise: in the picture supported
by duality all five of these consistent string theories seem to represent a perturbative expansion
at different points in moduli space
of a single underlying theory. However,  not all points in moduli space represent ten-dimensional
theories. In fact, there are points corresponding to an eleven-dimensional vacuum, since the
strong coupling limit of ten-dimensional Type IIA string theory is an eleven-dimensional theory
\cite{witten9503}. This eleven-dimensional theory is tentatively called $M$-theory; its low energy limit
is eleven-dimensional supergravity.

The use of duality in string/$M$-theory today is largely aimed at answering the question: "what is
$M$-theory?". However, as there are still many unanswered questions as to what
string theory really is about, this is of course a very early
stage for trying to answer such a question. One proposal is the Matrix Theory conjecture by
Banks et al. \cite{banks9610}: $M$-theory in the infinite momentum frame is described by a $U(N)$ Yang-Mills theory
in $1+0$ dimensions (with $N \rightarrow \infty$). 
The fundamental degrees of
freedom can be interpreted as the D0-branes of the Type IIA string theory. 
Another recent proposal is the Anti-de Sitter/conformal field theory correspondence of Maldacena
\cite{malda}. This conjecture states that $M$-theory compactified on $AdS_{d+1}$ ($d+1$-dimensional
Anti-de Sitter space) is dual to a conformal field theory living on the
$d$-dimensional boundary of $AdS_{d+1}$. This correspondence satisfies an interesting holography
principle: the bulk degrees of freedom can be identified with the degrees of freedom living on the
boundary of spacetime with (at most) one degree of freedom per Planck area \cite{susskind9805}. 

There are many examples of dualities in
the literature. Some have been known for a long time - e.g. in quantum field theory - whereas some have
only very recently been discovered - e.g. in string theory. To make the discussion a little more
concrete, we shall list some basic examples (which reveal properties
that will reappear a number of times
in this thesis).
\begin{itemize}

 \item A simple duality (and one which was already noted by Dirac \cite{dirac}) is that of
electric-magnetic duality. To describe it, begin with the source-free Maxwell equations. The
equations of motion for the field strength are
\beq
\pa_{\m}F^{\m\n}=0
\eeq
and the Bianchi identities are 
\beq
\pa_{\m}*F^{\m\n}=0\ ,
\eeq  
where $*F^{\m\n}=\half\epsilon^{\m\n\l\r}F_{\l\r}$ is the dual field strength.    
These equations are invariant under 
\beq\label{emsym}
F^{\m\n} \rightarrow *F^{\m\n}\ , \; \; *F^{\m\n} \rightarrow -F^{\m\n}\ ,
\eeq
which interchange the equations of motion with the Bianchi identities. Concretely, 
the equations of motion follow from the standard action $\frac{1}{4}\int d^{4}x F_{\m\n}F^{\m\n}$,
while the Bianchi identity is just the divergence condition that follows from the fact that
$F_{\m\n}=\pa_{\m}A_{\n}-\pa_{\n}A_{\m}$. In terms of
electric and magnetic fields we have $E_{i}=F_{0i}$ and $B_{i}=\half \epsilon_{ijk}F_{jk}$, so this
symmetry (\ref{emsym}) is the same as the discrete symmetry:
\beq
{\bf E}\rightarrow {\bf B}\ , \; \; {\bf B}\rightarrow -{\bf E}\ .
\eeq
To extend this duality to the Maxwell equations with sources, we have to add both electric and
magnetic sources in which case
\beq
\pa_{\m}F^{\m\n}=j^{\n}_e
\eeq
and the Bianchi identities are 
\beq
\pa_{\m}*F^{\m\n}=j^{\n}_m\ .
\eeq 
As shown by Dirac \cite{dirac}, a quantum theory of both electric and magnetic charges is
only consistent if the Dirac quantization condition is satisfied. It is
interesting to see how this condition can be derived. If we have a monopole, of charge $q_m$, at the origin
and surround it with a two-sphere, then
\beq
\int_{{\bf S}^2}\ F = q_{m}\ .
\eeq 
Globally we cannot have $F=dA$, since if this were true the flux would vanish by Stokes'
Theorem. But we can write $F=dA$ except along a Dirac string. An electric
charge $q_{e}$ will, when moving along a closed path $P$, acquire a phase in its wavefunction
\beq
\exp \left( iq_{e}\int_{P}A \right) = \exp \left( iq_{e}\int_{S}F \right)\ , 
\eeq
provided $S$ does not intersect the Dirac
string ($S$ is a surface that has $P$ as its boundary). When the path
$P$ is contracted to a vanishing circle but which still circumnavigates the Dirac string, 
the phase must be
\beq
\exp \left(iq_e\int_{{\bf S}^2}F\right)= \exp \left(iq_eq_m\right)
\eeq
and equal to 1 since the Dirac string should be non-physical. 
Therefore, we have the Dirac quantization condition:
\beq
q_eq_m=2\pi n\ ,
\eeq
where $n$ is integer (it is fascinating that this result not only applies to point particles; in
ten-dimensional string theory, this result readily generalizes to the above-mentioned D-branes).
Such a duality offers an explanation for why electric charge is observed to be quantized.

\item The two-dimensional sine-Gordon model is dual to the massive Thirring model \cite{coleman}. The
relation between the coupling constants in the two theories is 
\beq
\frac{\b^2}{4\pi}=\frac{1}{1+g/\pi}\ ,
\eeq
so that weak coupling in the sine-Gordon model ($\b$ small) corresponds to strong coupling ($g$
large) in the massive Thirring model -- and vice versa. This is also an example where a fundamental
object is dual to a solitonic object since the soliton of the sine-Gordon model can be interpreted
as the fundamental fermion field of the Thirring model. More concretely, the fermion field $\psi$ 
can be written as the following vertex operator of the boson field $\phi$ \cite{mandelstam}:
\beq
\psi(x)=:e^{-2\pi i\b^{-1}\int_{-\infty}^{x}dz\dot{\phi}(z)+\half i\b\phi(x)}:\ ,
\eeq
(here $\dot{\f}$ is the derivative of $\f$ with respect to time and :: means normal ordering).
Note that this is an exact and derived duality. 
 
\item The Ising model (on a square lattice) exhibits a so-called
Kramers--Wannier duality \cite{kramers}. In this model
the partition function $Z(K)$ is a function of the temperature $T$ and
the strength $J$ of the
nearest neighbour interaction. We introduce the quantity $K=J/(k_BT)$ for convenience. 
The partition function can be calculated exactly (Onsager's solution) and
is equal to the partition function of the dual lattice theory $Z^*$ if the coupling constants are
related according to
\beq
\sinh 2K=\frac{1}{\sinh 2K^*}\ .
\eeq 
Thus, weak coupling ($K\rightarrow 0$) in one theory is dual to strong
coupling ($K^*\rightarrow \infty$) in the dual theory.

\item Four-dimensional ${\cal N}=4$ non-Abelian supersymmetric
  Yang-Mills theory is conjectured to exhibit a
Montonen-Olive duality \cite{olive}. With gauge group $U(n)$ this theory is in fact self-dual, so
the dual theory at the other expansion point is identical
to the old one (meaning, among other things, that the actions are equal). 
The bosonic part of the Lagrangian of the ${\cal N}=4$ theory contains a Yang-Mills term
proportional to $1/g^2$, and a $\theta$-term. On the complex coupling constant
\beq
\t=\frac{\theta}{2\pi}+\frac{4\pi i}{g^2}\ ,
\eeq
the conjectured $SL(2,{\bf Z})$-duality acts as
\beq
\t\rightarrow \frac{a\t+b}{c\t+d}\ .
\eeq
Here $a,b,c,d\in{\bf Z}$ and $ad-bc=1$. The transformation
$\t\rightarrow -1/\t$, for $\theta=0$, is
seen to be a strong-weak coupling duality. 
\end{itemize}

What we learn from these examples is that the moduli space can conveniently be thought of
as a manifold covered by an atlas of different
perturbative expansions (or "patches"). Thus, a given description - in terms of fields,
action etc. - generally depends on the particular patch. In this
picture the transition functions
correspond to the duality transformations. 
When we ask a question like "what is string theory?", we are really asking what is the
correct description of this moduli space? Can it only be described in terms of different patches
or perturbative regions connected by duality transformations, or is there a more fundamental
description of the theory?

Let us add a comment about the validity of duality. In nearly all cases
studied so far, duality has the status of a conjecture. To really prove
duality, say
$S$-duality, we need to understand non-perturbative effects. Establishing that a pair of theories
are really dual can only be done by solving them exactly, or by finding a field redefinition that
brings one theory into the other.
Examples where this {\sl can} be done are the 
sine-Gordon/Thirring model pair of theories and the Ising model.
Duality therefore typically enters in as a working hypothesis: if we have strong evidence leading us
to  believe that two theories actually are dual, then by studying the strong coupling regime of one
theory in terms of the other weakly coupled theory, we are likely to learn something new and often
interesting. 

We might also add that in some cases it is only certain limits of the dual theories that are
known or can even be described in terms of a perturbative theory. The Type IIA string theory is for
example well understood in the limit of small string coupling and it is conjectured to be described in
terms of an eleven-dimensional theory in the limit of very large coupling. However, what happens in
between we do not really know. 

The thesis is organized as follows. 
In the second and third chapter original results of the author (and
collaborators) are used to gain insight into the use of duality in
some familiar theories.

In Chapter 2 we study Seiberg-Witten duality of topological
field theories. After reviewing the key facts about topological field theory we describe the Donaldson
and Seiberg-Witten theories as (dual) approaches to the study of four-manifolds. The last section of this
chapter is based on \cite{ols} and the dimensionally reduced versions of these theories are derived. 

In Chapter 3 we consider in detail the $T$-duality of two-dimensional sigma models away from the
conformal point. This chapter is primarily based on \cite{HO}, \cite{HOS} and \cite{OS}. It is conjectured
that the relation $[T,R]=0$ should hold true between the RG flow
(generated by $R$) and $T$-duality of such models. This
has been demonstrated to be satisfied at one-loop in bosonic and heterotic sigma models and also at
two-loop for models with ${\cal N}=0,1,2$ supersymmetry with a purely metric background. Demanding, on the
other hand, a priori that
$[T,R]=0$, one can essentially determine the exact (at least to the orders in $\a'$ considered) RG flow
of the various models. This has also been shown to apply to models
that are $S$-dual \cite{ritz9710}.   

Chapter 4 is a short review of duality in ten-dimensional string theory. We review the manner in which all
five consistent string theories can (at least in principle) be related. For consistency such dualities in
string theory should imply a number of dualities in field theory. As
an example, the Montonen-Olive
duality can be understood as coming from a duality in Type II string theory compactified to four
dimensions. 

Finally, Chapter 5 contains our discussion.

Appendix A describes the Kaluza-Klein reduction of certain tensors which are necessary for the
computations in Chapter 3; Appendix B contains a list of tensors which are important for the
computation of a two-loop beta function in Chapter 3.

\chapter{Duality in Topological Field Theories}

While it may be impossible to prove any of the nontrivial duality
relations in quantum field theory and string theory directly, one can infer
evidence for certain dualities by examining the consequences in some simple
models. 

There exist quantum field theories which are of a very simple kind
and apparently
of limited applicability in physics, namely the topological 
quantum field theories \cite{witten88a}. 
From a physical point of view, one might simply categorize these theories 
as trivial
since they describe a situation in which there are no propagating degrees
of freedom the only observables being (global) topological invariants.

But physically it is still useful to study topological field theories.
For example, a key to studying, say,  $S$-duality is to examine quantities/states in the full theories
which are such that some of their properties can be reliably calculated at both strong and weak coupling. 
The BPS states comprise 
one such set of examples (because of supersymmetry non-renormalization theorems).
It turns out that frequently BPS states of the full original
theory make up the complete physical spectrum of a simplified theory, namely a topological "twist" of
the original theory. It is in this connection that topological field theories allow one to
explore consequences of duality.
     
Also, from a mathematical standpoint, these theories are in no way trivial
as they lead to important results (for example in relation to Donaldson theory of
four-manifolds \cite{donaldson1}). Hence, topological field
theories can be expected to offer an excellent testing ground for certain dualities since
the results can in principle be checked independently of 
any field theory formulation (an important example is the test of Montonen-Olive duality
in ${\cal N}=4$ supersymmetric Yang-Mills theory by Vafa and Witten \cite{vafa9408}). Moreover,
if we believe that the duality conjectures are correct, new and important results in
mathematics may emerge.

In this chapter we will consider the topological field theories which can be
obtained from the ${\cal N}=2$ supersymmetric Yang-Mills theory in four dimensions
by a simple "twisting" procedure. The ${\cal N}=2$ theory has
two dual descriptions: one (relevant at weak coupling) in which the fundamental degrees
of freedom are the gauge particles of $SU(2)$ and another (which is relevant at
strong coupling) in which the fundamental degrees of freedom are monopoles
and dyons of a $U(1)$ theory. Twisting these two quantum field theories,
one obtains a dual pair of topological field theories relevant for the
description of Donaldson theory where the weak coupling description
opens the possibility of a
perturbative approach to this theory, while the strong coupling description
reveals interesting non-perturbative properties. 

To set the scene, we start
by giving a short review of topological field theory (an excellent review of
topological field theory can be found in
\cite{danny}).  In the following section we
show how Donaldson theory appears after a twisting of the ${\cal N}=2$ theory at weak
coupling and describe the observables which can be viewed as topological
invariants of smooth four-manifolds. We then present the dual formulation,
the Seiberg-Witten theory, and its salient points. Finally, we consider a
version of this Seiberg-Witten duality in three and two dimensions \cite{ols}. The
dimensionally reduced actions are derived and some results that could be
relevant for studying the so-called Hitchin equations on Riemann surfaces \cite{hitchin}
are presented.  

\section{Topological Field Theory}

The study of topological quantum field theory started in 1988 with the
work of Witten \cite{witten88a} who constructed a simple quantum field theory
that is now known as Donaldson-Witten theory. Witten observed that a twisted
version of ${\cal N}=2$ supersymmetric Yang-Mills theory in four dimensions has no
local degrees of freedom, but only global degrees of freedom; these are the
topological invariants. 
This theory provided the physical interpretation (that was speculated to
exist by Atiyah) of Donaldson theory which is a mathematical theory that
through the study of the instanton solutions of Yang-Mills theory had provided an
important advance in the topology of four-manifolds.

Later that year Witten formulated two other
topological field theories, namely the topological sigma model in two
dimensions \cite{witten88b} and the Chern-Simons theory in 
three dimensions \cite{witten88c}. All these theories are
related to different invariants which have been studied in the mathematics
literature. The Donaldson-Witten theory can be related to Donaldson
invariants (which we will define later) of four-manifolds and, in its
three-dimensional version, to the so-called Casson invariants \cite{danny}. The Witten
approach to Chern-Simons theory on the other hand can be related to the
Jones polynomial. This is a polynomial invariant of knots and
links in three dimensions \cite{witten88c}.   
Finally, the topological sigma models can be related to the so-called Gromov-Witten 
invariants and quantum cohomology \cite{danny}. 

One can rather naturally distinguish two types of topological quantum
field theories: the Witten type (or cohomological type) and the
Schwarz type \cite{witten88c}
(or quantum type). In the following we will mainly concentrate on the Witten type.

To define what a topological field theory is, we start with the
following objects. 
Let $X$ be a Riemann manifold with metric $g_{\m\n}$ and let $\F$
denote any set of fields on $X$ with an action $S(\F)$. Operators which are
functionals of these fields are denoted by $\op(\F)$ and a vacuum expectation
value of a product of fields is formally defined by the functional integral
\beq\label{corrfct}
\langle\op_{\a}\op_{\b}\cdots\op_{\g}\rangle=\int D[\Phi]
\op_{\a}(\F)\op_{\b}(\F)\cdots\op_{\g}(\F)e^{-S(\F)}\ ,
\eeq
where $D[\Phi]$ denotes the path integral measure.
A quantum field theory on $X$ is "topological" if there is a set of
"operators" which are invariant under arbitrary deformations of the
metric $\delta_g$, in the sense
that,
\beq\label{metricdef}
\delta_g\langle\op_{\a}\op_{\b}\cdots\op{\g}\rangle=0\ ,
\eeq
i.e. the expectation values of products of observables are topological
invariants. 

Our interest here will be focused on 
the so-called smooth invariants, that is
quantities which are invariant under diffeomorphisms (a diffeomorphism $f$ of
$X$ is a map  $f: X\rightarrow X'$ for which both $f$ and $f^{-1}$ are
$C^{\infty}$) of the base manifold $X$; phrased differently, that they are constant on a diffeomorphism
equivalence class of manifolds. The correlation functions in (\ref{metricdef}) are of this kind as are the
Donaldson invariants which we will discuss later.
\footnote{In {\sl mathematical} terms a topological invariant is a quantity
which is invariant under homeomorphisms of the base manifold $X$ (a
homeomorphism is a map $f: X\rightarrow X'$ for which both $f$ and $f^{-1}$
are continuous), or phrased differently, it is constant on a homeomorphism class of manifolds. 
So a smooth
invariant is not necessarily a topological invariant.}
  
Now we are in a position to define what a Schwarz and a Witten type topological
field theory is. 
 
A Witten type theory is topological since these theories have an
energy-momentum tensor which is BRST exact:
\beq\label{EMT}
T_{\m\n}=\{ Q, V_{\m\n}(\F,g)\}\ ,
\eeq 
where $V_{\m\n}$ is a symmetric functional (with ghostnumber equal to -1) of the fields and the metric. 
$Q$ is the
nilpotent BRST-like operator ($Q^2=0$) corresponding to some symmetry
$\delta$ of the theory that keeps the action
invariant 
(usually, $\delta$ is a combination of a so-called shift symmetry
$\delta\Phi=\epsilon$ and a gauge symmetry). 
Henceforth
$Q$ is simply called the BRST operator, and the energy-momentum tensor is $T_{\m\n}=\delta S/\delta
g_{\m\n}$. The notation used is such that the BRST variation
$\delta\Phi$  of any field
$\Phi$ is
\beq\label{deltadef}
\delta\Phi=-i\{ Q,\Phi \}\ ,
\eeq
expressing that $Q$ is the generator of the symmetry $\delta$.
For $\Phi$ bosonic the expression in (\ref{deltadef}) is a commutator and for
$\Phi$ fermionic it means an anti-commutator. The topological nature of the
theory then follows from the fact that any BRST closed operator
$\op_\a$ ($\{ Q,\op_{\a}\}=0$) satisfying 
$\delta_g\op_{\a}=\{ Q, R_\a\}$ has vanishing variation under the path
integral:
\bea\label{dgo}
\delta_g\langle\op_{\a}\rangle&=&
\int D[\Phi]
\left(\delta_g\op_{\a}-\delta_gS\cdot\op_{\a}\right) e^{-S(\F)}\nonumber\\
&=&\int D[\Phi] \left( \{ Q,R_{\a}\}-\{ Q,V\}\op_{\a} \right)\nonumber\\
&=&\langle\{ Q, R_{\a}-V\cdot\op_{\a}\}\rangle=0\ .
\ena
In deriving this, 
we have assumed that the measure $D[\Phi]$ is invariant under the
symmetry and that the vacuum is BRST invariant (for this implies 
$\langle\{ Q,X\}\rangle=0$ for any functional $X$). Also, we shall generally be assuming
that the BRST operator $Q$ is metric independent. 

What are then the natural
observables in the Witten type theory? In such a theory, any
BRST closed operator satisfying 
$\delta_g\op_{\a}=\{ Q, R_\a\}$ will be an observable because of (\ref{dgo}). 
Furthermore, adding
a BRST exact term to an observable will not change its expectation value,
and the observables can therefore be identified with cohomology classes of the
BRST operator, much like in string theory \cite{gsw}. It is simple to generalize
to any correlator
$\langle\op_{\a}\op_{\b}\cdots\op_{\g}\rangle$ and show that it is independent of
arbitrary deformations of the metric. Such a correlator is then a
topological invariant, though it might actually be trivial
in most cases.

Note that
a way to ensure that the energy-momentum tensor is BRST exact, is to require
BRST exactness of the quantum action itself
\beq\label{s=qf}
S_q(\F)=\{ Q, \Psi(\F,g)\}\ .
\eeq
This will often be assumed in the following (on the right hand side of 
(\ref{s=qf}) one can always add a metric independent term without
destroying the topological nature of the theory. However, if one does not
want that term to influence the moduli space
probed by the theory, then it should be not only metric independent but also topological
- in the sense that it is locally a total derivative).

A Schwarz type theory is topological since in such a theory
the classical action
$S_c(\F)$ and the operators $\op$ are independent of the metric. 
The quantum action - including ghosts for gauge fixing - is then of the form $S_q=S_c+\{Q_B,V \}$,
where $Q_B$ is now the standard field theory BRST operator. At least formally, one can then conclude that
(\ref{EMT}) holds and therefore also (\ref{metricdef}) as in the Witten type case. 
Celebrated
examples are the Chern-Simons gauge theories \cite{witten88c} and the BF theories \cite{danny}. 

We now turn to Witten type theories.

The most basic invariant is simply the partition function: as the identity operator is
always an observable we can conclude from Eq. (\ref{dgo}) that the partition function of
the theory is invariant under deformations of the metric
\beq
\delta_g\langle 1\rangle =\delta_gZ=0\ ,
\eeq 
which implies that $Z$ is a topological invariant. 
Moreover,
and very importantly, the partition function is independent also of the coupling
constant. To show this, we assume that the coupling constant $\l$ appears in the
action as
$S'/\l^2$. The variation of $Z$ with respect to $\l^2$ is:
\bea
\delta Z=\delta \int D[\Phi]e^{-S'/\l^2}&=&\delta(-\frac{1}{\l^2})
\int D[\Phi]e^{-S'/\l^2}\cdot S' \nonumber\\
&=&\delta(-\frac{1}{\l^2})\int D[\Phi]e^{-S'/\l^2}\cdot
\{Q, \Psi\}\cdot \l^2\nonumber\\ 
&=&\delta(-\frac{1}{\l^2})\cdot \langle\{Q, \Psi\}\rangle\cdot \l^2=0\ .
\ena
This means that, at least formally, we can evaluate $Z$ in the weak coupling
limit ($\l\rightarrow 0$) -- or the strong coupling limit ($\l\rightarrow
\infty$) for that matter -- meaning that the semi-classical approximation is
exact. Assuming that the observables do not
depend on the coupling constant, the same is of course true for any correlation function.

That topological field theories are simple and almost trivial from  a physical viewpoint
can be illustrated by the following considerations: in a Witten-type
theory any bosonic field will have a BRST (or $Q$-) superpartner, or
schematically
\beq
\{ Q,{\rm field}\} = {\rm ghost}\ ,
\eeq
which, since physical states should be annihilated by $Q$, must be interpreted
as ghosts. Thus, the total number of degrees of freedom is zero and the
physical phase space is zero-dimensional. Secondly, in such a theory the
energy of any physical state is zero:
\beq
\langle H\rangle
 = \langle\int T_{00}\rangle = \langle\int \{ Q, V_{00}\} \rangle=0\ ,
\eeq
and a topological field theory therefore has no dynamical excitations! 

The way we introduced topological field theories above was rather ad
hoc. While topological field theories might seem to be rather trivial from a
physical viewpoint, they are certainly not trivial from a mathematical
viewpoint.  Witten type theories for example are related to the study of
different moduli spaces which play an important role in topology.  
A typical (but certainly not any) moduli problem can be formulated in quantum field
theoretic terms by using the paradigm of "fields, equations and symmetries"
\cite{witten91a}. As an example, Donaldson theory -- which we will
describe later -- can be viewed as the study
of the moduli space of Yang-Mills instantons. The fields here are the gauge
potentials $A_{\m}^a(x)$ and the equations are the self-duality equations
$F_{\m\n}=*F_{\m\n}$,
where
$F_{\m\n}$ is the field strength
$F_{\m\n}=\pa_{\m}A_{\n}-\pa_{\n}A_{\m}+[A_{\m},A_{\n}]$ and $*F_{\m\n}$ is the Hodge
dual $*F_{\m\n}=\half \epsilon_{\m\n\l\r}F^{\l\r}$.  
The symmetries are of course just the gauge symmetries 
$\delta A_{\m}^a=-D_{\m}\kappa^a$. Finally, the moduli space can be
described as the space of instanton solutions modulo the gauge symmetries.
This moduli space ${\cal M}_k$ is characterized by the instanton number 
$k$, which is minus the second Chern number:
\beq
k=-c_{2}(F)=\frac{1}{8\pi^2}\int_{X}{\rm Tr}(F\wedge F)\ .
\eeq

Before jumping to the elusive four-dimensional
world, we would like to describe a topological field theory that appears
naturally in two dimensions - namely the topological sigma model. In this
case the "fields" can be identified with maps $\phi: \Sigma \rightarrow K$,
where $\Sigma$ is a two-dimensional surface and $K$ is a K\"ahler
manifold, which has even real dimension.
The "equations" state that $\f^I$ is a holomorphic map, concretely
$\pa_{\bar{z}}\phi^I=0$, with $(z,\bar{z})$ being coordinates on
$\Sigma$ and
$I=1,2,\ldots ,{\rm dim}K/2$. However, there are no "symmetries". The action of
a topological sigma model with K\"ahler target space is
\cite{witten88b}:
\bea
S&=&2\int_{\Sigma}d^2\sigma \left[
g_{I\bar{J}}\pa_{+}\phi^{I}\pa_{-}\phi^{\bar{J}}
-\frac{i}{2}\r_{+}^{\ I}D_{-}\chi^{\bar{J}}g_{I\bar{J}} 
-\frac{i}{2}\r_{-}^{\ \bar{J}}D_{+}\chi^{I}g_{I\bar{J}}\right. \nonumber\\ 
&&\left. -\frac{1}{4}\chi^{I}\chi^{\bar{I}}\r_{+}^{\ J}\r_{-}^{\ \bar{J}}
R_{I\bar{I}J\bar{J}} \right]\ ,
\ena
where $g_{I\bar{J}}=g_{\bar{J}I}$ is the K\"ahler metric and $R_{I\bar{I}J\bar{J}}$ is the
Riemann tensor; $D_{\pm}$ is the covariant derivative pulled back from $K$ to $\Sigma$:
\beq
D_{\pm}\chi^{I}=\pa_{\pm}\chi^{I}+\pa_{\pm}\phi^{K}\Gamma^{I}_{KL}\chi^{L}\ .
\eeq
Here $\Gamma^{I}_{KL}$ is the Christoffel connection.
The action has a symmetry generated by left-moving and right-moving charges:
\bea
\delta\chi^{I}&=&\delta\chi^{\bar{I}}=0, \; \; 
\delta\phi^{I}=i\epsilon\chi^{I}, \; \; 
\delta\phi^{\bar{I}}=i\tilde{\epsilon}\chi^{\bar{I}}\ ,\nonumber \\
\delta\r_{+}^{\
I}&=&2\tilde{\epsilon}\pa_{+}\phi^{I}-i\epsilon g^{I\bar{S}}\pa_{S}g_{K\bar{S}}
\chi^{S}\r_{+}^{\ K}\ ,\nonumber \\
\delta\r_{-}^{\
\bar{I}}&=&2\epsilon\pa_{-}\phi^{\bar{I}}-i\tilde{\epsilon}g^{S\bar{I}}
\pa_{\bar{S}}g_{S\bar{K}}\chi^{\bar{S}}\r_{-}^{\ \bar{K}}\ ,
\ena
with $\epsilon$ and $\tilde{\epsilon}$ being two independent and anticommuting parameters.
For
$\epsilon=\tilde{\epsilon}$ there is a single standard BRST operator $Q$ with $Q^2=0$. 
The ghost numbers of the fields $\phi, \rho$ and $\chi$ are $U=0, -1$ and $1$
respectively.  This theory can be constructed by twisting the
${\cal N}=2$ supersymmetric sigma model \cite{witten88b}. However in two dimensions the
twisting is not unique. For $K$ a Calabi-Yau manifold there are two possible twistings
that give rise to the so-called $A$- and $B$-models and they are related by mirror
symmetry of the target manifold \cite{witten9112}.
The model described above (and for $K$ a Calabi-Yau manifold) 
is in these terms called the $A$-model.

The observables in topological sigma models are constructed as follows
\cite{witten88b}. If $A_{(p)}=A_{i_1 \cdots i_p}d\f^{i_1}\wedge \cdots \wedge d\f^{i_p}$ 
is a $p$-form on
$K$ then one constructs the operator (we are now using real coordinates on $K$):
\beq
\op^{(0)}_A = A_{i_1\cdots i_p}\chi^{i_1}\cdots \chi^{i_p}\ ,
\eeq
that obeys
\beq\label{AdA}
\{ Q,\op^{(0)}_A\} =-\op^{(0)}_{dA}\ , 
\eeq
with $d$ the exterior derivative on $K$ and $\delta = -i\epsilon\{Q, \cdot\}$; $\op^{(0)}_A$
can be viewed as a zero-form on $\Sigma$.
Thus, according to (\ref{AdA}) BRST cohomology classes of 
operators are in one-to-one correspondence with the de Rham cohomology classes of
$K$: $\{ Q, \op_{A}\}=0$ if and only if $A$ is closed and 
$\op_{A}= -\{ Q,\op_{B}\}$ if and only if $A=dB$, that is $A$ is exact. Choosing $A$ to be closed, one
then recursively solves the two equations
\beq\label{sigmadesc}
d\op^{(0)}_A = i\{ Q, \op^{(1)}_A\}, \; \; \; 
d\op^{(1)}_A = i\{ Q, \op^{(2)}_A\}\ . 
\eeq
Here we find 
\beq
\op^{(1)}_A = ipA_{i_1\cdots i_p}\pa_{\a}\f^{i_1}\chi^{i_2}\cdots
\chi^{i_p}d\sigma^{\a},
\; \; \;
\op^{(2)}_A = -\frac{1}{2}p(p-1)
A_{i_1\cdots i_p}\pa_{\a}\f^{i_1}\pa_{\b}\f^{i_2}\chi^{i_3}\cdots
\chi^{i_p}d\sigma^{\a}\wedge d\sigma^{\b}\ ,
\eeq
and they can be seen as respectively a one- and a two-form on $\Sigma$ with local coordinates
$\sigma^{\a}$.  Thus we have
three classes of observables on
$\Sigma$. The first class consists of operators of the form
\beq\label{op0}
\op^{(0)}_A (P)\ ,
\eeq
where $P$ is a point in $\Sigma$. The second class consists of operators like
\beq\label{op1}
\int_{C}\op^{(1)}_A\ , 
\eeq
with $C$ a one-cycle in $\Sigma$; because of 
(\ref{sigmadesc}) this operator only depends on the homology
class of $C$. Finally the third class are operators of the form:
\beq\label{op2}
\int_{\Sigma}\op^{(2)}_A\ .
\eeq
The first of these operators is BRST closed because of (\ref{AdA}). 
The two other
operators (\ref{op1}) and (\ref{op2}) are BRST closed because of
(\ref{sigmadesc}). The correlation functions consisting of products of operators like
(\ref{op0}), (\ref{op1}) and (\ref{op2}) gives topological invariants as in
(\ref{metricdef}) - more precisely the so-called Gromov-Witten
invariants
\cite{danny}.

\section{Donaldson-Witten Theory}

What has now become known as Donaldson-Witten theory originated as a topological field theory
constructed by Witten in 1988 \cite{witten88a}. Motivated by work of Atiyah
and Floer, Witten showed that a certain twisting of ${\cal N}=2$ supersymmetric
Yang-Mills theory yields a topological field theory, which is
precisely such
that the vacuum expectation values of certain observables are Donaldson invariants of
four-manifolds. 
Such invariants were introduced by Donaldson in 1983 \cite{donaldson1} as an
important tool in the classification of four-dimensional (differentiable)
manifolds.

As such, an important motivation for studying Donaldson-Witten theory is its
relation to the classification problem of four-dimensional (differentiable)
manifolds. Here, the goal is to classify all differentiable manifolds up to
diffeomorphisms, the more general classification problem being to classify
all topological manifolds up to homeomorphisms.

\begin{figure}[htb]
\begin{center}
\mbox{
\epsfxsize=12cm
\epsffile{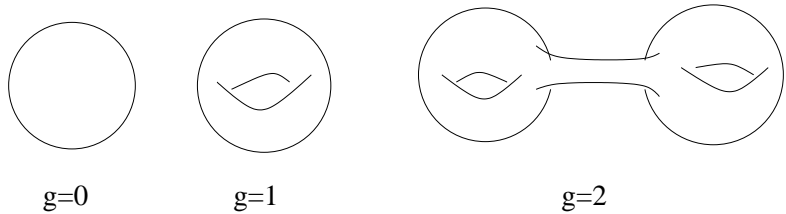}
}
\end{center}
\caption{The genus expansion of Riemann surfaces.}
\label{genus}
\end{figure}

It is well known that the classification problem is rather trivial in two
dimensions. Any compact, orientable surface, i.e. a Riemann surface, is
homeomorphic to a sphere with $g$ handles, and two such surfaces are 
homeomorphic exactly if they have the same number of
handles. Topologically, 
Riemann surfaces are
therefore classified by a single integer, the genus, see fig. \ref{genus}. 

In higher dimensions there is unfortunately no such simple classification 
(there is a partial classification for 
$D\geq 5$, see e.g. \cite{danny}). Especially in four
dimensions the situation is much more complicated -- and this is the dimension relevant
for Donaldson theory. That there can be no corresponding "list" of four-manifolds can be
demonstrated by looking at a very basic invariant of any
manifold, namely the fundamental group $\pi_1$ (this is the equivalence class of based
loops in $X$ -- for example $\pi_1({\bf S}^1)={\bf Z}$). Now, there is a theorem in
topology which states that any finitely representable group (this is a group generated by
finitely many elements that satisfy a finite number of relations) can appear as the
fundamental group of a four-dimensional manifold \cite{danny,dijk96a}. Moreover,
there is no algorithm to decide whether two such finitely representable groups are
isomorphic -- and therefore a classification similar to the one in two dimensions must
necessarily fail.  

A natural assumption in Donaldson theory is therefore that the four-manifold
$X$ is simply connected, that is the fundamental group vanishes:
$\pi_1(X)=0$. Hence, there can only be nontrivial 
$k$-dimensional homology cycles for $k=0,2,4$ (if
$\pi_1(X)$ is commutative then it is isomorphic to $H_1(X)$ \cite{nakahara}; in
particular if $\pi_1(X)=0$ then also $H_1(X)=0$ and by Poincar\'e
duality $H_3(X)=0$).

This being the case, it is natural to consider 
another important invariant the second cohomology group $H^2(X)$. 
For $\a,\b\in H^2(X)$ we can define the so-called intersection form that plays an
important role in Donaldson's work,
\beq
Q(\a,\b)=\int_{X}\a\wedge\b\ ,
\eeq
which is symmetric and non-degenerate (i.e. $Q(\a,\b)=Q(\b,\a)$ and $Q(\a,\b)=0$ for all $\a$
implies $\b=0$). This form can therefore be
diagonalized over ${\bf R}$. 
The intersection form $Q$ is called even if all its diagonal elements are
even and otherwise called odd. The importance of this invariant can be
appreciated by quoting a theorem of Freedman: 
a simply connected four-manifold $X$ with even intersection form $Q$ belongs
to a unique homeomorphism class, and if $Q$ is odd there are precisely two
non-homeomorphic $X$ with $Q$ as their intersection form \cite{freedman}. 
This of course means that the intersection form essentially determines the
homeomorphism class of a simply connected manifold $X$, and explains why the
intersection form is important for the study of topological four-manifolds.  

Donaldson theory, on the other hand, concerns mainly two things: (1) the study of topological
obstructions to the existence of a differentiable structure on a given topological
four-manifold and (2) the distinction between differentiable structures on a given
four-manifold. Phrased differently we can, given a topological four-manifold
$X$, ask: (1) does there exist one or more differentiable structures on $X$? and  (2) if
there is a differentiable structure, is it unique? 
One important theorem which was subsequently derived by Donaldson using the
theory of Yang-Mills instantons can now be stated: a compact smooth simply connected 
four-manifold, with positive
definite intersection form
$Q$ has the property that $Q$ is always diagonalizable over the integers to $Q={\rm
diag}(1,\ldots ,1)$ \cite{donaldson1}. This implies for example that no
simply connected four-manifold for which $Q$ is even and positive definite
has a smooth structure. 
The Donaldson invariants, which we will discuss later, are important because they
can distinguish between manifolds that have the same intersection form. So in mathematical terms
they are not topological invariants but rather smooth invariants: they
can distinguish homeomorphic non-diffeomorphic smooth manifolds.

After this mathematical interlude, we will turn to the field theory description of
Donaldson theory -- and in order to make the discussion more concrete we will start by
presenting the action of Donaldson-Witten theory and its symmetries.

We start with a four-manifold $X$ over which we have a non-Abelian connection
$A_{\m}$ transforming in the adjoint representation of $SU(2)$.
The Donaldson-Witten theory in four dimensions is then described by the
following topological action \cite{witten88a} (with $\m,\n=1,\ldots ,4$):

\begin{eqnarray}
S^{(4)}&=&\int_{X}d^{4}x\sqrt{g}\ {\rm Tr}[\frac{1}{4}F_{\m\n}
F^{\m\n}+ \frac{1}{4}F_{\m\n}
\tilde{F}^{\m\n}+\frac{1}{2}\phi D_{\m}D^{\m}\lambda
-i\eta D_{\m}\psi^{\m}+2iD_{\m}\psi_{\n}\chi^{\m\n} 
\nonumber \\ 
&&
-\frac{i}{2}\lambda[\psi_{\m},\psi^{\m}]
-\frac{i}{2}\phi[\eta ,\eta ]-\frac{1}{8}[\phi ,\lambda]^{2}]\ .
\label{dw}
\end{eqnarray}
We are here using the same notation as in \cite{ols} -- the term 
$\phi\left[\chi ,\chi\right]$ present in Witten's action \cite{witten88a} can be included
by adding to Eq.~(\ref{dw}) a $\delta$-exact term \cite{bau1}. This action can be
obtained as the BRST variation of 
\beq
V^{(4)}={\rm Tr}F_{\m\n}^{+}\chi^{\m\n}-
{\rm Tr}\frac{1}{2}B_{\m\n}
\chi^{\m\n}+\frac{1}{2}{\rm Tr}\psi_{\m}D^{\m}\lambda
-\frac{1}{4}{\rm Tr}(\eta\left[\phi ,\lambda\right] )\ ,
\label{v4}
\eeq
and the theory is therefore topological according to the discussion in the
previous section. $F_{\m\n}$ is the field strength,
$F_{\m\n}=\pa_{\m}A_{\n}-\pa_{\n}A_{\m}+[A_{\m},A_{\n}]$ and 
$F_{\m\n}^{+}$ is the self--dual part of $F_{\m\n}$, that is
$F_{\m\n}^{+}=\frac{1}{2}(F_{\m\n}+\tilde{F}_{\m\n})$
with $\tilde{F}_{\m\n}=
\frac{1}{2}\epsilon_{\m\n\gamma\delta}F^{\gamma\delta}$; $\chi$ is a self-dual two-form (that is
$\chi_{\m\n}=-\chi_{\n\m}$ and $\chi_{\m\n}=\half \epsilon_{\m\n\g\d}\chi^{\g\d}$) and
its BRST partner $B_{\m\n}$ has been integrated out of the action.  
The fields transform as
\be\label{dwalg}
\begin{array}{ll}
\delta A_{\m}=i\psi_{\m}\ , &  \mbox{} \\
\delta\psi_{\m}=- D_{\m}\phi\ , &  \mbox{} \\
\delta\phi=0\ , & \mbox{} \\
\delta\chi_{\m\n}= B_{\m\n}\ , & \delta\lambda=2i\eta\ , \\
\delta B_{\m\n}=0\ , & \delta\eta=\frac{1}{2}[\phi,\lambda]\ .
\end{array}
\label{dwbrst}
\ee
The corresponding ghost numbers of the fields $(A, \phi, \l, \eta, \psi, \chi)$ are
$U=(0,2,-2,-1,1,-1)$. While it might not be
obvious that this action is related to the moduli space of instantons, this can be argued
as follows \cite{witten88a}. The gauge field terms in the action are
\beq
\frac{1}{4}\int_{X}d^{4}x\sqrt{g}{\rm Tr}[F_{\m\n}
F^{\m\n}+F_{\m\n}\tilde{F}^{\m\n}]=
\frac{1}{8}\int_{X}d^{4}x\sqrt{g}{\rm Tr}(F_{\m\n}+\tilde{F}_{\m\n})
(F^{\m\n}+\tilde{F}^{\m\n})\ ,
\eeq
and vanishes only if $F_{\m\n}=-\tilde{F}_{\m\n}$, which are exactly the instanton
solutions. These classical minima dominate since, as mentioned previously, the partition function can be
evaluated at weak coupling.

\subsubsection*{Twisting of ${\cal N}=2$}

We will now describe how
this theory can be constructed as a twisting of standard ${\cal N}=2$ supersymmetric
Yang-Mills theory with gauge group $SU(2)$ (see \cite{witten88a,witten9403} for further
details).

Start with the usual ${\cal N}=2$ supersymmetric Yang-Mills theory on flat ${\bf R}^4$.
Four-dimensional Euclidean space has a symmetry (or rotation) group which is
$K={\rm Spin}(4)=SU(2)_L\times SU(2)_R$. The internal symmetry group
of the ${\cal N}=2$ theory
is $SU(2)_I\times U(1)_R$, where the first group is the isospin group and the last group
corresponds to the
$R$-symmetry of the ${\cal N}=2$ Lagrangian (that transforms the gluino field as
$\l_{\a}\rightarrow e^{i\g}\l_{\a}$ and
$\bar{\l}_{\dot{\a}}\rightarrow e^{-i\g}\bar{\l}_{\dot{\a}}$ for example). On 
${\bf R}^4$ the global symmetry group of the theory is accordingly:
\beq
H=SU(2)_L\times SU(2)_R\times SU(2)_I\times U(1)_R\ .
\eeq 
The twisting amounts to a redefinition of the rotation group. If
$SU(2)'_R$ is the diagonal subgroup of $SU(2)_R\times SU(2)_I$ then instead
of $K$ we take as rotation group 
\beq
K'=SU(2)_L\times SU(2)'_R\ ,
\eeq
leaving $U(1)_R$ as the entire internal symmetry group.
Now let us see what happens to
the transformations of the fields under this redefinition.  The ${\cal N}=2$ algebra
has a set of supercharges $Q_{i\a}$ and $\bar{Q}_{i\dot{\a}}$ which transform
under $H$ (the $U(1)$ charge will not be important in the following) as $(1/2,0,1/2)$ and
$(0,1/2,1/2)$ respectively. They satisfy 
\bea\label{N=2alg}
\{ Q_{i\a}, \bar{Q}_{\dot{\b}}^{j}\}&=&
2\s^{m}_{\a\dot{\b}}P_{m}\delta_{i}^{j}\ ,\nonumber\\
\{ Q_{i\a}, Q_{j\b}\}&=&\epsilon_{ij}\epsilon_{\a\b}Z\ ,
\ena 
where $Z$ is the central charge.
Under the new symmetry group 
\beq
H'=SU(2)_L\times SU(2)'_R\times U(1)_R\ ,
\eeq  
the supercharges will transform as $(1/2,1/2)\oplus (0,1)\oplus (0,0)$ (and this follows
from the fact that under $SU(2)$ we have:
$\underline{2}\otimes\underline{2}=\underline{3}\oplus
\underline{1}$). Here, the BRST-like
operator $Q$ that we introduced before is identified with the 
$(0,0)$ component of the supercharge - it is the scalar operator
$Q=Q^{\dot{\a}}_{\ \dot{\a}}$. 
What about the condition
$Q^2=0$? After twisting this follows directly from the supersymmetry algebra (\ref{N=2alg}), at least when the 
central charge vanishes. However, even with a non-vanishing central charge, the theory
continues to be topological, since it is enough that $Q^2$ vanishes up to a gauge transformation \cite{danny}.

It is now a rather straightforward
matter to see how the action in (\ref{dw}) appears as the twisting of
the ${\cal N}=2$ theory.
The ${\cal N}=2$ theory with fields in the adjoint representation of $SU(2)$ has the
following field content \cite{wess}: a gauge field
$A_{\m}$, a complex scalar field $B$, two Majorana spinors $\l_{i \a}$, $i=1,2$ (with
$\l_1$ and
$\l_2$ forming a doublet under $SU(2)_I$), and their conjugates
$\bar{\l}_{i\dot{\a}}$. 
The action, in Minkowski space with metric $(-+++)$, is \cite{wess}:
\bea
S&=&\frac{1}{g^2}\int d^4x {\rm Tr}\left[ -\frac{1}{4}F_{\m\n}F^{\m\n}
-i\bar{\l}_{i}^{\dot{\a}}\sigma_{\a\dot{\a}}^{\m}D_{\m}\l^{\a i}
-D_{\m}\bar{B}D^{\m}B\right. \nonumber\\
&&\left. -\frac{1}{2}[B, \bar{B}]^2-\frac{i}{\sqrt{2}}\bar{B} \epsilon_{ij}
[\l^{\a i}, \l^{j}_{\a}]+\frac{i}{\sqrt{2}}B\epsilon^{ij}
[\bar{\l}_{\dot{\a} i}, \bar{\l}^{\dot{\a}}_{\ j}]\right] \ .
\ena
Here the
Yang-Mills field strength is $F_{\m\n}=\pa_{\m}A_{\n}-\pa_{\n}A_{\m}+[A_{\m},A_{\n}]$ and
the covariant derivative is $D_{\m}\Phi=(\pa_{\m}+iA_{\m})\Phi$.
\footnote{Also $\s^{\m}=(-1,\vec{\s})$, and $\bar{\s}^{\m}=(-1,-\vec{\s})$ in terms of which 
$\s^{\m\n\ \b}_{\ \ \a}=\frac{1}{4}(\s_{\a\dot{\a}}^{\ \ \m}\bar{\s}^{\n\dot{\a}\b}
-\s_{\a\dot{\a}}^{\ \ \n}\bar{\s}^{\m\dot{\a}\b})$. The spinor indices are raised and lowered with the antisymmetric
tensor, $\epsilon_{12}=\epsilon^{21}=-1$.} The supersymmetry transformations are
\bea\label{susytrans}
\delta A_{\m}&=&-i\bar{\l}^{\dot{\a}}_{i}\sigma_{\m\a\dot{\a}}\eta^{\a i}
+i\bar{\eta}^{\dot{\a}}_{i}\sigma_{\m\a\dot{\a}}\l^{\a i}\nonumber\ , \\
\delta\l_{\a}^{\ i}&=&\sigma^{\m\n}_{\ \ \a\b}\eta^{\b i}F_{\m\n}
+i\eta_{\a}^{\ i}D+i\sqrt{2}\sigma^{\m}_{\ \a\dot{\a}}D_{\m}B\epsilon^{ij}
\bar{\eta}^{\dot{\a}}_{\ j}\nonumber\ , \\
\delta\bar{\l}_{\dot{\a} i}&=& \bar{\sigma}^{\m\n}_{\ \
\dot{\a}\dot{\b}}\bar{\eta}^{\dot{\b}}_{\ i}F_{\m\n}-i\bar{\eta}_{\dot{\a}i}D+
i\sqrt{2}\sigma^{\m}_{\ \a\dot{\a}}D_{\m}\bar{B}\epsilon_{ij}\eta^{\a j}\nonumber\ ,\\
\delta B &=& \sqrt{2}\eta^{\a i}\l_{\a i}\nonumber\ , \\
\delta \bar{B}&=& \sqrt{2}\bar{\eta}^{\dot{\a}}_{\ i}\bar{\l}_{\dot{\a}}^{\ i}\ ,
\ena
with $D=[B,\bar{B}]$.
In passing from
$SU(2)_L\times SU(2)_R\times SU(2)_I$ to
$SU(2)_L\times SU(2)'_R$, the quantum numbers of the various fields that appear in this
action are changed as:
\bea
A_{\m}\ (1/2,1/2,0) \ &\rightarrow& \ (1/2, 1/2)\nonumber \\
\l_{i\a}\ (1/2, 0, 1/2) \ &\rightarrow& \ (1/2, 1/2)\nonumber \\
\bar{\l}_{i\dot{\a}}\ (0,1/2,1/2) \ &\rightarrow& \ (0,0)\oplus (0,1)\nonumber \\
B\ (0,0,0) \ &\rightarrow& \ (0,0)\ . 
\ena
In practice one is replacing the isospin indices
$i,j,\ldots$ by an
$SU(2)_R$ index
$\dot{\a}$. For the fields this means that the gauge field is unchanged
($A_{\m}\rightarrow A_{\m}$) and $(B, \bar{B})$ is related to $(\f, \l)$
in the twisted theory.
$\l_{i\a}$ becomes a vector $\psi_{\a\dot{\a}}$ and finally $\bar{\l}_{i\dot{\a}}$ is a sum
of a scalar ($\eta$) and a selfdual two-form ($\chi_{\m\n}$). More concretely, we will
make the following identifications in the topological theory:
\bea
\psi_{\m}&=&\sigma_{\m\, \a\dot{\b}}\l^{\a\dot{\b}}\nonumber\ , \\
B&=& -\frac{i\f}{2\sqrt{2}}\ ,\nonumber \\
\bar{B}&=& \sqrt{2}\l\ ,
\ena
while the scalar $\eta$ and selfdual two-form $\chi^{\m\n}$ are identified through:
\bea
\eta&=&-\frac{1}{2}\bar{\l}^{\dot{\a}}_{\ \dot{\a}}\nonumber\ , \\
\bar{\l}_{(\dot{\omega}\dot{\b})}&=& -2\epsilon^{\a\b}(\sigma_{\m})_{\a \dot{\omega}}
(\sigma_{\n})_{\b\dot{\b}}\chi^{\m\n}\ ,
\ena
where $(\dot{\omega}\dot{\b})=\dot{\omega}\dot{\b}+ \dot{\b}\dot{\omega}$.
It is possible to see that these identifications will produce all terms appearing in the
topological action. 
After twisting, and rotating to Euclidean signature, the fermion
kinetic term in the ${\cal N}=2$ action for example will give rise to the fermion kinetic terms
\beq
-i\eta D_{\m}\psi^{\m}+2iD_{\m}\psi_{\n}\chi^{\m\n}\ ,
\eeq
present in the Donaldson-Witten action. (The term $\int F\tilde{F}$ is a topological term
that can be added for free, since it changes neither the energy-momentum tensor nor the
equations of motion). As for the  BRST algebra (\ref{dwalg}) one starts by setting
\bea
\eta^{\a i}&=&0\ , \nonumber\\
\bar{\eta}^{\dot{\a}\dot{\b}}&=&-\epsilon^{\dot{\a}\dot{\b}}\rho\ ,
\ena
where $\rho$ is an anticommuting parameter. The supersymmetry transformations
(\ref{susytrans}) then become identical to the BRST transformations given in
(\ref{dwbrst}) -- though multiplied with $\r$ on the right hand side, so that the 
symmetry $\d$ becomes bosonic.

\subsubsection*{Observables}

We will now discuss the relevant observables in the Donaldson-Witten theory. 
Such observables are cohomology classes of the BRST operator, that is
operators $\op$ (which we require to be gauge invariant) such that $\{ Q, \op\}=0$ modulo
exact operators
$\op'=\{ Q, R\}$. In practice the further condition that $\delta_g\op=\{ Q, R\}$
will be satisfied by having simply $\op$ independent of the metric on $X$.

We already have a non-trivial obvious candidate in the BRST algebra (\ref{dwbrst}). The
field $\f$ is BRST closed but not BRST exact; also it is
metric independent. A gauge invariant expression is

\beq
W_0(x)=\half{\rm Tr}\f^2(x)\ ,
\eeq
where $x$ is a point in $X$, and can be viewed as a zero form on $X$.

This enables us to define a class of topological invariants on $X$ as
\beq
\langle W_0(x_1)\cdots W_0(x_k)\rangle=
\int [D\Phi ]e^{-S}\prod_{i=1}^{k}W_0(x_i)\ .
\eeq
It is trivial to verify that this expression is metric independent,
following the discussion in the introduction. 

While it might seem that this correlator depends on the distinct points
$x_1,\ldots x_k$ this is in fact not so. Starting with $W_0(x)$, we
can show that
its derivative with respect to the coordinate $x^{\m}$ is zero in the BRST
sense:
\beq
\frac{\pa W_0}{\pa x^{\m}}=\half\frac{\pa}{\pa x^{\m}}({\rm Tr}\f^2(x))
={\rm Tr}\f D_{\m}\f=i\{ Q, {\rm Tr}\phi\psi_{\m}\}\ .
\eeq 
Picking two points $x$ and $x'$ in $X$ we then have 
\beq\label{2p}
W_0(x)-W_0(x')=i\{ Q, \int_{x'}^{x}{\rm Tr}\f\psi_{\m}dx^{\m}\}\ ,
\eeq
or in infinitesimal form:
\beq\label{dw0}
dW_0=i\{ Q, W_{1}\}\ ,
\eeq
where $W_1$ is the operator valued one-form $W_1={\rm Tr}\f\psi$. It then
follows directly from (\ref{2p}) that 
\beq
\langle (W_0(x_1)-W_0(x'_1))\cdot\prod_{i=2}^{k}W_0(x_k)\rangle=0\ ,
\eeq
which was what we initially set out to show.

Proceeding in this fashion we can generate a small tower of observables,
$W_k$, which can be viewed as $k$ forms on $X$, by solving the following set
of equations:
\bea\label{descent}
dW_1 &=& i\{ Q, W_2\}\ , \; \; dW_2=i\{ Q, W_3\}\nonumber\ ,\\
dW_3 &=& i\{ Q, W_4\}\ , \; \; dW_4=0\ ,
\ena
(the last equation follows trivially from the fact that $X$ is four-dimensional) which together with
Eq. (\ref{dw0}) are the so-called descent equations.  The explicit
form of the operators $W_k$ can be computed by recursion; for illustrational
purposes we will demonstrate how $W_2$ can be determined:
\bea
dW_1&=&{\rm Tr}d(\phi\wedge\psi)={\rm Tr}(d\phi\wedge\psi+ \phi\wedge d\psi)
\nonumber\\
&=&i\{ Q, {\rm Tr}(\half \psi\wedge\psi+i\phi\wedge F)\}
\nonumber\\
&\equiv&i\{ Q, W_2\}\ .
\ena
In the second line we used the BRST variation of the gauge field
strength that follows from the BRST algebra: $\delta
F_{\m\n}=i(D_{\m}\psi_{\n}-D_{\n}\psi_{\m})$. The complete list of operators that
one obtains in this way is easily found:
\bea
W_1&=&{\rm Tr}(\phi\wedge\psi)\ , \; \; W_2={\rm
Tr}(\half\psi\wedge\psi+i\phi\wedge F)\nonumber\ ,\\
W_3&=&i{\rm Tr}(\psi\wedge F)\ , \; \; W_4=-\half{\rm
Tr}(F\wedge F)\ .
\ena
Note that $W_4$ integrated over $X$ is just the familiar instanton number
apart from a trivial factor. By inspection the ghost numbers of $W_k$ are $U=4-k$,
which of course also follows directly from Eq. (\ref{descent}).

The relevance of the descent equations is the following. If $C$ is a circle
in $X$ then the operator
\beq
I_1(C)=\int_{C}W_1
\eeq
is BRST invariant, since (\ref{dw0}) implies:
\beq
\{ Q, I_1(C)\}=\int_{C}\{ Q, W_1\}=-i\int_{C}dW_0=0\ .
\eeq
Also, $I_1(C)$ only depends on the homology class of $C$ (that is if $C$ is a
boundary then this observable is trivial). This follows from the first
equation in (\ref{descent}). Namely, if $C$ is the boundary of a surface, $C=\pa\Sigma$,
then
\beq
I_1(C)=\int_CW_1=\int_{\Sigma}dW_1=i\{ Q, \int_{\Sigma}W_2\}\ .
\eeq
The observable is consequently trivial (in the BRST sense) if $C$ is a
boundary. Likewise, if $\Sigma$ is any surface in $X$ then
\beq
I_2(\Sigma)=\int_{\Sigma}W_2
\eeq
is BRST invariant; if $K$ is a three-dimensional cycle in $X$ then
\beq
I_3(K)=\int_{K}W_3
\eeq
is BRST invariant and finally
\beq
I_4(M)=\int_{X}W_4
\eeq
is BRST invariant (and it is as stated before proportional to the instanton number).
As is the case for $I_1(C)$, all operators $I_k(\Gamma)$ only
depend on the homology class of the cycle $\Gamma$.

\subsubsection*{Donaldson Invariants and Polynomials}

We are now in a position to define the Donaldson invariants.
A natural assumption in Donaldson theory is -- as we have mentioned
already -- that the four-manifold $X$ is simply connected, or that the
fundamental group vanishes, $\pi_1(X)=0$. Then, the only possible nontrivial homology
cycles are 
$k$-dimensional homology cycles for $k=0,2,4$. For $k=4$, $I_4(X)$ is basically the
instanton number which is a rather trivial invariant, so the interesting
cases are $k=0$ or $k=2$. For $k=0$ the relevant operator is just $W_0(x)$
and for $k=2$ it is $I_2(S)$ where $S$ is a two-dimensional
surface in $X$.

The Donaldson
polynomials \cite{donaldson1,donaldson3} can now be described as certain polynomials in the homology 
class of $X$ (in this subsection we are using the same notation as in \cite{moore9709}):
\beq
{\cal D}_{E}:\ H_{0}(X,{\bf R})\oplus H_{2}(X,{\bf R}) \rightarrow {\bf R}\ .
\eeq
Here $E$ is an $SU(2)$ bundle over $X$. Given that $p \in H_{0}(X,{\bf R})$ is defined as
having degree 4 and 
$S \in H_{2}(X,{\bf R})$ degree 2 (i.e. identical to the ghost numbers of the
aforementioned observables), such a polynomial of degree $n$ is expanded as:
\beq
{\cal D}_{E}(p,S)=\sum_{2r+4s=n}S^rp^sq_{r,s}\ ,
\eeq
such that $n$ is the dimension of the instanton configurations on $E$ and $q_{r,s}$ are
rational numbers which are defined by certain intersection numbers on
the moduli space
(the details of which are not important for the discussion)
\footnote{As an example, for $X$ complex two-dimensional projective space ${\bf P}^2$, Witten and Moore
found \cite{moore9709} a complete expression for the Donaldson polynomials with ${\cal D}_{n=2}=-3S/2$, ${\cal
D}_{n=10}=S^5-pS^3-13p^2S/8$, etc.}.

A generating function for the Donaldson polynomials can be obtained by summing
over all bundles $E$, that is, all possible instanton numbers:
\beq\label{doPhi}
\Phi^{X,g}(p,S)\equiv\sum_{r\geq 0,s\geq
0}\frac{S^r}{r !}\frac{p^s}{s !}q_{r,s}\ .
\eeq
The connection to Witten's topological field theory is as follows.
Previously, we defined the
observables $W_0(p)$ and $I_2(S)$. The main result of Witten's seminal work
\cite{witten88a} is that the Donaldson invariants can be identified with the following
correlation functions:
\beq\label{drs}
q_{r,s}\equiv \langle W_0(x_1)\ldots W_0(x_s)I_2(S_1)\ldots
I_2(S_{r})
\rangle\ ,
\eeq
computed in Donaldson-Witten theory,
or phrased differently that the generating function for the Donaldson polynomials is
identified with:
\beq
\Phi^{X,g}(p,S) \equiv \langle e^{W(p)+I_2(S)}\rangle\ . 
\eeq
Formally, these correlation functions are by construction topological
invariants. However, it is possible to show 
that this is only so when $b_{2}^{+}>1$
(where 
$b_{2}^{+}$ is the dimension of the space of self-dual two-forms on $X$)
\footnote{From the point of view of Donaldson theory, the main reason for requiring $b_2^{+}>1$ is that it implies
a nonsingular moduli space, see \cite{donaldson1,donaldson3} for further
discussion.}. We will therefore assume this to be the case. 
  
Now, it is natural to ask under what conditions
such correlation functions in Eq. (\ref{drs}) are trivial? 

Generically, and depending on the number of points and surfaces, such a
correlation function will vanish because the violation of the ghost number does not
match the number of zero modes in the path integral.  The ghost
number of $W_0$ is $U=4$ and of $I_2(S)$ it is $U=2$; it follows that
the total ghost number of the correlation function in (\ref{drs}) is
$2r+4s$ - and this should be equal to the dimension of the instanton
moduli space ${\cal M}_k$. This dimension, on the other hand, is for
$SU(2)$ \cite{nash}:
\beq
{\rm dim} {\cal M}_k =8k-\frac{3}{2}(\chi+\sigma)\ ,
\eeq
with $n$ the instanton number and $\chi$ and $\sigma$ the Euler
characteristic and signature of $X$ respectively. 
The Euler characteristic is computed
as the alternating sum $\chi(X)=\sum_k (-1)^k b_k$, with $b_k={\rm dim}H^k(X)$, and the
signature as the difference between positive and negative eigenvalues of the
intersection form $Q$,
$\s(X)=b_2^{+}-b_2^{-}$
where $b_2^{+}$ ($b_2^{-}$) are the number of positive (negative) eigenvalues
of $Q$ -- this definition of $b_{2}^{+}$ coincides
with the above-mentioned.
Then, on a simply connected four-manifold $X$ we find $\chi+\sigma=2+2b_{2}^{+}$,
which is an even number.

\begin{figure}[htb]
\begin{center}
\mbox{
\epsfxsize=12cm
\epsffile{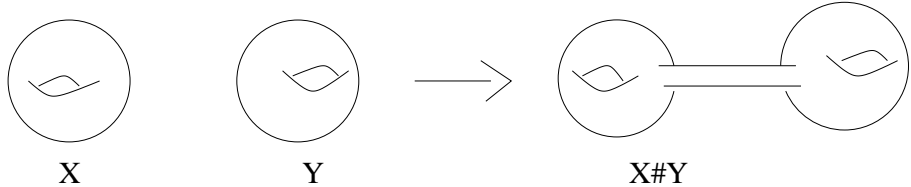}
}
\end{center}
\caption{The connected sum of two four-manifolds.}
\label{connected}
\end{figure}

The answer
to the question is therefore that the correlation function in (\ref{drs}) will
vanish unless
\beq
2r+4s={\rm dim} {\cal M}_k =8k-\frac{3}{2}(\chi+\sigma)\ .
\eeq 
This of course does not preclude that these invariants could be trivial for
another reason.
For example, they vanish on a manifold which is a connected sum $X\# Y$ with
$b_{2}^{+}>0$ on both $X$ and $Y$ \cite{donaldson1} 
(the connected sum $X\# Y$ of two four-manifolds is 
constructed by "cutting" out a three-ball of $X$ and $Y$ and then connecting them with
a "tube" $S^{3}\times I$, where $I$ is an interval, see
fig. \ref{connected}). 

Generally, however, the computation of the
invariants can be quite complicated since they require knowledge about a
space of $SU(2)$ instantons.

So historically, before the outcome of Seiberg-Witten theory, the Donaldson invariants where only known for few
manifolds (except where they are trivial) and on K\"ahler surfaces where that had been computed by Witten
\cite{witten9403} as correlation functions in an ${\cal N}=1$ Yang-Mills theory.

\section{Seiberg-Witten Theory in $D=4$}

So far we have presented a field theoretic approach to the Donaldson invariants which is
relevant at weak coupling ($g\rightarrow 0$) and can be obtained by
twisting the ${\cal N}=2$
theory. However, an important fact about ${\cal N}=2$ supersymmetric Yang-Mills theory is that
it is asymptotically free - it is weakly coupled in the ultraviolet limit and strongly
coupled in the infrared limit. So by analyzing the 
infrared behavior of the ${\cal N}=2$
theory, it should be possible to compute the
Donaldson invariants in a completely different way (since the
correlation functions in the topological theory are -- at least
formally -- independent of the coupling constant). 

A requisite for understanding this approach is provided 
by the work of Seiberg and Witten
\cite{witten9407,witten9408} in which they show that the infrared limit 
of the ${\cal N}=2$ theory is equivalent to a more tractable weak coupling limit of an Abelian theory.

These two theories can each be twisted to give topological quantum field theories. The
former is then related to Donaldson-Witten theory, or a theory of Donaldson
invariants. The latter should then be related to a much simpler Abelian
theory, or a theory of what is referred to as the Seiberg-Witten invariants. 

The general idea is therefore as follows: we have two dual moduli problems, one of
instantons (rather complicated) and one of Abelian monopoles (rather simple). 
Instead of computing the
Donaldson invariants from $SU(2)$ instanton solutions, one should be able 
to compute the same
invariants by using the solutions of the dual equations, which involve monopoles
of an Abelian $U(1)$ gauge theory.
\subsubsection*{The Seiberg-Witten Solution}

To understand the relation of the topological field theories to "physical"
theories, we will review a few facts about the solution of ${\cal N}=2$
supersymmetric Yang-Mills theory on ${\bf R}^4$ as described in
\cite{witten9407,witten9408}. Introductions to the Seiberg-Witten solution of the
${\cal N}=2$ theory can be found in
\cite{bilal9601,vecchia9803}. 

The pure
${\cal N}=2$ supersymmetric $SU(2)$ Yang-Mills action is:
\beq
S ={\rm Im\ tr}\int d^4x\frac{\t}{16\pi}\left[
\int d^2\theta W^\a W_\a 
+ \int d^2\theta
d^2\bar{\theta}\Phi^{\dagger}e^{-2gV}\Phi\right]\ ,
\eeq
here $\Phi$ is the chiral superfield, $V$ the vector superfield and $W_{\a}$ the spinor
superfield constructed from $V$; all fields are in the adjoint
representation of $SU(2)$, that is
$\Phi=\Phi^aT_a$ etc., where $\{ T_a\}$ is a set of generators of the
Lie algebra $su(2)$. 
Furthermore, $\tau$ is the complex coupling constant:
\beq
\tau = \frac{\theta}{2\pi}+\frac{4\pi i}{g^2}\ ,
\eeq
where $g$ is the Yang-Mills coupling and $\theta$ the QCD vacuum angle.
Classically, this theory
has a scalar potential
$V(\f)=\half{\rm tr}([\f^{\dagger}, \f ])$, $\f$ being the lowest scalar component of
$\Phi$. 
Unbroken supersymmetry requires
$V(\f)=0$, so the space of inequivalent vacua can be parametrized by a complex parameter
$u$, which is 
\beq\label{defu}
u=\langle {\rm tr}\ \f^2\rangle \ \; , \; \ \langle\f\rangle=\half a \s_3\ .
\eeq
$u$ is therefore a coordinate on the manifold of gauge inequivalent vacua, as it is easy to see
that one can always choose $\langle \f\rangle$ to be of the form in (\ref{defu}) with $a$ being a complex
constant.  The study of the ${\cal N}=2$ theory is basically the study of
the global structure of this moduli space and its singularities.

Classically, the moduli space is given by the complex plane -- or after
adding a point at infinity, the Riemann sphere. 
For $u\rightarrow \infty$, the theory becomes weakly coupled (because of
asymptotic freedom) and the $SU(2)$ gauge group is spontaneously broken down
to $U(1)$. For small $u$, where perturbation theory breaks down, the theory
gets strongly coupled and the gauge symmetry is $SU(2)$ at the
origin $u=0$. However, at $u=0$ the $W^{\pm}$ bosons become massless and there is no
description in terms of a Wilsonian effective action. So classically, the moduli space
looks like a Riemann sphere with two singularities at $u=0$ and $u=\infty$. 

\begin{figure}[htb]
\begin{center}
\mbox{
\epsfxsize=8cm
\epsffile{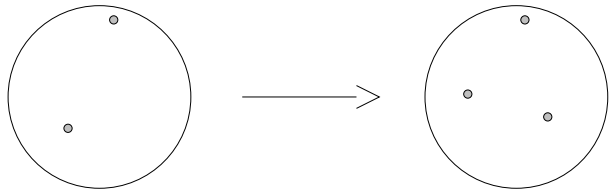}
}
\end{center}
\caption{The classical moduli space has singularities at $u=0,\infty$;%
the quantum moduli space at $u=\pm\L^2, \infty$.}
\label{moduli}
\end{figure}

According to the Seiberg-Witten solution, the quantum moduli
space looks like a Riemann sphere with singularities at $u=\Lambda^2,
-\Lambda^2$ and $\infty$, where $\Lambda$ is the scale of the ${\cal N}=2$
theory, see fig. \ref{moduli}.
But the $SU(2)$ gauge symmetry is never restored. Instead the
effective theory is that of an ${\cal N}=2$ supersymmetric Abelian gauge theory, which must be
of the general form
\beq\label{loweffsw}
S ={\rm Im}\int d^4x\frac{1}{16\pi}\left[
\int d^2\theta \frac{\pa^2{\cal F}}{\pa\Phi^2}W^\a W_\a 
+ \int d^2\theta
d^2\bar{\theta}\Phi^{\dagger}\frac{\pa{\cal F}}{\pa\Phi}\right]\ ,
\eeq
where ${\cal F}(\Phi)$ is the holomorphic prepotential, that determines the effective
coupling constant $\t$ as $\t=\pa^2{\cal F}/\pa\Phi^2$. What Seiberg
and Witten have achieved
is to determine
${\cal F}$ exactly in the quantum theory - which includes one-loop corrections and
instanton contributions - and thereby determined the complete low energy action of the
${\cal N}=2$ theory. 

At $u=\L^2$ the effective theory is an ${\cal N}=2$ supersymmetric Abelian gauge
theory coupled to a massless monopole (at $u=-\L^2$ it is coupled to a massless
dyon). And there is a ${\bf Z}_2$ symmetry ($u\rightarrow -u$) that relates the theories at these two
singularities, originating from the $U(1)_R$ symmetry of the classical action. 
These effective
theories are derived by the corresponding prepotential  which in turn is determined by the periods of
a meromorphic differential on the torus $\Sigma$ given by:
\beq\label{torus}
y^2=(x^2-\L^4)(x-u)\ .
\eeq
If $\a$ and $\b$ are the canonical basis homology cycles of the torus, and
\beq\label{swdiff}
\l=\frac{1}{\sqrt{2}\pi}\frac{x^2dx}{y(x,u)}\ 
\eeq
is the so-called Seiberg-Witten differential, then the result is as follows: the local coordinate
$a(u)$ around $u=\infty$ is:
\beq
a(u)=\int_{\a}\l\ ,
\eeq
while the coordinate $a_{D}(u)$ around $u=+\L^2$ is determined by:
\beq
a_D(u)=\int_{\b}\l\ .
\eeq
The upshot is that $a(u)$ and $a_D(u)$ are given by certain hypergeometric functions
(see e.g. \cite{bilal9601}), which in turn determine 
the exact prepotential ${\cal F}(a)$ according to:
\beq
a_{D}=\frac{\pa}{\pa a}{\cal F}(a)\ .
\eeq
The different low energy effective descriptions are connected by duality transformations. As an example,
in going from the description around $u=\infty$ to $u=+\L^2$ the effective complex coupling constant
$\t$ is changed by the $SL(2,{\bf Z})$-transformation $\t\rightarrow -1/\t$ (the full duality group is
actually $SL(2,{\bf Z})$, the same as the modular group of the torus $\Sigma$)
\footnote{This is not an exact duality of the theory. Instead the duality group is acting on the
various Lagrangian representations of the low energy effective behaviour of the theory.}. 

Why can this analysis be applied to a 
four-manifold $X$ in the topological theory? 
The reason is that in the twisted theory one can consider any
Riemann metric
$g$ on
$X$ since correlation functions are independent of the metric. In particular, we can
take the family of metrics $g_t=t^2g$, with $t>0$ and where $g$ is a fixed metric. Note that
$t$ large corresponds to large coupling constant.

For $t\rightarrow 0$ we get the Witten approach to Donaldson theory \cite{witten88a}. For
$t\rightarrow\infty$, on the other hand, one should expect that only the vacua of
${\bf R}^4$ are relevant (because the manifold now looks locally flat). 
Here, twisting the quantum theory near $u=\pm \L^2$ gives a topological quantum
field theory which is related to the moduli space of Abelian monopoles.
Actually,
it is possible to show \cite{moore9709} that for manifolds with $b_{2}^{+}>1$, only
contributions coming from $u=\pm \L^2$ are important - and that contributions away from
these singularities vanish as powers of
$t$ for $t\rightarrow\infty$. For $b_{2}^{+}=1$, there is a contribution to the
Donaldson invariants from an integral over the $u$-plane, which has been calculated
explicitly \cite{moore9709}, but we will only consider the case where
$b_{2}^{+}>1$, as it is much simpler to analyze.       

\subsubsection*{The Monopole Equations}

The theory around the monopole singularity is that of an ${\cal N}=2$ supersymmetric Abelian
gauge theory coupled to a massless hypermultiplet and the explicit
form of the associated topological Abelian field theory has been
found \cite{labas1} by twisting this theory as described earlier for the case of
Donaldson theory.  

While in Donaldson theory, one studies solutions to the instanton equations,
in the Seiberg-Witten approach one studies 
what has become known as the Seiberg-Witten monopole equations
\cite{witten9411} (see e.g. \cite{matilde,donaldson3} for a rather
mathematical introduction). 
The main feature is that they involve an Abelian gauge potential $A_{\m}$ and
a set of commuting Weyl spinors $M$ and $\overline{M}$, $\overline{M}$
being the hermitian conjugate of $M$. 

Strictly speaking, spinors can only be defined on manifolds which obey certain
conditions. (The second Steifel-Whitney class $w_2(X)\in H^2(X,{\bf Z}_2)$
should be trivial \cite{nakahara} implying for example that ${\bf C}P^2$ does not
admit spinors). But here one only needs a so-called Spin$^c$ structure, which can be
defined on any oriented four-manifold
\cite{matilde}
\footnote{A positive chirality spinor is, in mathematical terms, a section of the spinor bundle
$S^+$, this however might not be globally defined. The spinor $M$ appearing in the monopole
equations is a section of the Spin$^c$-bundle $S^+\otimes L^{1/2}$, where $L$ is the $U(1)$ line-bundle.
So in physical terms what we are dealing with are charged spinors.}. But we leave the
technical difficulties aside and assume that everything is working fine. 

Now, let
$X$ be an oriented, closed four-manifold with Riemann metric
$g_{\m\n}$. We choose Clifford matrices
$\G_{\m}$ on
$X$  (that is $\{ \G_{\m},\G_{\n}\}=2g_{\m\n}$) and define 
$\G_{\m\n}=\frac{1}{2}\left[ \G_{\m}, \G_{\n}\right]$. 
The conventions are such that $\G^{\m}=e^{\m}_{\a}\g^{\a}$, with vielbeins $e^{\a}_{\m}$
obeying $g_{\m\n}=e^{\a}_{\m}e^{\b}_{\n}\delta_{\a\b}$. A hermitian representation
of the Dirac matrices in flat ${\bf R^4}$ is given by 
\beq
\g^{0}=i\left(
\begin{array}{cc}0 & 1\\ -1 & 0\end{array}\right), \; \; \g^{k}=\left(
\begin{array}{cc}0 & \s^k\\ \s^k & 0\end{array}\right)\ , 
\eeq
with $k=1,2,3$. In this representation the chirality matrix takes the diagonal form,
\beq
\g_5=i^2\g^{0}\cdots \g^3=\left(
\begin{array}{cc}1 & 0\\ 0 & -1\end{array}\right)\ .
\eeq
The
Seiberg-Witten monopole equations are then:
\begin{eqnarray}\label{mp}
F^{+}_{\m\n}&=&-\frac{i}{2}\overline{M}\Gamma_{\m\n}M 
\nonumber\ , \\
D_{A}M&=& 0\ , 
\end{eqnarray}
where $D_{A}$ is the Dirac operator:
\beq
D_{A}=\G^{\m}D_{\m}=\Gamma^{\m}(\partial_{\m}+\omega_{\m}+iA_{\m})\ ,
\eeq 
which is twisted by the spin connection $\omega_{\m}
=\frac{1}{8}\omega_{\m ij}[\gamma^{i},\gamma^{j}]$ ($i,j=1,\ldots ,4$)
because we are now working on a
general (possibly non-flat) four-manifold. Note that there is a natural action of the gauge group
on the space of solutions to the equations in (\ref{mp}) by which $M$ is mapped to
$e^{i\s}M$ and $A_{\m}$ to $A_{\m}-\pa_{\m}\s$. This leaves the equations in (\ref{mp})
invariant.  

As discussed in \cite{labas1,carey} 
one can write a completely 
analogous topological field theory based on these monopole equations
(for a number of  reviews, see e.g. \cite{marino9701,labas2,labas9709}).
Let us use the notation of \cite{carey} where the topological action is

\be
S_{m}^{(4)}=\delta V_{m}^{(4)}\ ,
\ee
with ($\a,\b=1,\ldots ,4$):
\begin{eqnarray}
V_{m}^{(4)}&=&\int_{X}d^{4}x\sqrt{g}\left\{ \left[\na_{\alpha}\psi^{\alpha}
+\frac{i}{2}(\overline{N}M-\overline{M}N)\right]\lambda
-\chi^{\alpha\beta}(B_{\alpha\beta}
-F^{+}_{\alpha\beta}-\frac{i}{2}\overline{M}\Gamma_{\alpha\beta}M)\right.\nonumber \\
&&\left. -\overline{\mu}(\nu -iD_{A}M)-\overline{(\nu -iD_{A}M)}\mu\right\}\ .
\end{eqnarray}
The BRST algebra is:
\beq
\begin{array}{ll}
\delta A_{\alpha}=\psi_{\alpha}\ , &  \mbox{} \\
\delta\psi_{\alpha}=- \partial_{\alpha}\phi\ , &  \delta M=N\ , \\
\delta\phi=0\ , & \delta N=i\phi M\ , \\
\delta\chi_{\alpha\beta}= B_{\alpha\beta}\ , & \delta\mu=\nu\ , \\
\delta B_{\alpha\beta}=0\ , & \delta\nu=i\phi\mu\ , \\
\delta\lambda =\eta\ , & \delta\eta=0\ ,
\end{array}
\eeq
and as in the Donaldson theory, $\delta^{2}=0$ only up to a gauge transformation.
For instance, $\delta^{2}A_{\alpha}=-\partial_{\alpha}\phi$ which is the
variation of $A_{\alpha}$ under an infinitesimal gauge transformation
generated by $\phi$.
The corresponding ghost number assignments of the fields $(A,\f,\psi,M,N,\l,\eta,\chi,B,\m,\n)$ are 
$U=(0,2,1,0,1,-2,-1,-1,0,-1,0)$ -- with $(\l,\eta)$ the anti-ghost multiplet and Lagrange multiplier fields
$(\chi,B)$ and
$(\m,\n)$.
Using these transformation rules one finds
the following expression for the topological action
in four dimensions \cite{carey}:
\begin{eqnarray}\label{SWaction}
S_{m}^{(4)}&=&\int_{X}d^{4}x\sqrt{g}\left\{
\left[ -\Delta\phi+\overline{M}M\phi-i\overline{N}N\right]\lambda
-\left[ \na_{\alpha}\psi^{\alpha}
+\frac{i}{2}(\overline{N}M-\overline{M}N)\right]\eta
+2i\phi\overline{\mu}\mu\right. \nonumber \\
&&-\chi^{\alpha\beta}\left[ (\na_{\alpha}\psi_{\beta}
-\na_{\beta}\psi_{\alpha})^{+}
+\frac{i}{2}(\overline{M}\Gamma_{\alpha\beta}N
+\overline{N}\Gamma_{\alpha\beta}M)\right]\nonumber \\
&&+\frac{1}{4}(F_{\alpha\beta}^{+}
+\frac{i}{2}\overline{M}\Gamma_{\alpha\beta}M)^{2}
+\frac{1}{2}\overline{D_{A}M}D_{A}M\nonumber \\
&&\left. +\overline{(iD_{A}N-\Gamma^{\alpha}\psi_{\alpha}M)}\mu
-\overline{\mu}(iD_{A}N-\Gamma^{\alpha}\psi_{\alpha}M)\right\}\ ,
\end{eqnarray}
where the Lagrange multipliers $B$ and $\nu$ 
have been eliminated by their equations of motion, that is
\bea
B_{\a\b}&=&\half(F_{\a\b}^++\frac{i}{2}\overline{M}\G_{\a\b}M)\nonumber\ ,\\
\n&=&\half iD_AM\ ,
\ena
and the bar indicates
hermitian conjugation. In this form, it is clear that the dominant contribution to the
functional integral coming from the bosonic part of the action is given by the solutions
of the monopole equations (\ref{mp}). 

While in Donaldson theory we have a moduli space which is characterized by its instanton
number, the moduli in question is characterized by a monopole charge. 

The Weyl spinor $M$ being charged under $U(1)$, here we must consider a
$U(1)$ principle bundle over the four-manifold $X$ with an associated line bundle $L$. 
Topologically a $U(1)$ bundle
is characterized by the first Chern class
\beq
c_1(F)=[F/2\pi ] \in H^{2}(X, {\bf Z})\ .
\eeq
Introducing a basis $\Sigma_i$ of $H_2(X,{\bf Z})$ the monopole charges can then be
obtained as the total magnetic flux
\beq
m_i=\int_{\Sigma_{i}}c_1\ ,
\eeq
integrated over the surface $\Sigma_i$. 
Often the notation $x=-2c_{1}(L)$ is also used. The
moduli space -- with fixed monopole number
$x$ -- of solutions to the monopole equations modulo gauge transformations is
denoted by
${\cal M}_x$. The dimension of this moduli space can be determined by an index theorem
\cite{witten9411} and is
\beq\label{mpdim}
d=-\frac{2\chi +3\sigma}{4}+c_1(L)^2\ ,
\eeq
where again $\chi$ is the Euler characteristic and $\sigma$ is the signature of $X$. 
It is possible to show that the moduli space is a compact (and oriented) manifold
\cite{matilde}. Indeed,
$|M|$ is bounded by the scalar curvature of $X$ -- and there are accordingly no
square-integrable solutions on flat
${\bf R}^4$. 
The (virtual) dimension of the moduli space vanishes, i.e. $d=0$, exactly when:
\beq\label{x2}
x^2=2\chi+3\sigma\ , 
\eeq
and, because of compactness, the moduli space will then consist of a {\sl finite} number
of points denoted by
$P_{i,x}$, $i=1,\ldots t_x$. 
\subsubsection*{Seiberg-Witten Invariants}

Now, because of orientability, 
with each such point $P_{i,x}$ one can associate a sign $\epsilon_{i,x}=\pm 1$. For each
$x$ for which the virtual dimension is zero, i.e. for which Eq. (\ref{x2}) holds, one can
define an integer $n_x$ as
\beq
n_x=\sum_{i}\epsilon_{i,x}\ .
\eeq
These quantities $n_x$ are the celebrated Seiberg-Witten invariants. For $b_{2}^{+}>1$
they constitute a set of diffeomorphism invariants of four-manifolds \cite{matilde}
\footnote{In order to show topological invariance one has to show that the $n_{x}$ is constant on a path connecting two
metrics. Invariance can then fail if there are singularities and they appear if the gauge group does not act freely
on the space of solutions. A solution with $M=0$ has $F^{+}=0$, i.e. is an Abelian instanton, that can be identified
with an element in $H^2_{-}(X,{\bf R})$. Now, $F/2\pi\in H^{2}(X, {\bf Z})$, 
so $F/2\pi\in H^{2}(X,{\bf Z})\cap H^{2}_{-}(X, {\bf R})$ -- and this is generically empty for $b_2^{+}>1$
\cite{matilde}. To resume: when  $b_2^{+}>1$ there are no Abelian instantons.}. Because of a vanishing argument,
described later, a given four-manifold will only have a finite number of
$x's$ for which $n_{x}\neq 0$. On manifolds for which there are only trivial
solutions to the monopole equations, these invariants will of course all vanish. 

With these invariants at hand, a number of interesting results can be
obtained -- some completely new results and some
often in a much simpler way than with the Donaldson invariants. 

First of all, the partition
function of Donaldson-Witten theory can be calculated at strong coupling with the result
( see e.g. \cite{labas9709}):

\beq\label{SWpart}
Z=c\sum_{x}\delta(x^2-2\chi-3\sigma)\left[ n_x + i^{\Delta}n_x \right]\ ,
\eeq 
where $\Delta=(\chi+\sigma)/4$ and the constant $c$ in front can be fixed by
requiring agreement with the result at weak coupling \cite{witten9403} and is  a topological number:
\beq
c=2^{1+\frac{1}{4}(7\chi+11\sigma)}\ .
\eeq
The delta function in (\ref{SWpart}) means that only zero-dimensional
moduli spaces contribute. The second factor is a contribution coming from two parts: one
from the singularity at $u=+\L^2$ and one from the one at $u=-\L^2$. 

If the $u$-plane had more than these two singularities, the result in
(\ref{SWpart}) would have been radically different.

Secondly, it is natural to ask how these Seiberg-Witten invariants are related to the
Donaldson invariants? 
\footnote{The conjecture to be presented below in Eq. (\ref{dw=sw}) would seem to indicate that the
Seiberg-Witten invariants should contain more information than the Donaldson invariants since the
latter have been derived from a zero-dimensional moduli space ${\cal M}_x$ (and no
knowledge about the positive dimension moduli spaces has been used). But since correlation functions in the
topological theory are -- at least formally -- independent of the coupling constant we should expect the invariants
to contain exactly the same information.} 
In Donaldson theory -- at least for simply connected
$X$ -- there are two important observables, namely an operator $I(\Sigma)$ of ghost
number two (where $\Sigma$ is any two-dimensional homology cycle) and an operator of
dimension four $W_0$. 

The generating function for the Donaldson invariants is
\beq\label{generating}
\langle \exp \left( \sum_i\a_iI(\Sigma_i)+\lambda W_0\right)\rangle\ ,
\eeq 
where it is understood that one is summing over instanton numbers. Here
$\Sigma_i$ is a basis of
$H_2(X, {\bf Z})$, i.e. $i=1,\ldots ,{\rm dim}H_2(X, {\bf Z})$, and
$\l, \a_i$ are complex numbers.  

Using the same notation as in \cite{witten9411}, we define $v=\sum_{i}\a_i [\Sigma_i]$;
here $[\Sigma_i]$ is the element in $H^2(X, {\bf Z})$ which is Poincar\'e dual to
$\Sigma_i$;
$v^2=\sum_{i,j}\a_i\a_j \Sigma_i\cdot\Sigma_j$, where
$\Sigma_i\cdot \Sigma_j$ is the intersection number of $\Sigma_i$ and $\Sigma_j$
\footnote{The intersection number of $\Sigma_i$ and $\Sigma_j$ is the number of points
in $\Sigma_i\cap\Sigma_j$ counted with orientation.}.
Also
we define $v\cdot x=\sum_i\a_i(\Sigma_i, x)$ for any  $x\in H^2(X, {\bf Z})$.  

For manifolds of simple type, that is one for which the generating function in
(\ref{doPhi}) obeys $\pa^2\Phi/\pa\l^2 -4\Phi=0$, the following relation has been derived by Witten
\cite{witten9411}:
\bea\label{dw=sw}
\langle \exp \left( \sum_i\a_iI(\Sigma_i)+\lambda W_0\right)\rangle
&=&2^{1+\frac{1}{4}(7\chi+11\sigma)}\left( 
\exp \left(\frac{v^2}{2}+2\l\right) \sum_{x}n_{x}e^{v\cdot x}\right. \nonumber \\
&& \left. \; \; \; \; \; \; +i^{\Delta}\exp\left(
-\frac{v^2}{2}-2\l\right)\sum_{x}n_{x}e^{-iv\cdot x}\right)\ .
\ena
For a sketch of the derivation, see \cite{labas9709}.
As for the
partition function there is one important comment. The first term on the right hand side
\beq
\exp \left(\frac{v^2}{2}+2\l\right) \sum_{x}n_{x}e^{v\cdot x}\ ,
\eeq
is the contribution from the $u$-plane singularity at $u=\L^2$; the second term on the
right hand side
\beq
\exp\left(
-\frac{v^2}{2}-2\l\right)\sum_{x}n_{x}e^{-iv\cdot x}\ ,
\eeq
comes from the singularity at $u=-\L^2$. If the
vacuum structure had been different from the one predicted by Seiberg and Witten
\cite{witten9407,witten9408} the resulting relation would have been very different with additional
terms from other singularities.

Note that these results depend crucially on the connection to the quantum field theory formulation
of Donaldson theory and are not proved rigorously. So there is at least no mathematical proof of the conjectured
relation between Donaldson and Seiberg-Witten invariants.  

However, the form of (\ref{dw=sw}) agrees with a result proved by Kronheimer and Mrowka for manifolds of simple
type \cite{mrow}, in which the $n_x$ were unknown coefficients. 

Witten was able to fix the coefficient $2^{1+\frac{1}{4}(7\chi+11\sigma)}$ by requiring agreement with computations
on K\"ahler manifolds and other manifolds where the Donaldson invariants can be computed explicitly \cite{witten9403}.
It has later been shown that the resulting formula agrees with all cases, where the Donaldson invariants are known.

But one does not necessarily need the relation -- as conjectured from quantum field theory -- between Donaldson and
Seiberg-Witten invariants. Indeed, completely independent of this, the task of analyzing the solutions of the
monopole equations and the connected Seiberg-Witten invariants is a well-defined mathematical problem,
that has been shown to lead to many interesting results in topology. (And this 
is one reason why they have proven to be so important in the mathematics literature).

As an example, we could mention the celebrated proof of the Thom conjecture for embedded surfaces
in ${\bf CP}^2$ by Kronheimer and Mrowka \cite{kronheimer}. By studying the monopole equations on 
the four-manifold $X={\bf R}\times {\bf S}^1\times \Sigma$,
Kronheimer and Mrowka showed that if $\Sigma$ is an oriented two-manifold embedded in ${\bf CP}^2$
and representing the same homology class as an algebraic curve of degree $d$, then the genus of
$\S$ satisfies: $g\geq (d-1)(d-2)/2$. For further discussion, see \cite{matilde,donaldson3}.

As a further check Witten has been able to compute the invariants exactly on K\"ahler
manifolds, where they are non-vanishing \cite{witten9411}.

\subsubsection*{Vanishing Theorems}

Among the most important applications of the Seiberg-Witten equations are the so-called
vanishing theorems that follow from (\ref{mp}). 

From a strictly mathematical point of view, these vanishing theorems
can be rigorously derived from (\ref{mp}) without the use of physical
arguments. 

The derivation begins by defining
\beq
s_{\m\n}=F^{+}_{\m\n}+\frac{i}{2}\overline{M}\Gamma_{\m\n}M\ , \; \; 
k^{\a}=(D_{A}M)^{\a}\ . 
\eeq
A solution of the monopole equations obeys of course:
\beq
\int_{X}d^4x \sqrt{g}\left( \half|s|^2+|k|^2 \right)= 0\ .
\eeq
One can rewrite this in an interesting way by using that the square of the Dirac
operator is (the Lichnerowicz-Weitzenbock formula):
\bea
D_{A}^2&=&\Gamma^{\m}\Gamma^{\n}D_{\m}D_{\n}\nonumber\\
&=& \left(\half \left\{ \Gamma^{\m},\Gamma^{\n}\right\}
+\half [\Gamma^{\m},\Gamma^{\n}]\right)D_{\m}D_{\n}\nonumber \\
&=&D^{\m}D_{\m}+\frac{i}{2}\Gamma^{\m\n}F_{\m\n}-\frac{R}{4}\ ,
\ena
where $R$ is the scalar curvature, and using one of the Fierz
identities (details can be found in
\cite{thompson1}): 
\bea\label{va1}
\int_{X}d^4x \sqrt{g}\left( \half|s|^2+|k|^2 \right)&=&
\int_{X}d^4x\sqrt{g}\left(\half |F^{+}|^2+g^{\m\n}D_{\m}M\overline{D_{\n}M}\right.
\nonumber\\ 
&& \left. +\half |M|^4+\frac{1}{4}R|M|^2\right)\ .
\ena
An immediate consequence is the following vanishing theorem: if $(A,M)$ is a solution to
(\ref{mp}) and the scalar curvature of $X$ is positive, $R\geq 0$, then
\beq
F_{\m\n}^{+}=0\ , \; \; M=0\ ,
\eeq
i.e. the only solutions are from the Abelian instanton equations. This in turn implies
\cite{witten9411}
that a four-manifold $X$ with $b_{2}^{+}>0$ and non-vanishing Seiberg-Witten invariants,
that is $n_x\neq 0$ for some $x$, cannot have a metric with positive scalar curvature. 

As another application of such vanishing arguments, Witten has shown \cite{witten9411}
that the Seiberg-Witten invariants vanish on manifolds which are connected sums $X\# Y$
when $b_{2}^{+}>0$ on both $X$ and $Y$, see fig. \ref{connected}. 

The curvature scalar $R$ can be taken positive on such a tube and any solution of the
Seiberg-Witten equations can therefore be brought to vanish on it, at least when the
tube is taken to be very long. Then one can define a $U(1)$ action on the moduli space of
solutions. This is obtained by gauge transforming the solutions on $X$ with a constant
gauge transformation that keeps the fields on $Y$ fixed. The fixed points of this action
are solutions with
$M=0$ on $X$ or
$Y$, but for $b_{2}^{+}>0$ one can argue that the only such solutions are the trivial
ones. So we have a free action on a set of points and this set must therefore be empty.

In particular it follows that four-dimensional K\"ahler manifolds cannot be obtained as 
connected sums, since Witten showed \cite{witten9411} that K\"ahler manifolds have nontrivial invariants.

As stated previously, 
for a given four-manifold $X$, there will only be a finite number
$x$'s for which the Seiberg-Witten invariants $n_x\neq 0$. This is derived from the following vanishing
theorem \cite{witten9411}. Throwing away the $|DM|^2$-term in (\ref{va1}) we have,
\beq
\int_{X}d^4x\sqrt{g}\half |F^{+}|^2 \leq
- \int_{X}d^4x\sqrt{g}\left( \half |M|^4+\frac{1}{4}R|M|^2\right)\ .
\eeq 
Combined with the obvious inequality
\beq
\int_{X}d^4x\sqrt{g}\left( \half |M|^4+\frac{1}{4}R|M|^2\right)
\geq -\frac{1}{32}\int_{X}d^4x\sqrt{g}R^2\ ,
\eeq
we find
\beq
\int_{X}d^4x\sqrt{g}|F^{+}|^2 \leq 
\frac{1}{16}\int_{X}d^4x\sqrt{g}R^2\ .
\eeq
This shows that $\int_{X}d^4x\sqrt{g}|F^{+}|^2$ is bounded for a class $x$ with $n_x\neq 0$. The same
is true for $\int_{X}d^4x\sqrt{g}|F^{-}|^2$ since we have from (\ref{mpdim}):
\bea
\frac{1}{4}x^2=c_{1}(L)^2&=& \frac{1}{(2\pi)^2}\int_{X}d^4x\sqrt{g}F^2\nonumber\\
&=&
\frac{1}{(2\pi)^2}\int_{X}d^4x\sqrt{g}\left(|F^{+}|^2- |F^{-}|^2\right) =
\frac{2\chi+3\s}{4}\ .
\ena
One can argue that there are only finitely many line-bundles $L$ for which both $\int_{X}d^4x\sqrt{g}|F^{\pm}|^2$ are
bounded \cite{matilde} -- and hence for every four-manifold $X$, there will only be a finite number of
non-trivial invariants $n_x$.

A number of similar vanishing theorems can also be derived in the
lower-dimensional versions of Seiberg-Witten theory.

\section{Seiberg-Witten Duality in $D<4$}

Having considered the Seiberg-Witten duality applied to topological theories in four
dimensions, it becomes natural to ask what happens in lower dimensions?
Dimensionally reducing the four-dimensional theory by taking $X=X_{n}\times T^{4-n}$ for
$n=2, 3$ with the radius of the compact directions going to zero, one obtains
dimensionally reduced theories in three and two dimensions which by construction are
topological \cite{ols}.

Such dimensional reductions of Donaldson-Witten theory
have been known for a long time \cite{birm,chap} (and is
briefly reviewed below). As far as topological properties are
concerned, the analogous dimensional reductions of the four-dimensional
{\em dual theory} should provide new Abelian topological theories
which are duals of the dimensionally reduced
Donaldson-Witten theories. 

As in \cite{ols} we start by dimensionally reducing the Donaldson-Witten theory in four
dimensions with an action given in (\ref{dw}). Concretely, we take
$X$ to be a product manifold $X=Y\times {\bf S}^{1}$
with signature (++++) and assume that all fields are
$x^{0}$-independent. 
Here $Y$ is a compact and oriented three-manifold.
Furthermore, we
define $\chi^{i}\equiv \chi^{0i}$ such that
$\chi^{ij}=\epsilon^{ijk}\chi_{k}$.
This gives the three-dimensional action ($i,j,k =1,2,3$),
\begin{eqnarray}\label{3ddw}
S^{(3)}&=&\int_{Y}d^{3}x\sqrt{g}\, {\rm Tr}[\frac{1}{4}F_{ij}
F^{ij} + \frac{1}{2}F_{ij}\tilde{F}^{ij}
+\frac{1}{2}D_{i}\varphi_{0}D^{i}\varphi_{0}
-\frac{1}{2}[\varphi_{0},\phi ][\varphi_{0},\lambda ]
+\frac{1}{2}\phi D_{i}D^{i}\lambda \nonumber \\
&&-i\eta D_{i}\psi^{i}
-i\eta[\varphi_{0},\psi_{0} ]
+2i\epsilon^{ijk}(D_{i}\psi_{j})\chi_{k}
+2i[\varphi_{0},\psi_{i} ]\chi^{i} +2i\psi_{0}D_{i}\chi^{i}\nonumber \\
&&
-\frac{i}{2}\lambda[\psi_{0},\psi_{0}]
-\frac{i}{2}\lambda[\psi_{i},\psi^{i}]
-\frac{i}{2}\phi[\eta ,\eta ]
-\frac{1}{8}[\phi ,\lambda]^{2}]\ ,
\label{s3}
\end{eqnarray}
where we defined $A_{0}\equiv \varphi_{0}$ and 
$\tilde{F}_{ij}=\epsilon_{ijk}F_{0k}=-\epsilon_{ijk}D_{k}\varphi_{0}$.

The reduction to two dimensions is obtained by assuming that
the three manifold $X$ is a product  manifold of the form
$Y=\Sigma\times {\bf S}^{1}$ and $x^{1}$-independence of all
fields ($\mu,\nu =2,3$):
\begin{eqnarray}\label{2ddw}
S^{(2)}&=&\int_{\Sigma}d^{2}x\sqrt{g}\, {\rm Tr}[\frac{1}{4}F_{\mu\nu}F^{\mu\nu} 
+\frac{1}{2}D_{\mu}\varphi_{0}D^{\mu}\varphi_{0}
+\frac{1}{2}D_{\mu}\varphi_{1}D^{\mu}\varphi_{1}
+\frac{1}{2}[\varphi_{1},\varphi_{0}]^{2}\nonumber \\
&&-\frac{1}{2}[\varphi_{0},\phi ][\varphi_{0},\lambda ]
-\frac{1}{2}[\varphi_{1},\phi ][\varphi_{1},\lambda ]
+\frac{1}{2}\phi D_{\mu}D^{\mu}\lambda
-i\eta[\varphi_{0},\psi_{0}] -i\eta[\varphi_{1},\psi_{1}]\nonumber \\
&&-i\eta D_{\mu}\psi^{\mu}
+2i\epsilon^{\mu\nu}(D_{\mu}\psi_{\nu})\chi
+2i\epsilon^{\mu\nu}[\varphi_{1},\psi_{\mu}]\chi_{\nu}
-2i\epsilon^{\mu\nu}(D_{\mu}\psi_{1})\chi_{\nu}\nonumber \\
&&+2i[\varphi_{0},\psi_{1} ]\chi
+2i[\varphi_{0},\psi_{\mu} ]\chi^{\mu}
+2i\psi_{0}D_{\mu}\chi^{\mu}
-2i[\varphi_{1},\psi_{0}]\chi \nonumber \\
&&-
\frac{i}{2}\lambda[\psi_{0},\psi_{0}]
-\frac{i}{2}\lambda[\psi_{1},\psi_{1}]\nonumber \\
&&-\frac{i}{2}\lambda[\psi_{\mu},\psi^{\mu}]
-\frac{i}{2}\phi[\eta ,\eta ]
-\frac{1}{8}[\phi ,\lambda]^{2}\nonumber \\
&&+ \frac{1}{2}[\varphi_{0},\varphi_{1}]\epsilon^{\mu\nu}F_{\mu\nu}
+ \epsilon^{\mu\nu}D_{\mu}\varphi_{1}D_{\nu}\varphi_{0}]\ ,
\label{s2}
\end{eqnarray}
where we defined $A_{1}\equiv \varphi_{1}$ and $\chi_{1}\equiv \chi$.
Though rather complicated this action can be rewritten in a somewhat simplified form by
introducing the complex scalar field $\Phi=\varphi_{0}+i\varphi_{1}$. One can then write
the action as:
\begin{eqnarray}
S^{(2)}&=&\int_{\Sigma}d^{2}x\sqrt{g}\, {\rm Tr}
[\frac{1}{4}(F_{\mu\nu}- \frac{1}{2}i\epsilon_{\mu\nu}[\Phi ,\Phi^{*}])^{2}
+\frac{1}{2}D_{\mu}\Phi D_{\mu}\Phi^{*}\nonumber \\
&&-\frac{1}{2}[\varphi_{0},\phi ][\varphi_{0},\lambda ]
-\frac{1}{2}[\varphi_{1},\phi ][\varphi_{1},\lambda ]
+\frac{1}{2}\phi D_{\mu}D^{\mu}\lambda
-i\eta[\varphi_{0},\psi_{0}] -i\eta[\varphi_{1},\psi_{1}]\nonumber \\
&&-i\eta D_{\mu}\psi^{\mu}
+2i\epsilon^{\mu\nu}(D_{\mu}\psi_{\nu})\chi
+2i\epsilon^{\mu\nu}[\varphi_{1},\psi_{\mu}]\chi_{\nu}
-2i\epsilon^{\mu\nu}(D_{\mu}\psi_{1})\chi_{\nu}\nonumber \\
&&+2i[\varphi_{0},\psi_{1} ]\chi
+2i[\varphi_{0},\psi_{\mu} ]\chi_{\mu}
-2i\psi_{0}D_{\mu}\chi^{\mu}
-2i[\varphi_{1},\psi_{0}]\chi\nonumber \\
&&-\frac{i}{2}\lambda[\psi_{0},\psi_{0}]
-\frac{i}{2}\lambda[\psi_{1},\psi_{1}]\nonumber \\
&&-\frac{i}{2}\lambda[\psi_{\mu},\psi_{\mu}]
-\frac{i}{2}\phi[\eta ,\eta ]
-\frac{1}{8}[\phi ,\lambda]^{2}]\ .
\label{s2.2}
\end{eqnarray}
It is easy to check, that the resulting action is a 
BRST gauge fixing of the anti--self-duality
equation in four dimensions, $F_{\alpha\beta}=
- \frac{1}{2}\epsilon_{\alpha\beta\gamma\delta}F^{\gamma\delta}$,
reduced to two dimensions:
\bea\label{hitchin}
F_{\mu\nu}&=& \frac{1}{2}i\epsilon_{\mu\nu}\left[\Phi
  ,\Phi^{*}\right]\ ,
\nonumber\\
D_{\m}\Phi&=&0\ .
\ena
These equations have been studied, in the context of Riemann surfaces, by
Hitchin \cite{hitchin} (though he mainly concentrated on the case
where the gauge group is $SO(3)$, rather than $SU(2)$). 
The main points following from Hitchin's analysis are:
(1) that the moduli space of solutions modulo gauge transformations is a smooth
noncompact manifold ${\cal M}$ of dimension $12(g-1)$, where $g$ is the genus of the
Riemann surface; (2) that  there is a vanishing theorem related to the solutions of
(\ref{hitchin}), similar to the vanishing theorem of Donaldson theory in four dimensions
and (3) that it is possible to prove the uniformization theorem: that every
compact Riemann surface of genus $g\geq 2$ admits a metric of constant negative
curvature. It would be nice if one could give a simple proof of this uniformization
theorem by using the two-dimensional version of the Seiberg-Witten equations.

However, it would take us to far astray to go into detail with all this, but the main
point is that the moduli space of solutions to (\ref{hitchin}) is an object which has
some relevance in the mathematics literature.
Also Chapline and Grossman
\cite{chap} has been considering these equations, thereby indicating a connecting of 
conformal field
theory to  Donaldson theory. This seems to indicate a possible physical relevance of
these equations, however it is not clear whether their results have any signification 
relation to the analogous dimensionally reduced monopole equations in two dimensions
\footnote{The Hitchin equations also naturally appear in two-dimensional BF gravity \cite{danny} as equations of
motion of the zwei-bein and spin-connection. But we will not try to relate this to the two-dimensional 
monopole equations.}.

Now we can turn to the analogous dimensional reduction of the dual
theory, which is an Abelian gauge theory with action given in (\ref{SWaction}). 
By taking $X=Y\times {\bf S}^{1}$ as before the dimensionally reduced action becomes:
\begin{eqnarray}\label{SW3d}
S_{m}^{(3)}&=&\int_{Y}d^{3}x\sqrt{g}\left\{
\left[ -\Delta\phi+\overline{M}M\phi-i\overline{N}N\right]\lambda
-\left[ \na_{k}\psi^{k}
+\frac{i}{2}(\overline{N}M-\overline{M}N)\right]\eta
+2i\phi\overline{\mu}\mu\right. \nonumber \\
&&-2\chi^{k}\left[ -\partial_{k}\psi_{0}+\epsilon_{kij}(\na_{i}\psi_{j})
-\overline{M}\G_{k}N
-\overline{N}\G_{k}M\right]\nonumber \\
&&+\frac{1}{8}(F_{ij}-\epsilon_{ijk}\pa_{k}\varphi_{0}
-\epsilon_{ijk}\overline{M}\G_{k}M)^{2}
+\frac{1}{2}\overline{(D_{A}+\varphi_{0})M}(D_{A}+\varphi_{0})M\\
&&+\left. \overline{( i(D_{A}+\varphi_{0})N-(\G^{k}\psi_{k}-i\psi_{0})M)}\mu
-\overline{\mu}( i(D_{A}+\varphi_{0})N-(\G^{k}\psi_{k}-i\psi_{0})M)\right\}\nonumber\ ,
\end{eqnarray}
where $\chi^{i}\equiv \chi^{0i}$ and
$\G^k=e^k_{\hat{s}}\sigma^{\hat{s}}$, $\hat{s}=1,2,3$, are the Dirac
matrices in three dimensions.

Generally, computing Donaldson invariants on $Y\times {\bf S}^1$, with
the radius of the ${\bf S}^1$ going to zero, one would expect to
obtain invariants of the three-manifold $Y$. 

Using Donaldson-Witten theory, the partition function related to  (\ref{dw})
on $Y\times {\bf S}^1$ computes the so-called Rozansky-Witten
invariant of $Y$ \cite{rozansky}.
On the Seiberg-Witten side, 
it has been shown that the 
partition function of the three-dimensional theory (\ref{SW3d}) gives a Seiberg-Witten
version of the so-called Casson invariant \cite{carey} (as discussed
in \cite{marcos} this, however, only holds when $b_{1}(Y)>1$). A further discussion of this
three-dimensional case, which also discusses a non-Abelian version of the Seiberg-Witten
monopoles can be found in \cite{ohta9611}. 

The three-dimensional version of the monopole equations can be obtained from the local
minima of the classical part of the action in (\ref{SW3d}). These equations are
accordingly:
\begin{eqnarray}\label{mp3d}
F_{ij}-\epsilon_{ijk}\overline{M}\G_{k}M &=& 0\ , \nonumber\\
D_{A}M &=& 0\ , \\
\varphi_{0} &=& 0\ ,\nonumber
\end{eqnarray}
but of course they could have been derived directly by dimensionally reducing the
four-dimensional monopole equations (\ref{mp}). In (\ref{mp3d}) the last condition is
only necessary if we have a nontrivial solution. Otherwise, it can be replaced by the
condition
$d\varphi_{0}=0$. 

Similarly, making a reduction to two dimensions - with $Y=\Sigma\times {\bf S}^{1}$ - 
results in the following action:
\begin{eqnarray}\label{SW2d}
S_{m}^{(2)}&=&\int_{\Sigma}d^{2}x\sqrt{g}\left\{
\left[ -\Delta\phi+\overline{M}M\phi-i\overline{N}N\right]\lambda
-\left[ \na_{\mu}\psi^{\mu}
+\frac{i}{2}(\overline{N}M-\overline{M}N)\right]\eta
+2i\phi\overline{\mu}\mu\right. \nonumber \\
&&-2\chi^{\mu}\left[ -\partial_{\mu}\psi_{0}
+\epsilon_{\mu\nu}\partial_{\nu}\psi_{1}
-\overline{M}\G_{\mu}N
-\overline{N}\G_{\mu}M\right]
-2\chi\left[ \epsilon_{\mu\nu}(\na_{\mu}\psi_{\nu})
-\overline{M}\sigma_{1}N-\overline{N}\sigma_{1}M\right]\nonumber \\
&&+\frac{1}{8}(F_{\mu\nu}
-\epsilon_{\mu\nu}\overline{M}\sigma_{1}M)^{2}
+\frac{1}{4}\partial_{\mu}\varphi_{0}\partial_{\mu}\varphi_{0}
+\frac{1}{4}\partial_{\mu}\varphi_{1}\partial_{\mu}\varphi_{1}
+\frac{1}{2}\epsilon_{\mu\nu}\partial_{\mu}\varphi_{1}
\partial_{\nu}\varphi_{0}
\nonumber\\
&&+\frac{1}{2}\partial_{\mu}\varphi_{1}\epsilon_{\mu\nu}
\overline{M}\G_{\nu}M
+\frac{1}{2}\partial_{\mu}\varphi_{0}\overline{M}\G_{\mu}M
+\frac{1}{4}(\overline{M}\G_{\mu}M)^{2}
\nonumber \\
&&+\frac{1}{2}\overline{(D_{A}+\varphi_{0}+i\varphi_{1}\sigma_{1})M}
(D_{A}+\varphi_{0}+i\varphi_{1}\sigma_{1})M\nonumber \\
&& +\overline{( i(D_{A}+\varphi_{0}+i\varphi_{1}\sigma_{1})N-(\G^{\mu}\psi_{\mu}
-i\psi_{0}+\psi_{1}\sigma_{1})M)}\mu\nonumber \\
&&\left. -\overline{\mu}( i(D_{A}+\varphi_{0}+i\varphi_{1}\sigma_{1})N
-(\G^{\mu}\psi_{\mu}-i\psi_{0}+\psi_{1}\sigma_{1})M)\right\}\ ,
\end{eqnarray}
here $\G^{\m}=e^{\m}_{\hat{\r}}\s^{\hat{\r}}$, $\hat{\r}=2,3$, are the corresponding Dirac matrices in two
dimensions. This reduction gives rise to a two-dimensional topological theory, as 
one can check that the resulting two-dimensional action obeys 
$S_{m}^{(2)}=\delta V_{m}^{(2)}$. Here, $V_{m}^{(2)}$ is the
dimensional reduction of $V_{m}^{(4)}$, i.e.
\begin{eqnarray}
V_{m}^{(2)}&=&\int_{X}d^{2}x\sqrt{g}\left\{ \left[\na_{\mu}\psi^{\mu}
+\frac{i}{2}(\overline{N}M-\overline{M}N)\right]\lambda
-2\chi(2H
-\frac{1}{2}\epsilon_{\mu\nu}F_{\mu\nu}+\overline{M}\sigma_{1}M)
\right.\nonumber \\
&&-2\chi_{\mu}(2H_{\mu}+\epsilon_{\mu\nu}\partial_{\nu}\varphi_{1}
-\partial_{\mu}\varphi_{0}-\overline{M}\G_{\mu}M)\nonumber \\
&&\left. -\overline{\mu}(\nu 
-i(D_{A}+\varphi_{0}+i\varphi_{1}\sigma_{1})M)-
\overline{(\nu -i(D_{A}+\varphi_{0}+i\varphi_{1}\sigma_{1})M)}\mu\right\}.
\end{eqnarray}
We have defined $\chi\equiv\chi^{1}$, $H^{\mu}\equiv H^{0\mu}$ and $H\equiv H^{1}$. 

Computing Donaldson invariants on $\Sigma\times {\bf S}^1$, with the radius of the circle going to
zero, one would -- as in the three-dimensional case -- expect to obtain invariants of the
two-manifold $\Sigma$. However, when $\Sigma$ is a compact orientable surface its topology is
uniquely characterized by a single integer, the genus $g$, so any non-trivial
topological invariant will be a function of $g$ and hence contains 
at most as much information as the function $f(g)=g$. So (at least a priori) nothing interesting
seems to be obtained in this direction.

As for the monopole equations they are either inferred from reducing the
three-dimensional monopole equations further to two dimensions, or 
as the minima of the
classical part of the action (\ref{SW2d}), which is:
\begin{eqnarray}
S_{0}&=&\frac{1}{8}(F_{\mu\nu}-\epsilon_{\mu\nu}\overline{M}
\sigma_{1}M)^{2}+\frac{1}{4}(\partial_{\mu}\varphi_{1})^{2}
+\frac{1}{2}(\partial_{\mu}\varphi_{0})^{2}
+\frac{1}{4}(\overline{M}\G_{\mu}M)^{2}\nonumber \\
&&+\frac{1}{2}|D_{A}M|^{2}+\frac{1}{2}|\varphi_{0}M|^{2}
+\frac{1}{2}|\varphi_{1}\sigma_{1}M|^{2}\ .
\end{eqnarray}
The two-dimensional variant of the Seiberg-Witten equations are consequently as follows:
\begin{eqnarray}\label{mp2d}
F_{\mu\nu}-\epsilon_{\mu\nu}\overline{M}\sigma_{1}M &=& 0 \nonumber\ ,\\
D_{A}M &=& 0 \nonumber\ ,\\
\overline{M}\G_{\mu}M &=& 0\ ,\nonumber\\
\varphi_{0} &=& 0 \nonumber\ ,\\
\varphi_{1} &=& 0\ .
\end{eqnarray}
If $(A,M)$ is a trivial solution, then the last two conditions
can be replaced by $d\varphi_{0}=d\varphi_{1}=0$.

Though very similar to Hitchin's self-duality equations these equations of course
describe a totally different moduli space: the former is a moduli space of solutions to
a $U(1)$-problem while the latter is related to $SU(2)$ "instantons". 
However, as is the case in four dimensions, it should be possible to obtain -- in a
simple way -- result obtained by studying solutions of the Hitchin equations, in terms of the
moduli space corresponding to monopole equations as (\ref{mp2d}). To my knowledge, this has not been
done in the literature.

However, a number of vanishing theorems, similar to those previously considered
in four dimensions can be derived in this context of a two-dimensional surface $\Sigma$ 
\cite{ols,kronheimer}.  
In fact, it follows from Eq. (\ref{mp2d}), that if
$(A,M)$ is a solution of the two-dimensional monopole equations 
then the pair must obey the following identity \cite{ols}
\be
\int_{\Sigma}d^{2}x\sqrt{g}(\frac{1}{4}|F|^{2}+\overline{D^{\mu}M}D_{\mu}M+
\frac{1}{2}|\overline{M}\sigma_{1}M|^{2}+\frac{1}{4}R|M|^{2})=0\ ,
\ee
where $R$ is the scalar curvature.
If there is a metric so that $R$ is positive on $X$ then this implies
that $F_{\mu\nu}=0$ and $M=0$ are the only solutions. On a sphere,
for example, we are actually looking at flat Abelian connections.
One might therefore naively worry that on a surface of genus $g$ there are only trivial
solutions. However, a surface of genus $g\geq2$ admits a metric of constant 
{\sl negative} curvature and the argument does not apply. 
 
Without assuming any positivity of the scalar curvature one can also derive the
following inequality: 
\beq\label{f2r2}
\int_{\Sigma}d^{2}x\sqrt{g}\frac{1}{4}|F|^{2} 
\leq \frac{1}{32}\int_{\Sigma}d^{2}x\sqrt{g}R^{2}\ .
\eeq
Inserting this in (\ref{mp2d}) on gets an upper bound on $|M|^4$ and this implies that
the moduli space is compact as in four dimensions. 

Another variant of such vanishing arguments shows that if $\Sigma$ is a genus $g$ surface
(taken to be of unit area and constant scalar curvature $-4\pi(2g-2)$) then the first
Chern number is bounded as \cite{kronheimer}:
\beq
|c_1(\Sigma)|=|\frac{1}{2\pi}\int_{\Sigma}F|\leq 2g-2\ .
\eeq
This result (which basically is just (\ref{f2r2}) in another disguise) plays an important
role in the proof of the Thom conjecture by Kronheimer and Mrowka \cite{kronheimer}.

Finally, let us mention that some explicit solutions to 
the monopole equations on ${\bf R}^{2}$ have been constructed 
in \cite{nergiz}
and the solutions turn out to be vortex configurations.
They are singular, as are the analogous solutions, given 
by Freund \cite{freund}, in ${\bf R}^3$. As noted by Witten in \cite{witten9411},
the monopole equations admit no square-integrable solutions on flat 
${\bf R}^{n}, n\leq4$.\\

The Seiberg-Witten equations have been generalized to non-Abelian monopoles, mainly by
Labastida and Mari\~no and is reviewed in \cite{marino9701}. Furthermore, the Donaldson
invariants have been computed by, e.g., Moore and Witten on four-manifolds with
$b_{2}^{+}=1$ \cite{moore9709} and by Mari\~no and Moore on non-simply connected manifolds
\cite{moore2}. In the latter case the "invariants" are not really invariants since
they are not constant functions on the space of metrics but only piecewise constant.

The generalization to non-Abelian monopoles is especially interesting since
mathematicians are studying these to come up with a mathematical proof of 
the equivalence of Donaldson-Witten and Seiberg-Witten invariants. 
The idea being that both the instanton  and
the Abelian Seiberg-Witten moduli space appear as boundaries of a so-called non-Abelian $PU(2)$-moduli 
space and that some cobordism argument may then relate the two, see \cite{teleman1}.


\chapter{T-Duality in String Theory and in Sigma Models}

In this chapter we will be focusing on one duality, namely that of $T$-duality (a useful
reference is \cite{porra}), and we will be analyzing some consequences imposed by this duality
in a variety of sigma models (both bosonic, supersymmetric and heterotic models). These
consequences can be formulated as a certain relation between $T$-duality and the renormalization
group flow (operator $R$) of such models, namely that they commute: $[T,R]=0$. 

We will start by describing $T$-duality as a perturbative (order by order) symmetry of string
theory. Then we consider the restrictions of scale and Weyl invariance for
consistent string propagation. Such invariances are not mandatory for general
two-dimensional sigma models which we treat in the rest of the chapter but are related to the
renormalization group (RG) flow of such models.  Accordingly, in the following sections we
study the relation between
$T$-duality and RG flow -- as defined by the beta functions --in bosonic sigma models, and the
extend to which our "hypothetical" relation, $[T,R]=0$, determines the exact RG flow. Then
we treat the case of supersymmetric and heterotic sigma models in a simplified setting.   
In both cases it turns out that duality implies strong constraints on the RG flow.

\sect{Introduction}

$T$-duality is one of the most important dualities in string theory. It
was first discovered in the context of toroidal compactifications of
closed strings as an invariance under the change of compactification
radius from $R$ to $\alpha'/R$ \cite{kikkawa}. Later it was shown that
this symmetry appears not only in toroidal compactifications, but in
all target space backgrounds with isometries \cite{bus1,rocek}. 

The main property of string theory that enables a $T$-duality is that, in a space with compact
dimensions, strings can wrap around nontrivial loops. At the same time, the momentum of the
string must be quantized along these compact directions. $T$-duality is basically a
symmetry under interchange of these wrapping and momentum modes. 

To set the scene we start with the free string action describing a closed string moving
in flat Minkowski spacetime

\beq\label{maction}
S_0=\frac{1}{4\pi\alpha'}\int_{\Sigma} d^{2}\sigma[\partial_aX^{\mu}
\partial^aX_{\mu} + {\rm fermions}]\ ,
\eeq
where $T=\frac{1}{2\pi\alpha'}$ is the string tension. The parameter
$\a'$ plays the role of Planck's constant such that quantum mechanical
perturbation theory for strings is an expansion in $\a'$. The classical limit then corresponds
to $\a'$ small.  The integration is over the worldsheet $\Sigma$ which we take to be
without boundary and orientable - this is the description relevant for
closed strings. The local coordinates on this worldsheet are $(\s,\t)$ with $0\leq \s \leq
2\pi$ and periodic and with $-\infty < \t < \infty$.  In (\ref{maction}), the 
spacetime coordinates $X^{\mu}(\t, \s)$ describe the embedding of the string in
spacetime. The fermion terms  
depend on which kind of string theory we are considering. For the bosonic string such model
dependent terms are absent.

In the case of the superstring,
the action contains fermionic degrees of freedom $\psi^{\mu}$ residing on the worldsheet. 
As we go around the periodic direction, the fermions can either be periodic
or anti-periodic thereby giving rise to a total of four different sectors as the left- and
right-moving modes are treated independently. 
The Ramond (R) fermion is periodic, while the Neveu-Schwarz
(NS) fermion is anti-periodic. The NS-NS sector contains massless spacetime bosons,
namely a graviton ($g_{\mu\nu}$), an antisymmetric tensor ($b_{\mu\nu}$)
and a scalar dilaton ($\phi$). The fundamental string is charged
under the two-form $b_{\mu\nu}$. The R-R sector also contains bosons,
which are antisymmetric tensor fields $C_{p}$. Such fields will not play any role in this
chapter, but are relevant for the understanding of non-perturbative string theory as discussed
in Chapter 4. For completeness, we should mention that the R-NS and NS-R sectors gives the
spacetime fermions.

Two classical symmetries of the action in (\ref{maction}) will be important in the
following: the action is invariant under diffeomorphisms
of the worldsheet, which is just a change in the coordinates $(\s, \t)
\rightarrow (\s', \t')$ and also invariant under local Weyl symmetry by which is meant a
scaling of the worldsheet metric according to $h_{ab}\rightarrow 
e^{2\omega(\s, \t)}h_{ab}$. On a flat worldsheet this is a conformal
symmetry in which case it implies that the classical energy-momentum tensor
vanishes:
\beq\label{EMtensor}
T_{ab}=-\frac{2\pi}{\sqrt{h}}\frac{\d S_0}{\d h^{ab}}=0\ .
\eeq

We have started with a theory of strings moving in 26-dimensional
spacetime (ten-dimensional for the superstrings).
To make the connection with the real (observed) world, we should consider
compactifications of string theory. This means that we take our
spacetime to be a product $M^{26-d}\times K^{d}$ where usually 
$M^{26-d}$ is flat $(26-d)$-dimensional Minkowski space and $K^d$ is some
compact manifold. The simplest
compactification is compactification on a circle, $K^{d}={\bf S}^1$.  
Choosing $X^{25}$ as the coordinate curcumnavigating the circle, we must identify
$X^{25}\sim X^{25}+2\pi R$ since the wave function should be single
valued. Hence, the center of mass momentum is quantized 
in units of $1/R$ along this direction:
$P^{25}=n/R$. 
The string can wind around the circle any number of
times so that if we go along the compactified dimension the string coordinate
$X^{25}$ does not have to come back to itself, rather 
\beq\label{2pcond}
X^{25}(\t, \s+2\pi)=X^{25}(\t, \s)+2\pi mR\ , \; \; m\in {\bf Z}\ .
\eeq
The two integers $(n,m)$ are called momentum and
winding modes respectively and are conserved charges. 

Let us analyze what the appearance
of these modes implies for the mode expansion of the fields $X^{\m}$. The equation of motion
following from the action (\ref{maction}) is a wave equation that implies that the fields can
be written as a sum of a left and a right-moving part:
\beq
X^{\m}=X^{\m}_{L}(\t+\s)+X^{\m}_{R}(\t-\s)
\eeq
and the periodicity of the $\s$-coordinate implies that the mode expansions are
\bea\label{mode}
X^{\m}_{L}&=&x^{\m}_{L}+\a' p^{\m}_{L}(\t+\s)
+i\sqrt{\a'}\sum_{n\neq 0}\frac{1}{n}\a^{\m}_{n}e^{-in(\t+\s)}\nonumber\ , \\
X^{\m}_{R}&=&x^{\m}_{R}+\a' p^{\m}_{R}(\t-\s)
+i\sqrt{\a'}\sum_{n\neq 0}\frac{1}{n}\tilde{\a}^{\m}_{n}e^{-in(\t-\s)}\ ,
\ena
where $x^{\m}_{L}+x^{\m}_{R}=x^{\m}_{CM}$ is the center of mass position and
$p^{\m}_{L}+p^{\m}_{L}=P^{\m}$ is the total center of mass momentum. With the periodicity
conditions (\ref{2pcond}) on the $X^{25}$ coordinate we can then write
\beq
X^{25}=x^{25}_{CM}+\a' \frac{n}{R}\t + mR\s + {\rm osc.}
\eeq
It then follows that
\bea\label{p25}
p_{L}^{25}&=&\frac{1}{2}\left( \frac{n}{R}+\frac{mR}{\a'}\right)\ , \nonumber \\
p_{R}^{25}&=&\frac{1}{2}\left( \frac{n}{R}-\frac{mR}{\a'}\right)\ .
\ena
Now, the spectrum of the string theory
is determined by the mass-shell condition \cite{gsw}: 
\bea\label{2spectrum}
E^2&=&P^2+(\frac{n}{R}-\frac{mR}{\alpha'})^2+\frac{4}{\alpha'}N_L\nonumber \\
&=&P^2+(\frac{n}{R}+\frac{mR}{\alpha'})^2+\frac{4}{\alpha'}N_R\ ,
\ena
where $P$ is the momentum in the noncompact directions and $N_{L}$ and $N_{R}$ are the
oscillator levels of the string.  

It follows from (\ref{2spectrum}) that the spectrum 
is invariant under the interchange of momentum and winding modes 
($n\leftrightarrow m$), if we change $R$ according to
\beq\label{2rdual} 
R \rightarrow \alpha'/R\ .
\eeq
This symmetry of string theory is called target space duality (or
$T$-duality for short). 
The result in (\ref{2rdual}) means that compactification on a small radius
$(R/\sqrt{\alpha'}\ll 1)$ is equivalent to compactification on a large
radius $(R/\sqrt{\alpha'} \gg 1)$. 
This is of
course very interesting and different from the situation in field theory since a particle will
have a spectrum which is basically only determined by the momentum in the compact direction -
there is no such thing as a winding number! 

Also $T$-duality seems to suggest that
there is a minimum length scale in string theory: by using one or the other formulation we can
restrict to $R \geq\sqrt{\a'}$.

Another important interpretation of $T$-duality is that it can be
seen as a parity transformation on the right-moving coordinates: in
terms of the left- and
right-moving momentum (\ref{p25}), the
$T$-duality transformation is simply:
\beq
p^{25}_{L}\leftrightarrow p^{25}_{L}, \; \; \; p^{25}_{R}\leftrightarrow -p^{25}_{R}\ .
\eeq
Without changing the spectrum, one can also change the oscillator
modes according to
\beq
\a_{n}^{25} \leftrightarrow \a_{n}^{25} , \; \; \; 
\tilde{\a}_{n}^{25} \leftrightarrow -\tilde{\a}_{n}^{25}\ . 
\eeq
This means -- because of (\ref{mode}) -- that $T$-duality can be viewed as the transformation
\beq\label{Tparity}
X^{25}=X^{25}_{L}+X^{25}_{R} \; \; \rightarrow X'^{25}=X^{25}_{L}-X^{25}_{R}\ ,
\eeq
which is a spacetime parity transformation that acts only on the
right-movers (in the case of superstrings, this would be supplemented
with the transformation of the worldsheet fermion $\psi_{R}\rightarrow
-\psi_{R}$, in the directions where duality is performed). 

So far
we have considered free string theory. However, the $T$-duality transformation also acts
nontrivially on the string coupling constant, which is determined by the dilaton as $g=\langle
e^{\f}\rangle$. The transformation can be determined by noting that the 25-dimensional coupling constant
$g_{25}$ is related to the 26-dimensional coupling by
\beq
g^2=2\pi Rg_{25}^2\ ,
\eeq
and must be invariant under duality (since the $T$-duality only acts in the 26'th direction), so
\beq
g^2 \rightarrow g'^2=\frac{2\pi\a'}{R}g^2_{25}\ .
\eeq
In conclusion $T$-duality must act as
\beq\label{2tdual}
R \rightarrow R'=\alpha'/R\ , \; \; g \rightarrow g'=\frac{g\sqrt{\alpha'}}{R}\ .
\eeq
These transformations were first derived by Buscher \cite{bus1} by requiring duality to be a
symmetry of the full string theory with
interactions.
Moreover, since
the coupling constant is basically unchanged, this duality
transformation maps the weak coupling limit of one theory to the weak
coupling limit of another theory. As an example, it is seen that this procedure yields a
duality between Type IIA string theory compactified on a circle of radius $R$ and Type IIB theory
compactified on a circle of radius $\alpha'/R$ \cite{dine89,dai89}.\\ 

The description relevant for 
strings moving in a more general background than a flat Minkowski
spacetime can be obtained given a nontrivial vacuum expectation value for the 
massless NS-NS bosons $g_{\mu\nu}, b_{\m\n}$ and $\phi$. These can
be incorporated by using the vertex operators of the massless
fields \cite{gsw}. This gives the following 
generalization of (\ref{maction}) as the worldsheet action describing a string
moving in a curved background:
\beq\label{sigmaction}
S=\frac{1}{4\pi\alpha'}\int_{\Sigma} d^2\sigma\sqrt{h}\,\left[ h^{ab}g_{\mu\nu}(X)
+\frac{i\epsilon^{ab}}{\sqrt{h}}
b_{\mu\nu}(X)\right] \partial_{a}X^{\mu}\partial_{b}X^{\nu}
+\frac{1}{4\pi}\int_{\Sigma} d^2\sigma\sqrt{h}\, R^{(2)}\phi(X)\ .
\eeq
Here $h_{ab}$ is the worldsheet metric, $g_{\mu\nu}(X)$ is the background metric,
$b_{\mu\nu}(X)$ is the antisymmetric tensor and $\phi(X)$ is the dilaton field.
$R^{(2)}$ is the worldsheet Ricci scalar, so the last term is only
relevant 
on a curved worldsheet. 
$X^{\mu}(\s)$ gives a map $\S \rightarrow M$ from the worldsheet into
the target space $M$ and can conveniently be thought of as defining local 
coordinates on $M$. $M$ is then some Riemann manifold with metric
tensor $g_{\mu\nu}$.

Historically, an action of the form (\ref{sigmaction}) arise in the context of what is called
a {\sl non-linear sigma model}. In contrast to the Minkowski space action (\ref{maction}), this
action (\ref{sigmaction}) is no longer quadratic in
$X^{\mu}$ and accordingly it describes an interacting two-dimensional field theory. 


However, not every such background -- as defined by $g_{\mu\nu}$,
$b_{\mu\nu}$ and $\phi$ -- yields a consistent string
theory. The values of these background fields are restricted by demanding
local scale invariance or conformal invariance.
\subsubsection*{Scale and Weyl Invariance}

Let us start by considering invariance under global scale transformations.
Classically (\ref{maction}) and (\ref{sigmaction}) are invariant under scale transformations. This
is not necessarily true in the corresponding quantum theory as is seen by considering the scale transformation of the
flat worldsheet metric:
\beq
\delta_{\epsilon}h_{ab}=\epsilon h_{ab}\ .
\eeq
The resulting change in the partition function is
\beq
-\frac{\epsilon}{2\pi}\int d^2\sigma \langle T^{a}_{\ a}(\sigma)\rangle\ ,
\eeq
where $T^{ab}$ is the energy-momentum tensor. Scale invariance therefore 
requires $T^{a}_{\ a}=\partial_{a}{\cal O}^a$, for ${\cal O}^a$ some local
operators.  In classical field theory scale invariance is insured if the coupling constants are
dimensionless (the Lagrangian contains no dimensionfull parameters). 
However divergences in the
quantum theory gives rise to a non-vanishing renormalization group (RG) beta function with
the consequence that the effective coupling depends on the length scale. For the sigma
models in question, the relevant beta functions (or actually beta functionals)
are 
\beq\label{betadef}
\b^{g}_{\m\n}\equiv\L\frac{d}{d\L}g_{\m\n}\ ,\;
\b^{b}_{\m\n}\equiv\L\frac{d}{d\L}b_{\m\n}\ ,\;
\b^{\f}\equiv\L\frac{d}{d\L}\f\ ,
\eeq
where $\L$ is the appropriate renormalization scale parameter. 
Note that Eq. (\ref{betadef}) describe a change in
geometry.

The
renormalization properties of the non-linear sigma model in (\ref{sigmaction}) has been studied
extensively in the literature, see e.g.
\cite{freed,fried}. 

In string theory, scale invariance alone does not ensure consistency. For this also Weyl
invariance is required. 
Classically, the two first terms in (\ref{sigmaction}) are also invariant under Weyl rescalings of
the worldsheet metric, while the last term breaks it explicitly. 
Consider now a Weyl
transformation of the worldsheet metric
\beq
\delta_{\omega} h_{ab}=2\omega(\sigma)h_{ab}\ .
\eeq
The resulting change in the partition function is
\beq
-\frac{1}{2\pi}\int d^2\sigma\sqrt{h(\sigma)}\omega(\sigma)
\langle T^a_{\ a}(\sigma)\rangle \ .
\eeq
So Weyl invariance requires $T^a_{\ a}=0$, which evidently is a stronger condition than scale
invariance. In the
classical theory, given by (\ref{maction}), this trace is identically zero. Indeed, this follows from the fact that the
classical energy momentum tensor vanishes (\ref{EMtensor}). In the quantum theory
$T^{a}_{\ a}$ does not vanish and it implies that there is an anomaly in the local worldsheet
symmetry. This is the so-called Weyl anomaly. The propagation of the string is only consistent
if this anomaly vanishes (in the light-cone gauge it can
for example be shown that the theory
is only Lorentz invariant if the anomaly is vanishing \cite{ooguri9612}).

As a side remark we note that $T^{a}_{\ a}$ ought to vanish on a flat
worldsheet where we have conformal invariance. This 
determines 
\beq
T^a_{\ a}=a_1R^{(2)}\ ,
\eeq
where $a_1$ is a constant and $R^{(2)}$ is the Ricci scalar of the worldsheet. The
computation of $a_{1}$ is straightforward and the result is \cite{pol9411}:
\beq\label{a1}
a_1 = -\frac{c}{12}\ ,
\eeq
where $c$ is the total central charge of the worldsheet conformal field theory. So
Weyl invariance is the same as demanding that the total central charge
$c$ is zero. This in turn determines the dimension of flat spacetime: the bosonic string
propagates in $D=26$ dimensions and the superstring in $D=10$ dimensions.\footnote{The total
central charge of the bosonic string conformal field theory is the sum of the matter central
charge and ghost central charge:
$c=c_{X}+c_{g}$. The ghost conformal field theory is a $bc$-theory with central charge
$c_g=-26$, so
$c=c_{X}-26$. Each free field $X^{\m}$ adds $+1$ to the total central charge and consequently there must
be 26 of such fields for the theory to be Weyl invariant \cite{gsw}.}

The sigma model in (\ref{sigmaction}) is - as indicated above - not
conformally invariant for all backgrounds. The necessary conditions on the
background can be calculated using dimensional regularization in
$2+\epsilon$ dimensions and calculating those terms of the action that
violate the symmetry at the quantum level in the limit $\epsilon \rightarrow
0$. 
The possible values of the couplings are then determined by requiring the vanishing of these terms.
In terms of the so-called Weyl anomaly coefficients $\bar{\b}^g_{\m\n}$, $\bar{\b}^b_{\m\n}$
and $\bar{\b}^{\f}$ one finds \cite{callan}:
\beq\label{wanomaly}
T^{a}_{\ a}=-\frac{1}{2\alpha'}
(\bar{\beta}^{g}_{\mu\nu}\partial_aX^{\mu}\partial^aX^{\nu}
+\bar{\beta}^{b}_{\mu\nu}
\frac{\epsilon^{ab}}{\sqrt{h}}\partial_aX^{\mu}\partial_bX^{\nu}
+\alpha'\bar{\beta}^{\phi}R^{(2)})\ .
\eeq
The Weyl anomaly
coefficients can obtained using
the weak coupling expansion of the sigma model i.e., as a perturbative expansion in
$\a'$ . To first order in $\a'$ these equations have the following
form \cite{tsey1,tsey2}:  
\bea
\bar{\beta}^{g}_{\mu\nu}&=&\alpha'R_{\mu\nu}
+2\alpha'\nabla_{\mu}\partial_{\nu}\phi-\frac{\alpha'}{4}H_{\mu\lambda\rho}
H_{\nu}^{\ \lambda\rho}\label{barg}\ ,\\
\bar{\beta}^{b}_{\mu\nu}&=&-\frac{\alpha'}{2}\nabla^{\rho}H_{\rho\mu\nu}
+\alpha'\nabla^{\rho}\phi H_{\rho\mu\nu}\label{barb}\ ,\\
\bar{\beta}^{\phi}&=&\frac{D-26}{6}-\frac{\alpha'}{2}\nabla^2\phi
+\alpha'\nabla_{\omega}\phi\nabla^{\omega}\phi
-\frac{\alpha'}{24}H_{\mu\nu\lambda}H^{\mu\nu\lambda}\label{barphi}\ ,
\ena
where $D$ is the spacetime (or target space) dimension and $H$ is the field strength of the
$b$-field:
\beq
H_{\mu\nu\lambda}=3\pa_{[\m}b_{\n\l ]}=\pa_{\mu}b_{\nu\lambda}+
\pa_{\nu}b_{\lambda\mu}+
\pa_{\lambda}b_{\mu\nu}\ .
\eeq
The Weyl anomaly coefficients differ from the beta functions in (\ref{betadef}) as follows:
\bea\label{weyldef}
\bar{\beta}_{\mu\nu}^g&=&\beta_{\mu\nu}^g+2\alpha '\nabla_\mu
\partial_\nu\phi\ ,\\ 
\bar{\beta}_{\mu\nu}^b&=&\beta_{\mu\nu}^b
+\alpha '{H_{\mu\nu}}^\lambda
\partial_\lambda\phi\ ,\\ 
\bar{\beta}^\phi&=&\beta^\phi
+\alpha '(\partial_\mu\phi)^2\ .
\ena
Comparing the result in (\ref{a1}) with (\ref{wanomaly}), we see
\footnote{According to the so-called Curci-Paffuti Theorem \cite{curci}, $\bar{\b}^{g}_{\m\n}=\bar{\b}^b_{\m\n}=0$ implies that 
$\bar{\b}^{\f}$ is constant and therefore can be interpreted as a
central charge. I thank H. Dorn for pointing out this reference.} that a background
$(g,b,\f)$ with $\bar{\b}^{g}_{\m\n}=\bar{\b}^{b}_{\m\n}=0$ describes a conformal field theory
of central charge $c=6\bar{\b}^{\f}$. Hence, from (\ref{wanomaly}) we
conclude that the
requirement of conformal invariance or consistent string propagation 
is tantamount to
the condition that the Weyl anomaly coefficients associated with the
background must vanish, i.e.  
\beq\label{weylvanish}
\bar{\beta}_{\mu\nu}^g=\bar{\beta}_{\mu\nu}^b=\bar{\beta}_{\mu\nu}^\phi=0\ .
\eeq
There is, in fact, a simple physical interpretation of this condition. The equation
$\bar{\beta}_{\mu\nu}^g=0$ looks like an Einstein equation (i.e. $R_{\m\n}=0$) with source terms
coming from dilaton and antisymmetric tensor fields. Also, the equation
$\bar{\beta}_{\mu\nu}^b=0$ is similar to the Maxwell equation (i.e. $\nabla F=0$) generalized to
the antisymmetric tensor field.  

Higher order terms in
$\alpha'$ gives stringy corrections to these Einstein-like equations. At
two-loop there is -- as we shall discuss later -- a
correction term to the metric beta function of the form 
\beq
\frac {\a'}{2}R_{\m\a\b\g}R_{\n}^{\ \a\b\g}\ ,
\eeq
which gives a correction to the Einstein equations.

In General Relativity, the empty space Einstein equation can be derived
as the equation of motion from an action of the form 
\beq\label{standard}
S_E \sim \int d^Dx\sqrt{g}R\ .
\eeq
An obvious question then is whether it is possible to 
give a physically sensible
interpretation of all the equations in (\ref{weylvanish}) as equations
of motion of an effective low energy action? The answer is
affirmative and to lowest order in
$\alpha'$ that effective action is \cite{gsw}:
\beq\label{effaction}
S'=\frac{1}{2\kappa_0^{2}}\int d^{D}x\sqrt{g}e^{-2\phi}
\left[
\frac{-2(D-26)}{3\alpha'}+R-\frac{1}{12}H_{\mu\nu\lambda}H^{\mu\nu\lambda}
+4\partial_{\mu}\phi\partial^{\mu}\phi\right]\ ,
\eeq
where $1/\k_{0}^2$ is the gravitational constant.
This is the effective action that is expected to appear after integrating out
all the massive modes of the string and as such it is a {\sl low energy} effective action. The
action is written in terms of the so-called string metric $g_{\m\n}$ and has a nonstandard
normalization of the $R$ term. The action can be written in the standard form (as in Eq.
(\ref{standard})) by using instead the Einstein metric $\tilde{g}_{\m\n}=e^{-2K\f}g_{\m\n}$,
where $K$ is a constant to be determined below. 
More generally, with two metrics related by a spacetime
Weyl transformation as
$\tilde{g}_{\m\n}=\Omega^2(X)g_{\m\n}$, the Ricci scalars are related
as \cite{wald} (p. 446):
\beq\label{rtilde}
\tilde{R}=\Omega^{-2}\left( 
R-2(D-1)\nabla^2\ln\Omega-(D-2)(D-1)(\nabla_{\l}\ln\Omega)(\nabla^{\l}\ln\Omega)
\right)\ .
\eeq
For $\Omega=e^{-K\f}$ this is 
\beq
\tilde{R}=e^{2K\f}\left( 
R+2K(D-1)\nabla^2\f-K^2(D-2)(D-1)(\nabla\f)^2
\right)\ ,
\eeq
so to cancel the unwanted dilaton factor in (\ref{effaction}) we should take 
$K=-2/(D-2)$. With this choice one finds the 
following expression for the effective action in terms of the Einstein metric:
\beq\label{Eeffaction}
S'_{E}=\frac{1}{2\kappa_0^{2}}\int d^{D}x \sqrt{\tilde{g}}
\left[
\frac{-2(D-26)}{3\alpha'}e^{4\f/(D-2)}+\tilde{R}-\frac{1}{12}H_{\mu\nu\lambda}
H^{\mu\nu\lambda}e^{-8\f/(D-2)}
-\frac{4}{D-2}\partial_{\mu}\phi\partial^{\mu}\phi\right]\ . 
\eeq
The vanishing of the Weyl anomaly coefficients (\ref{weylvanish}) can then be derived as the
equations of motion of the effective action (\ref{effaction}). Define the Lagrange
density ${\cal L}$ as
\beq
S'=\frac{1}{2\kappa_0^{2}}\int d^{D}x{\cal L}\ .
\eeq
The Euler-Lagrange equations for the antisymmetric tensor (that is $\frac{\pa {\cal L}}{\pa
b_{\m\n}} - \nabla^{\omega}\frac{\pa {\cal L}}{\pa (\nabla_{\omega}b_{\m\n})}=0$) becomes after
using the small calculation,
\beq
\frac{\pa {\cal L}}{\pa \nabla_{\omega}b_{\m\n}}=-\frac{1}{12}\sqrt{g}e^{-2\phi}
\frac{\pa H^2}{\pa \nabla_{\omega}b_{\m\n}}= -\frac{1}{6}\sqrt{g}e^{-2\f}
H^{\l\k\s}\frac{\pa H_{\l\k\s}}{\pa \nabla_{\omega}b_{\m\n}}=
-\frac{1}{2}\sqrt{g}e^{-2\f}H_{\omega\m\n}\ ,
\eeq
that
\beq
0 = \nabla^{\omega}\left(\sqrt{g}e^{-2\f}H_{\omega\m\n}\right)
=\sqrt{g}e^{-2\f}\left( -2\nabla^{\omega}\f H_{\omega\m\n} +
\nabla^{\omega}H_{\omega\m\n}\right)\ . 
\eeq
This implies the equation $\bar{\b}^b_{\m\n}=0$. The variation of the action with respect to
$\phi$ is:
\beq
\delta_{\f}S'=-\frac{1}{2\kappa_0^{2}}\int d^{D}x\sqrt{g}e^{-2\phi}
\left[
\frac{-4(D-26)}{3\alpha'}+2R-\frac{1}{6}H_{\mu\nu\lambda}H^{\mu\nu\lambda}
+8\partial_{\mu}\phi\partial^{\mu}\phi + 8\nabla^2\f \right]\cdot \delta\f 
\eeq
and it implies the following equation of motion for the dilaton:
\beq\label{eq1}
0 = \frac{2(D-26)}{3\a'}-R+\frac{1}{12}H^2+4(\nabla\f)^2-4\nabla^2\f\ .
\eeq
The variation with respect to the metric background is more complicated. First we need the
following relation:
\beq\label{trik}
\int d^{D}x \sqrt{g}e^{-2\f}(\nabla\f)^2= \int d^{D}x \sqrt{g}e^{-2\f}\left( 
\nabla^2\f-(\nabla\f)^2\right) \ ,
\eeq
which is obtained by a partial integration and using that the metric is covariantly constant. 
We then choose to write the action in (\ref{effaction}) as,
\bea
S'&=&\frac{1}{2\kappa_0^{2}}\int d^{D}x e^{-2\phi}
\left[
-\sqrt{g}\frac{2(D-26)}{3\alpha'}+\sqrt{g}R+4\sqrt{g}g^{\m\n}\nabla_{\mu}\nabla_{\n}\phi
-4\sqrt{g}g^{\m\n}\nabla_{\m}\f\nabla_{\n}\f\right. \nonumber\\
&& \left. -\frac{1}{12}\sqrt{g}g^{\m\n}g^{\l\k}g^{\r\s}H_{\m\l\r}H_{\n\k\s}\right]\ , 
\ena
in which form the dependence on the metric is explicit.  
The variation of this expression with respect to the metric is 
\bea
\delta_{g}S'&=&\frac{1}{2\kappa_0^{2}}\int d^{D}x \sqrt{g}e^{-2\phi}
\left[\left(
\frac{2(D-26)}{6\alpha'}g_{\m\n}+R_{\m\n}-\half g_{\m\n}R
-2g_{\m\n}\nabla^2\phi +4\nabla_{\m}\nabla_{\n}\f \right.\right.\nonumber\\
&&\left. \left. +2g_{\m\n}(\nabla\f)^2-4\nabla_{\m}\f\nabla_{\n}\f 
\right)\delta g^{\m\n}
-\frac{1}{12}\left(
-\frac{1}{2}g_{\m\n}H^2+3H_{\m}^{\ \k\l}H_{\n\k\l}\right)\delta g^{\m\n}\right]\ , 
\ena
where we have used that
\beq
\delta (\sqrt{g})=-\frac{1}{2}\sqrt{g}g_{\m\n}\delta g^{\m\n}\ ,
\eeq
and that 
\beq
\delta (\int d^{D}x \sqrt{g}e^{-2\f}R) = 
\int d^{D}x \sqrt{g}e^{-2\f}(R_{\m\n}-\frac{1}{2}g_{\m\n}R)\delta g^{\m\n}\ . 
\eeq
After a partial integration, 
the equation of motion that follows for the metric is:
\bea\label{eq3}
0&=& g_{\m\n}\frac{D-26}{3\a'}+R_{\m\n}-\frac{1}{2}g_{\m\n}R-2g_{\m\n}\nabla^2\f
+2\nabla_{\m}\nabla_{\n}\f+2g_{\m\n}(\nabla\f)^2 \nonumber \\
&& +\frac{1}{24}g_{\m\n}H^2
-\frac{1}{4}H_{\m}^{\ \k\l}H_{\n\k\l}\ .
\ena
This equation together with (\ref{eq1}) is not immediately related to the vanishing of the Weyl
anomaly coefficients. But in fact adding Eq. (\ref{eq3}) and $-\half g_{\m\n}$ times Eq.
(\ref{eq1}) gives:
\beq
0=R_{\m\n}-\frac{1}{4}H_{\m}^{\ \k\l}H_{\n\k\l}+2\nabla_{\m}\nabla_{\n}\f\ ,
\eeq
which is the same as $\bar{\b}^g_{\m\n}=0$. Similarly, taking the trace of Eq. (\ref{eq3}) and
adding $(1-D/2)$ times  Eq. (\ref{eq1}) gives
\beq
0=\frac{2(D-26)}{3\a'}-2\nabla^2\f-\frac{1}{6}H^2+4(\nabla\f)^2\ ,
\eeq
which implies $\bar{\b}^{\f}=0$, so
that indeed the equations of motion of (\ref{effaction})
imply that the Weyl anomaly coefficients vanish. 

\subsubsection*{$T$-Duality of Sigma Models}

So far we have focused on $T$-duality in string theory and on the sigma
model description of string propagation in curved backgrounds.
While it is true that consistent string propagation requires
conformal invariance on the worldsheet (and hence vanishing of the Weyl
anomaly coefficients), this is not a requirement for the existence of a
$T$-duality in general. 
As the set of
conformal
backgrounds is just a small subset of all possible backgrounds, it seems natural
to extend the action of duality to all such backgrounds.

Following this philosophy, we will henceforth be studying the properties
of the sigma model (which we first take to be bosonic) away from the
conformal point. First we derive the $T$-duality transformations of any
bosonic sigma model with a target space isometry and then we study the
implications of this duality on the renormalization group flow of the
model.

In order to derive the duality transformations -- which we will do in a truncated model
where $\f$ is identically zero -- relating different sigma model backgrounds,  we will assume
that the target space has an Abelian isometry which can be represented as a translation in a
coordinate
$X^0$ in the target space. It is then simple to see that we can choose
``adapted'' coordinates $\{X^{0}, X^{i}\}$ such that the background
fields are independent of $X^{0}$.

To be more precise, we start by assuming that the action (\ref{sigmaction}) is
invariant under an isometry in the target space. The corresponding
Killing vector is denoted by $k^{\mu}$
\beq\label{isometry}
\delta_{\epsilon}X^{\mu}=\epsilon k^{\mu}\ .
\eeq
A straightforward generalization of this is to have several
Killing vectors forming an Abelian or maybe non-Abelian group. We will
not discuss that generalization here.

Requiring invariance of the action under the Killing symmetry gives us some conditions on
the background fields. The first is that the Lie derivative of the metric with respect to the
Killing vector vanishes:
\beq
{\cal L}_kg_{\mu\nu}=\nabla_{\mu}k_{\nu}+\nabla_{\n}k_{\m}=0\ .
\eeq
Also, there must exist some vector $\omega_{\mu}$ such that
\beq
{\cal L}_kb_{\mu\nu}=\partial_{\mu}\omega_{\nu}-\partial_{\nu}\omega_{\mu}\ .
\eeq
In adapted coordinates, the isometry acts in the
$X^{0}$-direction as a translation $X^{0}\rightarrow
X^{0}+\epsilon$. So $k=k^{\mu}\frac{\partial}{\partial X^{\mu}}
=\frac{\partial}{\partial X^{0}}$ and the Lie derivative is
then just the derivative with respect to $X^{0}$. It is then obvious
that the background metric is independent of
$X^{0}$. There would seem to be a problem though since the Lie derivative of
the antisymmetric tensor is generally not zero but equal to the
exterior derivative of a one-form, but actually there is no problem since the action
obviously has a gauge symmetry under which
\beq
b\rightarrow b+d\lambda
\eeq
with $\lambda$ some one-form. In the gauge where
\beq
{\cal L}_k\lambda=-\omega\ ,
\eeq
we find
\beq
{\cal L}_k(b+d\lambda)={\cal L}_kb+d({\cal L}_k\lambda)
=d\omega+d(-\omega)=0\ ,
\eeq
so that all background fields are actually independent of $X^{0}$. In
the following we will assume that such adapted coordinates have been
chosen. 

To find the dual model we will gauge the isometry in the target space (\ref{isometry}) by
introducing a gauge field $A_{\mu}$ that transforms as $\delta A_{\mu}
=-\partial_{\mu}\epsilon$ \cite{alv9410}. Adding a Lagrange multiplier term forces
the gauge field to be a pure gauge. The gauged action is then
\bea\label{gaction}
S_{gauged}&=&\frac{1}{4\pi\alpha'}\int_{\Sigma}(g_{\mu\nu}\delta^{ab}
+i\epsilon^{ab}b_{\mu\nu})D_aX^{\mu}D_bX^{\nu}\nonumber\\
&&+\frac{i}{4\pi\alpha'}\int_{\Sigma}\tilde{X}^{0}(\partial_aA_b-
\partial_bA_a)\epsilon^{ab},
\ena
where we have defined $D_aX^{\mu}=\partial_{a}X^{\mu}+k^{\mu}A_a$, and $\tilde{X}^0$ acts as 
a Lagrange multiplier. 
The
dual theory is obtained by integrating the $A$-field
\beq
A_{a}=-\frac{1}{k^{2}}(k^{\mu}g_{\mu\nu}\partial_aX^{\nu}
+i{\epsilon_a}^{b}\partial_b\tilde{X}^0+i{\epsilon_a}^{b}k^{\mu}b_{\mu\nu}\partial_bX^{\nu})\ ,
\eeq
which is inserted back into the action (\ref{gaction}) and then fixing the gauge with the
condition
$X^{0}=0$. Subsequently, the following expression for the dual action is obtained:
\bea
\tilde{S}=\frac{1}{4\pi\alpha'}\int_{\Sigma}&d^2\sigma& 
[\frac{1}{g_{00}}\partial_a\tilde{X}^0
\partial^a\tilde{X}^0+2\frac{b_{0i}}{g_{00}}\partial_a\tilde{X}^0\partial^aX^i
+(g_{ij}-\frac{g_{0i}g_{0j}-b_{0i}b_{0j}}
{g_{00}})\partial_aX^i\partial^aX^j\nonumber\\
&&+i\epsilon^{ab}(\frac{g_{0i}}{g_{00}}\partial_a\tilde{X}^0\partial_bX^i+
(b_{ij}-\frac{g_{0i}b_{0j}-b_{0i}g_{0j}}{g_{00}})\partial_aX^i\partial_bX^j)] \ .
\ena
So the dual background - which is what we intended to derive - is
given by 
\bea\label{dualb}
\tilde{g}_{00}&=&{1\over g_{00}}\ ,\
\tilde{g}_{0i}={b_{0i}\over g_{00}}\ ,\ \tilde{b}_{0i}={g_{0i}\over
g_{00}}\ , \nonumber\\ \tilde{g}_{ij}&=&g_{ij}
-{g_{0i}g_{0j}-b_{0i}b_{0j}\over g_{00}}\ ,\\
\tilde{b}_{ij}&=&b_{ij}-{g_{0i}b_{0j}-b_{0i}g_{0j}\over g_{00}}\ .\nonumber
\ena
This expression was first derived by Buscher \cite{bus1}.
There is also a dilaton shift \cite{bus1}
\beq\label{dilatonshift}
\tilde{\phi}=\phi-\frac{1}{2}\ln g_{00}\ .
\eeq
As is clear from the action (\ref{sigmaction}), this is, however, a one-loop effect. Basically, this effect
follows from the fact that the $T$-duality acts nontrivially on the coupling
constant. Indeed, the string coupling is determined by the expectation value of the dilaton
field, since $g= \langle e^{\f}\rangle$, and it therefore follows from (~\ref{2tdual}) that 
\beq
\f'=\f-\half\ln (R^2/\a')\ .
\eeq
\subsubsection*{Duality and Renormalization Group Flow}

Having considered the renormalization group flow (as encoded by the beta functions) of bosonic
sigma models and the $T$-duality of such models, it remains to connect these two subjects. 

More generally, 
one could enquire as to what restrictions can be imposed by duality
in quantum field theory and
string theory?  An important tool in that direction is
that of a moduli space, which typically will parametrize a family of
quantum field theories. A point in moduli space is then given by a set of
parameters $(\lambda_1, \lambda_2, \ldots)$. Consider for example the
moduli space of toroidal (${\bf S}^1$) compactifications of bosonic 
string theory. Such a
parameter is then the radius $R$ of the compactifying circle. What
$T$-duality tells us in this case is that the point $\lambda=R$ in parameter
space determines the same (bosonic) string theory as that of $\lambda'=1/R$. 
(or that the moduli space should be modded out by ${\bf Z}_2$). 
So in this respect duality symmetry acts as a
transformation in moduli (or parameter) space while leaving the partition function
invariant.

Another transformation that also acts naturally on this parameter space, is the one of the
renormalization group. 

In quantum field theory a central concept is that of regularization and
renormalization. After proper renormalization the couplings of the
theory will be functions of a renormalization scale $\L$. The flow
of the parameters is then determined by the renormalization group
(RG). This RG also acts naturally in the parameter space as it
determines the change in the renormalized parameters as
we change the renormalization scale.
\footnote{Actually, since the RG is
encoded in the beta functions which can be viewed as tangent vectors
on the parameter space, it would be more correct to say that the RG acts
on the tangent space $T{\cal M}$ of the parameter space ${\cal M}$.}

With both the $T$-duality and the RG acting as motions in the parameter space it becomes natural to
study the interrelation between the two operators (early related work was done by 
L{\"u}tken \cite{lut}, in which he studied some constraints on the RG flow from duality in the Ising model
and the quantum Hall effect).

The requirement that duality symmetry and the RG be mutually consistent is
formulated as follows. In full generality, we will assume that we have a
system with a parameter space parametrized by a set of couplings, $g^i$.
The duality symmetry $T$ (which later actually will mean the usual
target space duality) acts on the parameter space according to
\beq
Tg^i\equiv\tilde{g}^i=\tilde{g}^i(g)\ ,
\eeq
and connects equivalent points in parameter space -- as the $R\rightarrow
\a'/R$ symmetry of string theory. The renormalization group flow, on the
other hand, is given in terms of a set of beta functions
\beq
Rg^i\equiv \beta^i(g)=\L\frac{d}{d\L}g^i\ ,
\eeq  
where $\L$ is the renormalization scale parameter. More generally, if
$F(g)$ is any function on the parameter space then the action of these
operations are 
\bea
TF(g)&=&F(\tilde{g}(g))\nonumber\\
RF(g)&=&\frac{\delta F(g)}{\delta g^j}\cdot \beta^j(g)\ .
\ena
The hypothesis that we will bring forward is that the
consistency requirement governing the relation between duality and the RG
is expressed by \cite{HO}:
\beq\label{haagolsen}
[T,R]=0\ ,
\eeq
which simply asserts that the duality transformations and the RG flow 
are mutually commuting
operators on the parameter space of the theory. 
Eq. (\ref{haagolsen}) is seen to be equivalent to the following relations
between beta functions in the original and the dual theory:
\beq\label{haagolsen2}
\beta^i(\tilde{g})=\frac{\delta\tilde{g}^i}{\delta g^j}\cdot
\beta^j(g)\ ,
\eeq
or that the beta functions should transform as a contravariant
vector under duality. 
In a quantum field theory with
duality, the existence - and the consistency - of the duality symmetry
should not depend on the renormalization of the model, so what (\ref{haagolsen}) 
really means is that the duality is a {\sl quantum symmetry} of the quantum field theory.
The relation
in (\ref{haagolsen}) was first explicitly formulated in \cite{HO}.
Its consequences, and more generally, the conditions on the RG flow from duality, have been
investigated in many different contexts and cases. We will list here a few important
contributions:
$T$-duality (in the context of a bosonic sigma model at one-loop) was first considered in the
seminal paper of Haagensen
\cite{haag}. This analysis was subsequently extended to two-loop in a bosonic sigma model with
purely metric background in
\cite{HO} and
\cite{HOS}. In the first of these papers, it was shown that $[T,R]=0$ continues to be
valid at two-loop; in the second paper it was shown that assuming $[T,R]=0$ determines the
form of the two-loop beta function. The heterotic sigma models were considered at 
one-loop in \cite{OS}. After cancellation of anomalies it was found that the consistency conditions
are exactly satisfied.

Damgaard and Haagensen \cite{damg9609} studied spin systems with Kramers-Wannier symmetry. In
this case the duality group is ${\bf Z}_2$ and the restrictions were not strong enough to
completely determine the RG flows. However, a similar consistency condition on the beta function
$\b(K)$ implies that there must be a first or higher order phase transition at the self-dual
point $K=K^*$.   

Quantum Hall systems with $SL(2,{\bf Z})$ symmetry were studied by Burgess and L\"utken in
\cite{burg9611}. Here it was shown that requiring a certain symmetry 
to be commuting with the RG flow (supplemented with some additional mild assumptions), were
enough to completely determine the $c$ function - and in particular the RG beta function.

Ritz \cite{ritz9710} considered constraints on the RG flow in ${\cal N}=2$ $SU(2)$
supersymmetric models, which exhibit an $S$-duality. Assuming (supplemented with some
additional mild assumptions) that the RG flow commutes with a certain subgroup of $SL(2,{\bf
Z})/{\bf Z}_2$, the exact non-perturbative beta function could be determined. Happily enough,
the result coincides with the Seiberg-Witten solution. Similar work was done by Latorre and
L\"utken \cite{latorre9711}, in which they also address the question of RG flow constraints in ${\cal N}=0$
gauge theories.

More recent work by Balog, Forg\'acs, Mohammedi, Palla and Schnittger
\cite{balog9806} that gives a proof of
$[T,R]=0$ at one-loop (for both Abelian and non-Abelian dualities) 
from first principles will be described in the last part of this chapter.

\section{Bosonic Models at One-Loop}

In this section we will consider the restrictions on RG flow imposed by
duality in bosonic sigma models at one-loop. We will follow
\cite{HO,haag}. 

The target space of the sigma
model is taken to be
$D$-dimensional. After going to adapted coordinates as described in the
previous section, the relevant action is 
\bea
S=\frac{1}{4\pi\alpha'}\int_{\Sigma}&d^2\sigma& [g_{00}(X)\partial_aX^0
\partial^aX^0+2g_{0i}(X)\partial_aX^0\partial^aX^i
+g_{ij}(X)\partial_aX^i\partial^aX^j\nonumber\\
&&+i\epsilon^{ab}(2b_{0i}(X)\partial_aX^0\partial_bX^i+
b_{ij}\partial_aX^i\partial_bX^j)]\ , 
\ena
with all background fields independent of $X^0$ and $i,j=1, \ldots ,D-1$. 
The corresponding one-loop duality transformations can be found in (\ref{dualb}). 

For $\Sigma$ a curved worldsheet, we must include another background
coupling, namely that of the dilaton $\phi(x)$ \cite{tsey1}. After renormalization
this and the other couplings will flow as encoded in the RG beta
functions:
\beq
\b^{g}_{\m\n}\equiv\L\frac{d}{d\L}g_{\m\n}\ ,\;
\b^{b}_{\m\n}\equiv\L\frac{d}{d\L}b_{\m\n}\ ,\;
\b^{\f}\equiv\L\frac{d}{d\L}\f\ .
\eeq
In studying the relation between duality and the RG flow we might as well
use the Weyl anomaly coefficients which for this model are \cite{tsey2}:
\bea
\bar{\beta}_{\mu\nu}^g&=&\beta_{\mu\nu}^g+2\alpha '\nabla_\mu
\partial_\nu\phi\ +\nabla_{(\mu}W_{\nu)}\ ,\\ 
\bar{\beta}_{\mu\nu}^b&=&\beta_{\mu\nu}^b
+\alpha '{H_{\mu\nu}}^\lambda
\partial_\lambda\phi\ +H_{\mu\nu}^{\ \ \lambda}W_{\lambda}
+\nabla_{[\mu}L_{\nu ]}\ ,\\ 
\bar{\beta}^\phi&=&\beta^\phi+\alpha' 
(\partial_{\mu}\phi)^2+\nabla^{\mu}\phi W_{\mu}\ .
\ena
Here $L_{\m}, W_{\m}$ are target tensors depending on the metric and
antisymmetric tensor.
Under a target diffeomorphism we have e.g. that the metric beta function
change as $\b^{g}_{\m\n}\rightarrow \b^g_{\m\n}+\nabla_{(\m}\xi_{\n)}$,
which reflects the fact that the beta functions are ambiguous. However,
quite the opposite is true for the Weyl anomaly coefficients: a target
reparametrization is accompanied by $W_{\m}\rightarrow W_{\m}-\xi_{\m}$,
and in the end the Weyl anomaly coefficients are invariant \cite{tsey2}. 
This also implies that 
for the backgrounds that we consider we can assume that
the target tensors $W_{\mu}=L_{\mu}=0$, which we therefore will assume henceforth.

In the notation of the last section, the couplings $g^i$ are 
$g^i=\{ g_{\mu\nu}, b_{\mu\nu}, \phi\}$. Since we study the RG
flow as generated by the Weyl anomaly coefficients, the action of this
flow on any functional $F[g,b,\phi ]$ is given by
\beq
RF[g,b,\phi ]=\frac{\delta F}{\delta
g_{\mu\nu}}\cdot\bar{\beta}^g_{\mu\nu}+
\frac{\delta F}{\delta b_{\mu\nu}}\cdot\bar{\beta}^b_{\mu\nu}+
\frac{\delta F}{\delta\phi}\cdot\bar{\beta}^{\phi}\ ,
\eeq
while the action of the duality operation $T$ is
\beq
TF[g,b,\phi ]=F[\tilde{g},\tilde{b},\tilde{\phi}]\ .
\eeq
In the previous section we postulated that the condition for the mutual consistency of
renormalization group flow and duality can be formulated as the requirement
that the RG flow and duality operations commute in parameter space:
\beq
[T,R]=0\ .
\eeq
We will start by assuming that this is true at the one-loop order we are considering and then
see to what extend this determines the RG flow. Later we will show -- by simplifying the
resulting equations -- that this condition is in fact satisfied at one-loop.

Now, using the R operation on each side of the "classical" duality
transformations in (\ref{dualb}), we see that this is equivalent to the following
set of consistency conditions:
\bea\label{cc}
\bar{\beta}^{\tilde{g}}_{00}&=&-\frac{1}{g_{00}^2}
\bar{\beta}^{g}_{00}\nonumber\ ,\\
\bar{\beta}^{\tilde{g}}_{0i}&=&-\frac{1}{g_{00}^2}
(b_{0i}\bar{\beta}^{g}_{00}-\bar{\beta}^{b}_{0i}g_{00})\nonumber\ ,\\
\bar{\beta}^{\tilde{b}}_{0i}&=&-\frac{1}{g_{00}^2}
(g_{0i}\bar{\beta}^{g}_{00}-\bar{\beta}^{b}_{0i}g_{00})\ ,\\
\bar{\beta}^{\tilde{g}}_{ij}&=&\bar{\beta}^{g}_{ij}-\frac{1}{g_{00}}
(\bar{\beta}^{g}_{0i}g_{0j}+\bar{\beta}^{g}_{0j}g_{0i}
-\bar{\beta}^{b}_{0i}b_{0j}-\bar{\beta}^{b}_{0j}b_{0i})+
\frac{1}{g_{00}{^2}}(g_{0i}g_{0j}-b_{0i}b_{0j})
\bar{\beta}^{g}_{00}\nonumber\ ,\\
\bar{\beta}^{\tilde{b}}_{ij}&=&\bar{\beta}^{b}_{ij}-\frac{1}{g_{00}}
(\bar{\beta}^{g}_{0i}b_{0j}+\bar{\beta}^{b}_{0j}g_{0i}
-\bar{\beta}^{g}_{0j}b_{0i}-\bar{\beta}^{b}_{0i}g_{0j})+
\frac{1}{g_{00}{^2}}(g_{0i}b_{0j}-b_{0i}g_{0j})
\bar{\beta}^{g}_{00}\ ,\nonumber
\ena
where on the left hand side we denote by e.g. $\bar{\b}^{\tilde{g}}_{\m\n}$ the Weyl anomaly
coefficient computed as a functional of the dual geometry. These relations were first presented in
\cite{haag}. Here it was shown that the known beta functions satisfy the relations up to a target
space diffeomorphism and that they are exactly satisfied for the Weyl anomaly
coefficients. If, on the other hand, we start by demanding that the
consistency relations should be satisfied, how much can be determined?
These conditions are very restrictive and indeed it was shown in
\cite{haag} that they (essentially) determine the one-loop beta
functions -- to precisely what extend will be clear later. 

A simple argument shows that at loop order $\ell$, the tensor structures
$T_{\mu\nu}$ which can appear in the beta functions must scale as
$T_{\mu\nu}(\Omega g, \Omega b)=\Omega^{1-\ell}T_{\mu\nu}(g,b)$, with
$\Omega$ a constant. This is because the $\ell$
loop counterterm has $\a'^{\ell}$ as a factor and we can express this as,
\beq
\b_{\m\n}^{(\ell)}(\frac{1}{\a'}g)= \a'^{\ell -1}T_{\m\n}(g)\ ,
\eeq
so that a scaling $g\rightarrow \Omega g$ of the metric is the same as
scaling $\a'\rightarrow \Omega^{-1}\a'$.

This implies that to order \oa, i.e. $\ell=1$, the only tensors
which can possibly appear are
\begin{eqnarray}
\b_{\m\n}^{g}&=&\a'\left( A_1\, R_{\m\n}+A_2\, H_{\m\l\r}H_{\n}^{\; \l\r}
+A_3\, g_{\m\n}R+A_4\, g_{\m\n}H_{\a\b\g}H^{\a\b\g}\right)\ ,\nonumber\\
\b_{\m\n}^{b}&=&\a'\left(A_5\, \nabla^{\l}H_{\m\n\l}\right)\ ,
\end{eqnarray}
since these scale as $\Omega^0$ under global scalings of the
background fields, are generally covariant and have the correct symmetry under $(\m
\leftrightarrow \n)$. Here, $H$ is
the exterior derivative of the
$b$-field,
$H_{\m\n\l}=3\pa_{[\m}b_{\n\l ]}$, and the constants $A_1,\ldots ,A_5$ can be computed from one-loop
Feynman diagrams using dimensional regularization; see for example \cite{callan}. It was,
however, shown in
\cite{haag} that requiring
$[T,R]=0$ and choosing $A_1=1$ uniquely determines $A_2=-1/4$, $A_5=-1/2$
and $A_3=A_4=0$. Note that the overall factor of the beta function cannot
be determined since the relation $[T,R]=0$ is a linear relation at
this order. As was also shown in \cite{haag}
the transformation of the dilaton under duality,
$\tilde{\f}=\f-\half\ln g_{00}$ can be determined from the consistency
conditions (\ref{cc}) on $g_{\m\n}$ and $b_{\m\n}$. This gives the remaining consistency
condition for the dilaton beta function, 
\beq
\tilde{\bar{\beta}}^{\phi}= \bar{\beta}^\phi-\frac{1}{2g_{00}}
\bar{\beta}_{00}\ .
\eeq
From this condition one can determine the dilaton beta function up to a global constant.
Hence we see that requiring $[T,R]=0$ at this one-loop order determines all beta functions
(but only up to a global factor):
\bea\label{betaol}
\beta^{g}_{\mu\nu}&=&\alpha'R_{\mu\nu}
-\frac{\alpha'}{4}H_{\mu\lambda\rho}
H_{\nu}^{\ \lambda\rho}\\
\beta^{b}_{\mu\nu}&=&-\frac{\alpha'}{2}\nabla^{\rho}H_{\rho\mu\nu}\\
\beta^{\phi}&=&C-\frac{\alpha'}{2}\nabla^2\phi
-\frac{\alpha'}{24}H_{\mu\nu\lambda}H^{\mu\nu\lambda}\ .
\ena      
A standard procedure for verifying the consistency conditions such as the
ones which appear in (\ref{cc}) is to decompose the background fields
(i.e. the metric and antisymmetric tensor) into a form which singles out the
direction ($X^0$) in which the duality transformation is performed. 
This is the Kaluza-Klein reduction where the metric is decomposed as:
\beq\label{metric}
g_{\mu\nu}=\left( \begin{array}{cc}a & av_i\\ av_i &\bar{g}_{ij}+av_iv_j
\end{array}\right)\ ,
\eeq
such that $g_{00}=a, g_{0i}=av_i$ and
$g_{ij}=\bar{g}_{ij}+av_iv_j$. With this procedure the antisymmetric tensor components are
$b_{0i}\equiv w_i$ and $b_{ij}$. In this representation, the duality
transformations take the rather simple form $a\rightarrow 1/a$,
$v_i\leftrightarrow w_i$ and
$b_{ij}\rightarrow b_{ij}+w_iv_j-w_jv_i$. However, the task of actually
verifying the consistency conditions is still very lengthy and the
fact that the duality group is just ${\bf Z}_2$ is rather obscured by the form of the consistency
condition in (\ref{cc}).
\footnote{The $T$-duality group that results
from compactifying on $T^k$ is $O(k,k,{\bf Z})$ \cite{narain86,maharana}
and compatification on ${\bf S}^1$ therefore has a duality group $O(1,1,{\bf
Z})={\bf Z}_2$.}  

In \cite{HO} it was suggested that one should express these conditions
in the tangent frame of the target space where the ${\bf Z}_2$ structure is
evident. This tangent frame is determined by the local vielbeins, i.e.
$e_{\m}^{\ a}e_{\n}^{\ b}\delta_{ab}=g_{\m\n}$, ($a=\hat{0},\a$ with $\a=1,\ldots D-1$), and a
specific solution corresponding to the decomposition of the metric in (\ref{metric}) is:
\beq
e_{\m}^{\ a}=\left( \begin{array}{cc}e_{0}^{\ \hat{0}} 
& e_{0}^{\ \a}\\ e_{i}^{\ \hat{0}} 
&e_{i}^{\ \a}\end{array}\right)\
=\left( \begin{array}{cc}\sqrt{a}
& 0\\ \sqrt{a}v_i 
&\bar{e}_{i}^{\ \a}
\end{array}\right)\ ,
\eeq
where the notation is $\bar{e}_{i}^{\ \a}
\bar{e}_{j}^{\ \b}\delta_{\a\b}=\bar{g}_{ij}$. 

In the tangent frame, the Weyl anomaly coefficients have the components:
\beq\label{tf}
\bar{\b}^g_{ab}=e_{a}^{\ \m}e_{b}^{\ \n}\bar{\b}^g_{\m\n}, \ \
\bar{\b}^b_{ab}=e_{a}^{\ \m}e_{b}^{\ \n}\bar{\b}^b_{\m\n}\ , 
\eeq
where $e_a^{\ \m}$ is the inverse vielbein, namely
\beq
e_{a}^{\ \m}=\left( \begin{array}{cc}e_{\hat{0}}^{\ 0} 
& e_{\hat{0}}^{\ i}\\ e_{\a}^{\ 0} 
&e_{\a}^{\ i}\end{array}\right)\
=\left( \begin{array}{cc}1/\sqrt{a}
& 0\\ -v_{\a} 
&\bar{e}_{\a}^{\ i}
\end{array}\right)\ .
\eeq
The ${\bf Z}_2$ symmetry is now easily seen: one uses (\ref{cc}) and express them in
the tangent frame through (\ref{tf}). We are going to illustrate this with two
examples; the $\hat{0}\hat{0}$ and the $\a\b$ components:
\beq
\bar{\b}^{\tilde{g}}_{\hat{0}\hat{0}}=
\tilde{e}_{\hat{0}}^{\ 0}\ \tilde{e}_{\hat{0}}^{\ 0}\
\bar{\b}^{\tilde{g}}_{00}=a\ \bar{\b}^{\tilde{g}}_{00}=
-{1\over a}\ \bar{\b}^{g}_{00}=-e_{\hat{0}}^{\ 0}\ e_{\hat{0}}^{\ 0}\
\bar{\b}^{g}_{00}=-\bar{\b}^{g}_{\hat{0}\hat{0}}\ ,
\eeq
and
\bea
\bar{\b}^{\tilde{g}}_{\a\b}&=&\tilde{e}_{\a}^{\ 0}\
\tilde{e}_{\b}^{\ 0}\ \bar{\b}^{\tilde{g}}_{00}+
\tilde{e}_{\a}^{\ 0}\ \tilde{e}_{\b}^{\ i}\
\bar{\b}^{\tilde{g}}_{0i}+
\tilde{e}_{\a}^{\ i}\ \tilde{e}_{\b}^{\ 0}\
\bar{\b}^{\tilde{g}}_{i0}+
\tilde{e}_{\a}^{\ i}\ \tilde{e}_{\b}^{\ j}\
\bar{\b}^{\tilde{g}}_{ij}\nonumber\\
&=&\bar{e}_{\a}^{\ i}\ \bar{e}_{\b}^{\ j}\
w_iw_j\ \bar{\b}^{\tilde{g}}_{00}-
\bar{e}_{\a}^{\ j}\ \bar{e}_{\b}^{\ i}\ w_j\ 
\bar{\b}^{\tilde{g}}_{0i}-
\bar{e}_{\a}^{\ i}\ \bar{e}_{\b}^{\ j}\ w_j\ \bar{\b}^{\tilde{g}}_{i0}+
\bar{e}_{\a}^{\ i}\ \bar{e}_{\b}^{\ j}\
\bar{\b}^{\tilde{g}}_{ij}\nonumber\\
&=&-\frac{1}{a^2}\bar{e}_{\a}^{\ i}\ \bar{e}_{\b}^{\ j}\
w_iw_j\ \bar{\b}^{g}_{00}+
\frac{1}{a^2}\bar{e}_{\a}^{\ j}\ \bar{e}_{\b}^{\ i}\ w_j
(w_i\bar{\b}^{g}_{00}-\bar{\b}^{b}_{0i}a)\nonumber\\
&&+\frac{1}{a^2}\bar{e}_{\a}^{\ i}\ \bar{e}_{\b}^{\ j}\ w_j
(w_i\bar{\b}^{g}_{00}-\bar{\b}^{b}_{0i}a)+
\bar{e}_{\a}^{\ i}\ \bar{e}_{\b}^{\ j}\left(
\bar{\beta}^{g}_{ij}\right.\nonumber\\
&&\left. -\frac{1}{a}
(\bar{\beta}^{g}_{0i}av_j+\bar{\beta}^{g}_{0j}av_i
-\bar{\beta}^{b}_{0i}w_j-\bar{\beta}^{b}_{0j}w_i)+
\frac{1}{a^2}(a^2v_iv_j-w_iw_j)
\bar{\beta}^{g}_{00}\right)\nonumber\\
&=&\bar{e}_{\a}^{\ i}\ \bar{e}_{\b}^{\ j}\
v_iv_j\ \bar{\b}^{g}_{00}-
\bar{e}_{\a}^{\ j}\ \bar{e}_{\b}^{\ i}\ v_j\ 
\bar{\b}^{g}_{0i}-
\bar{e}_{\a}^{\ i}\ \bar{e}_{\b}^{\ j}\ v_j\ \bar{\b}^{g}_{i0}+
\bar{e}_{\a}^{\ i}\ \bar{e}_{\b}^{\ j}\
\bar{\b}^{g}_{ij}\nonumber\\
&=& \bar{\b}^{g}_{\a\b}\ .
\ena
The complete set of consistency conditions is \cite{HO}:
\bea\label{tfcc}
\bar{\b}^{\tilde{g}}_{\hat{0}\hat{0}}
&=&-\bar{\b}^g_{\hat{0}\hat{0}}\ ,\nonumber\\
\bar{\b}^{\tilde{g}}_{\hat{0} \a}&=& \bar{\b}^{b}_{\hat{0} \a}\ \ ,\ \
\bar{\b}^{\tilde{b}}_{\hat{0} \a}= \bar{\b}^{g}_{\hat{0} \a}\ ,\nonumber\\
\bar{\b}^{\tilde{g}}_{\a\b}&=& \bar{\b}^{g}_{\a\b}\ \ ,\ \
\bar{\b}^{\tilde{b}}_{\a\b}= \bar{\b}^{b}_{\a\b}\ .
\ena
In this form the ${\bf Z}_2$ duality covariance is now manifest.
An obvious question is why the consistency conditions have such a simple
form when formulated in the tangent frame?
  
The task of verifying the consistency conditions (\ref{cc}) in the form (\ref{tfcc}) is
now very simple. At this point, one needs to express the geometrical tensors which
appear in the beta functions in the tangent frame; this has been done
for the Riemann tensor, and contractions thereof, in Appendix A. Let us
show by explicit calculations that the consistency condition for the
metric: $\bar{\b}^{\tilde{g}}_{\hat{0}\hat{0}}=-
\bar{\b}^{g}_{\hat{0}\hat{0}}$, is satisfied. Using the expressions in
Appendix A we find
\bea
\bar{\b}^{g}_{\hat{0}\hat{0}}&=&
e_{\hat{0}}^{\ 0}\ e_{\hat{0}}^{\ 0}\
\bar{\b}^{g}_{00}\nonumber\\
&=&\frac{1}{a}\left( \a' R_{00}-\frac{1}{4}\a' H_{0\l\r}H_{0}^{\ \l\r}+2\a' 
\na_{0}\pa_{0}\f\right)\nonumber\\
&=&\a' \left( -\frac{1}{2}\bar{\na}_ia^i-\frac{1}{4}a_ia^i
+\frac{a}{4}F_{ij}F^{ij}-\frac{1}{4a}G_{ij}G^{ij}+a^i\pa_i\f\right)\ ,
\ena
while the dual expression is
\bea
\bar{\b}^{\tilde{g}}_{\hat{0}\hat{0}}&=&
\a' \left( \frac{1}{2}\bar{\na}_ia^i-\frac{1}{4}a_ia^i
+\frac{1}{4a}G_{ij}G^{ij}-\frac{a}{4}F_{ij}F^{ij}-a^i\pa_i\tilde{\f}\right)\ ,
\ena
where we have left the dilaton shift undetermined. It is now clear that this consistency
condition is satisfied if and only if the transformation of the dilaton is
$\tilde{\f}=\f-\half\ln a$ as claimed in Eq. (\ref{dilatonshift}). While we did not gain much
here by formulating the consistency condition in the tangent frame, the other condition,
$\bar{\b}^{\tilde{g}}_{\a\b}=\bar{\b}^{g}_{\a\b}$, proves much easier to verify in the tangent
frame; with the help of the expressions in the Appendix A, we find
\bea
\bar{\b}^{g}_{\a\b}&=&\bar{e}_{\a}^{\ i}\bar{e}_{\b}^{\ j}\left[ 
\bar{R}_{ij}-\half\bar{\na}_ia_j-\frac{1}{4}a_ia_j-\frac{a}{2}F_{ik}F_{j}^{\ k}
-\frac{1}{4}v_iv_jG_{kl}G^{kl}-\half v_{(i}G_{j)k}G^{kl}v_l\right.\nonumber\\
&& -\frac{1}{4}v_{(i}
H_{j)kl}G^{kl}-\half (\frac{1}{a}+v_mv^m)G_{i}^{\ k}G_{jk}+\half v^kv^mG_{ik}G_{jm}
-\half H_{km(i}G_{j)}^{\ k}v^m\nonumber\\
&&\left.-\frac{1}{4}H_{ikm}H_{j}^{\
km}+2\bar{\na}_i\pa_{j}\f \right]\ .
\ena
The dual Weyl coefficient is:
\bea
\bar{\b}^{\tilde{g}}_{\a\b}&=&\bar{e}_{\a}^{\ i}\bar{e}_{\b}^{\ j}\left[ 
\bar{R}_{ij}+\half\bar{\na}_ia_j-\frac{1}{4}a_ia_j-\frac{1}{2a}G_{ik}G_{j}^{\ k}
-\frac{1}{4}w_iw_jF_{kl}F^{kl}-\half w_{(i}F_{j)k}F^{kl}w_l\right.\nonumber\\
&& -\frac{1}{4}w_{(i}
\tilde{H}_{j)kl}F^{kl}-\half (a+w_mw^m)F_{i}^{\ k}F_{jk}+\half w^kw^mF_{ik}F_{jm}
-\half \tilde{H}_{km(i}F_{j)}^{\ k}w^m\nonumber\\
&&\left.-\frac{1}{4}\tilde{H}_{ikm}\tilde{H}_{j}^{\
km}+2\bar{\na}_i\pa_{j}\f -\bar{\na}_ia_j\right]\ ,
\ena
(note that $a_i \rightarrow -a_i$ under duality at this order in $\a'$).
Inserting that 
\beq
\tilde{H}_{ijk}=H_{ijk}+G_{ij}v_k+w_jF_{ik}+G_{jk}v_i
+w_kF_{ji}+G_{ki}v_j+w_iF_{kj}\ ,
\eeq
and after straightforward but also rather tedious calculations, it is
found that the two coefficients match: 
$\bar{\b}^{\tilde{g}}_{\a\b}= \bar{\b}^{g}_{\a\b}$ as promised.

Finally, there is the question whether scale invariant models are mapped to dual models which
are also scale invariant. This is however, easily seen since the relation corresponding to
(\ref{tfcc}) between the beta functions is \cite{HO}:
\bea
\b^{\tilde{g}}_{\hat{0}\hat{0}}&=&-\b^g_{\hat{0}\hat{0}}
+\a'\nabla_{(\hat{0}}\cs_{\hat{0})}\nonumber\ , \\
\b^{\tilde{g}}_{\hat{0} \a}&=& \b^{b}_{\hat{0} \a}
-\a'H_{\hat{0}\a}^{\ \ \g}\cs_{\g}\ \ , \ \
\b^{\tilde{b}}_{\hat{0} \a}= \b^{g}_{\hat{0} \a}
-\a'\nabla_{(\hat{0}}\cs_{\a)}\nonumber\ , \\
\b^{\tilde{g}}_{\a\b}&=& \b^{g}_{\a\b}
-\a'\nabla_{(\a}\cs_{\b)}\ \ ,\ \
\b^{\tilde{b}}_{\a\b}= \b^{b}_{\a\b}
-\a'H_{\a\b}^{\ \ \g}\cs_{\g}\ , 
\ena
where $\cs_{a}=-\half e_{a}^{\ \m}\pa_{\m}\ln g_{00}$ generates a target space diffeomorphism,
and $(ab)=ab+ba$. So, if the original model has
$\b^g_{ab}=\b^{b}_{ab}=0$ then the same is true in the dual
model -- after performing a target diffeomorphism generated by
$-\cs_{a}$.

\bigskip

So far we have demonstrated that our relation $[T,R]=0$ is satisfied at one-loop order for the
bosonic sigma model. However in string theory the natural restriction from duality is that the
effective action should be invariant under $T$-duality. It therefore becomes natural to
investigate further the relation between these two conditions.
\subsubsection*{The Effective Action}

At any loop order the effective action should be such that the equations of motion are
identical to the requirement of vanishing Weyl anomaly
coefficients. At one-loop order the effective action is given by
\beq\label{oneloopS}
S=\a'\int d^{D}x\sqrt{g}e^{-2\phi}
\left[
R-\frac{1}{12}H_{\mu\nu\lambda}H^{\mu\nu\lambda}
+4\partial_{\mu}\phi\partial^{\mu}\phi\right]\ ,
\eeq
where we have chosen the additive constant in the dilaton beta function to be $C=0$ and we are
using string metric with conventions such that $2\k_{0}^2=1$.

Using the one-loop expressions for the beta functions (\ref{betaol}) this effective action can
be written as
\beq
S=RV\equiv \frac{\delta V}{\delta g^i}\cdot \bar{\b}^i\ ,
\eeq
where $V=2\sqrt{g}\exp (-2\phi)$ -- which we will call the measure factor -- and where the dot
implies a functional integration. Our relation relating the RG flow and duality states that
$[T,R]=0$, while the natural requirement from string theory is that the low energy effective
action should be invariant under duality (since duality should just be a reformulation of the
same theory). However, these two conditions are not equivalent -- which is maybe not so strange
since in string theory we will not have any RG flow: cancellation of the Weyl anomaly requires
that the Weyl anomaly coefficients vanish for a consistent string theory. 

First, $[T,R]=0$ obviously does not imply that the low energy action is invariant under
duality, since using the above relation we have
\beq
TS=T(RV)=R(TV)=RV=S\ ,
\eeq
only if the measure factor is duality invariant: $TV=V$. Can we go the other way around, that
is, if both the effective action and the measure factor are invariant under duality: $TS=S$ and
$TV=V$, can we then conclude that duality and RG commute $[T,R]=0$? Using these two conditions
we can only conclude that 
\beq
[T,R]V=T(RV)-R(TV)=TS-RV=0\ ,
\eeq
which of course is not the same as $[T,R]=0$ as an operator equation. This can also be
demonstrated by looking at the one-loop Weyl anomaly coefficients. If we use the standard
duality transformations at one-loop then it is easy to see that they obey $TV=V$ together with
$TS=S$, but if we change the RG flow by multiplying $\bar{\b}^b_{\m\n}$ with two,
say, then duality and RG flow will no longer commute: $[T,R]\neq 0$. 

So, in conclusion, while the natural requirement in string theory is that the low energy
effective action is invariant, the natural condition in studying duality of sigma models is
that duality and RG flow commute (which must be the same as saying that duality is a quantum
symmetry). Furthermore, these conditions are not equivalent: if we are able to find a duality
transformation $T$ that keeps the low energy action
invariant at some order in $\a'$, then we cannot be sure that this $T$ will obey $[T,R]=0$. The
main reason is that the form of the action in (\ref{oneloopS}) does not determine the Weyl
anomaly coefficients uniquely: while any term $z^i$ obeying
$z^i\cdot (\delta V/\delta g^{i})=0$ can be added to the
beta function without changing the form of the action, but with
\beq
\bar{\b}^i(\tilde{g})+\tilde{z}^i(\tilde{g})=\frac{\delta \tilde{g}^i}{\delta g^j}\cdot 
(\bar{\b}^j(g)+z^j(g))\ .
\eeq
This is clearly not of the form in Eq. (\ref{haagolsen2}) and the flow is therefore not
covariant under duality.
\subsubsection*{Scheme (In)dependence}

In the discussion we have implicitly been using the assumption that the low energy action is
given as $S=(\delta V/\delta g^i)\cdot\bar{\b}^i$. This form of the action is only valid in a
specific scheme and has no invariant meaning (scheme ambiguity is here defined as
being equivalent to the possibility of having different field redefinitions). What is
invariant, or at least expected to be true in any scheme, is that the variation of the action with
respect to a coupling
$g^i$ is linear in the Weyl anomaly coefficients
\beq\label{inv?}
\frac{\delta S}{\delta g^i}=G_{ij}\cdot \bar{\b}^j\ ,
\eeq
with $G_{ij}$ invertible. Then the equations of motion will imply that the Weyl anomaly
coefficients vanish (which is one way of defining what the low energy effective action should
be). 
However, even formulated this way, it is clear that $T$-duality invariance of the effective
action does not generally imply that the Weyl anomaly coefficients transform covariantly. This
is because $G_{ij}$ is by construction such that equations of motion implies vanishing Weyl
anomaly coefficients; then one could imagine multiplying one $\bar{\b}^i$ with 2 and $G_{ij}$
with $1/2$ and (\ref{inv?}) would still be satisfied.

This conclusion, that invariance of the background effective action does not imply $[T,R]=0$
seems to be in conflict with the claims made in \cite{kalop9705}. In
this paper the bosonic, supersymmetric and heterotic sigma model
effective action is studied at two-loop order. A set of "corrected"
duality transformations at this order is then found by requiring
$T$-duality invariance of the effective action. We would add that this set of duality
transformations is not guaranteed to satisfy $[T,R]=0$, but we have not tried explicitly
to verify whether actually $[T,R]$ does vanish or not.  

Before turning to the case of purely metric background at two-loop, we
would like to make a further comment about scheme independence. 

One could enquire as to whether the relation $[T,R]=0$ makes any sense
independently of field redefinitions. That is, 
if we imagine that we have one scheme with duality
transformations that keeps the low energy action invariant and
satisfying $[T,R]=0$, can we then find the duality
transformations in any other scheme such that they preserve the
consistency condition and keeps the low energy action invariant?

This is in fact possible. As shown by Haagensen \cite{haag2} 
(to order \oaa) one can explicitly construct the set of
transformations keeping the effective action invariant in any other scheme.

On this background, we turn now to the case of two-loop corrections.

\section{Bosonic Models at Two-Loop}
              
In this section we show that bosonic sigma models have a two-loop beta
function which is of the form $\beta^{(2)}_{\mu\nu}=\lambda
R_{\mu\alpha\beta\gamma}{R_\nu}^{\alpha\beta\gamma}$ where
$\l$ is a constant which cannot be determined by the analysis \cite{HOS}.
Furthermore, it turns out that the "classical" $T$-duality transformations gets modified at
this order. This has also been discussed in \cite{tseytlina,forgacs}.
 
It is known \cite{metsaev} that the two-loop beta function is scheme
independent using the standard set of subtraction schemes determined by minimal and nonminimal
subtractions of the one-loop divergent structure $R_{\m\n}$. However,
this is only \cite{metsaev} true if we consider purely metric
backgrounds, so in order not to complicate the discussion we will concentrate on such
backgrounds. That is we take
\beq
g_{\mu\nu}=\left( \begin{array}{cc}a & 0\\ 0 &\bar{g}_{ij}
\end{array}\right)\ ,
\eeq
and $b_{\m\n}=0$ which implies that the dual background is torsionless also. 
Furthermore, we define $a_i\equiv \pa_i\ln a$ and $q_{ij}\equiv \bar{\nabla}_i a_j+\half
a_ia_j$. 

For such a background one-loop $T$-duality transformations amounts to $a\rightarrow 1/a$ while
$\bar{g}_{ij}$ is left unchanged. Our relation $[T,R]=0$ is satisfied
at one-loop as elaborated upon
in the last section -- so according to our general philosophy we only need to determine the corrections to $T$
such that this relation is also satisfied at two-loop. The resulting duality transformations have been
derived, using geometrical and duality arguments alone (i.e. without using any Feynman diagram calculations),
in \cite{HOS}: 
\bea\label{twoloopdual}
\ln\tilde{a}&=&-\ln a+\l\a' a_{i}a^{i}\ ,\nonumber\\
\tilde{g}_{ij}&=&g_{ij}=\bar{g}_{ij}\ , \nonumber \\
\tilde{\phi}&=&\phi -\frac{1}{2}\ln a +\frac{\l}{4}\a' a_{i}a^{i}\ ,
\ena
here $\lambda$ is an arbitrary constant which is left undetermined by the consistency
conditions. The transformation of the dilaton has been found by
requiring that the measure factor is invariant in order to have a
duality invariant low energy action also.

This set of corrected duality transformations were first derived by Tseytlin 
\cite{tseytlina}, motivated by the fact that the "classical" (i.e. one-loop) 
duality transformations
did not keep the effective two-loop action invariant (he found $\l=1/2$). As a puzzle we should add that in
that paper Tseytlin derived another set of duality transformations that keeps the string effective action
invariant - with the classical relation
$\ln\tilde{a}=-\ln a$ and $g_{ij}$ not invariant - but it turns out that
$[T,R]=0$ is not satisfied for these transformations. However, this is of course not in
contradiction with our general hypothesis that at any order in $\a'$ should $T$ be such that 
it commutes with the RG flow. 

Using $[T,R]=0$ on both sides of (\ref{twoloopdual}) we can as usual derive the consistency
conditions which are to be satisfied by the beta functions:
\bea\label{cc2loop}
\frac{1}{\tilde{a}}\tilde{\bar{\beta}}_{00} &=&
-\frac{1}{a}\bar{\beta}_{00}+2\l\alpha'\left[
a^{i}\partial_{i}\left( \frac{1}{a}\bar{\beta}_{00}\right)
-\frac{1}{2} a^{i}a^{j}\bar{\beta}_{ij}\right]\ ,\nonumber\\
\tilde{\bar{\beta}}_{ij}&=& \bar{\beta}_{ij}\ ,\nonumber\\
\tilde{\bar{\beta}}^{\phi}&=& \bar{\beta}^\phi-\frac{1}{2a}
\bar{\beta}_{00}+\frac{\l}{2}\alpha'\left[ a^{i}\partial_{i}\left(
\frac{1}{a}\bar{\beta}_{00}\right)
-\frac{1}{2} a^{i}a^{j}\bar{\beta}_{ij}\right]\ .
\ena
Scaling arguments alone determine the maximal set of tensors which can appear as
counterterms at this loop order as follows. 
Using that 
under $g\rightarrow \Omega g$ (with
$\Omega$ a global scaling-constant) the metric beta function must
scale as $T_{\m\n}(\Omega g)=\Omega^{-1}T_{\m\n}(g)$ 
constrains such possible counterterms to having the form
\bea\label{tenterms}
\b_{\m\n}^{(2)} &=& A_1\, \nabla_\mu\nabla_\nu R
+A_2\,\nabla^2 R_{\mu\nu}+
A_3\, R_{\mu\alpha\nu\beta}R^{\alpha\beta} \nonumber\\
&&+A_4\, R_{\mu\alpha\beta\gamma}{R_\nu}^{\alpha\beta\gamma}+
A_5\, R_{\mu\alpha}{R_\nu}^\alpha+A_6\, R_{\mu\nu}R \nonumber\\
&&+A_7\, g_{\mu\nu}\nabla^2 R +A_8\, g_{\mu\nu}R^2 +
A_9\, g_{\mu\nu}R_{\alpha\beta}R^{\alpha\beta}\nonumber\\
&&+A_{10}\,g_{\mu\nu}
R_{\alpha\beta\gamma\delta}R^{\alpha\beta\gamma\delta}\ ,
\ena
(the total number of possible terms at this order is 18, but only ten of these are linearly
independent, which can be shown by using the first and second Bianchi identity -- see Appendix
B). By using that the one-loop Weyl anomaly coefficient satisfies the one-loop
consistency condition, one can prove \cite{HOS} that the two-loop beta functions in the original
and dual backgrounds are related by
\beq\label{nodd}
\tilde{\b}_{ij}^{(2)}=\b_{ij}^{(2)}-\frac{1}{4}\l a_{(i}\pa_{j)}
(a^ka_k)\ .
\eeq
Under the duality transformation -- which at this order is just 
$a\rightarrow 1/a$ -- the possible tensor structures can be decomposed
into even and odd tensors:
\beq
\b_{ij}^{(2)}=E_{ij}+O_{ij}\ , \; \; \;
\tilde{E}_{ij}=E_{ij}\ , \; \; \; \tilde{O}_{ij}=-O_{ij}\ .
\eeq
Using (\ref{nodd}) we then 
arrive at the condition:
\beq\label{odd}
O_{ij}=\frac{1}{8}\l a_{(i}\pa_{j)}(a^ka_k)\ .
\eeq
We have to find the linear combination -- if there is any -- of the ten tensors in (\ref{tenterms})
which satisfy this relation. First, of course, we need to perform a Kaluza-Klein reduction
on these terms, which can be determined by the help of the expressions in Appendix A:
\bea
&(1)&:\ \nabla_{i}\nabla_{j}R = \bar{\nabla}_{i}\bar{\nabla}_{j}
(\bar{R}-{q_ n}^{n})\ ,\nonumber\\
&(2)&:\ \nabla^{2}R_{ij} = (\bar{\nabla}^{2}
+\frac{1}{2}a_{k}\bar{\nabla}^{k})
(\bar{R}_{ij}-\frac{1}{2}q_{ij})
-\frac{1}{4}a_{i}a_{j}{q_ n}^{n}\nonumber\\
&&\hspace{18mm}-\frac{1}{4}a^{k}a_{(i}\left(\bar{R}_{j)k}-\frac{1}{2}q_{j)k}
\right)\ ,\nonumber\\
&(3)&:\ R_{i\alpha j\beta}R^{\alpha\beta} = \frac{1}{4}q_{ij}{q_ n}^{n}
+\bar{R}_{injm}(\bar{R}^{nm}-\frac{1}{2}q^{nm})\ ,\nonumber\\
&(4)&:\ R_{i\alpha\beta\gamma}{R_ j}^{\alpha\beta\gamma} =
\frac{1}{2}q_{ik}{q_ j}^{k}+\bar{R}_{iknm}\bar{R_ j}^{knm}\ ,\nonumber\\
&(5)&:\ R_{i\alpha}{R_j}^\alpha =
\bar{R}_{ik}{\bar{R}_j}^{\;\, k}-{1 \over 2}\bar{R}_{k(i}{q_{j)}}^k+
{1 \over 4}q_{ik}{q_j}^k\ ,\label{kaluza}\\
&(6)&:\ R_{ij}R = (\bar{R}_{ij}-{1 \over 2}q_{ij})(\bar{R}-{q_n}^n)\ ,
\nonumber\\
&(7)&:\ g_{ij}\nabla^{2}R = \bar{g}_{ij}\left[ \half a^{k}\partial_{k}
(\bar{R}-{q_ m}^{m})\right.\nonumber\\
&&\hspace{21mm}\left. \phantom{ {a\over b}}
+\bar{\nabla}^{k}\partial_{k}(\bar{R}-{q_ m}^{m})
\right]\ ,\nonumber\\
&(8)&:\ g_{ij}R^{2} = \bar{g}_{ij}\left(\bar{R}-{q_ m}^{m}\right)^{2}\ ,
\nonumber\\
&(9)&:\ g_{ij}R_{\alpha\beta}R^{\alpha\beta} = \bar{g}_{ij}
\left[ \frac{1}{4}({q_ m}^{m})^{2}+(\bar{R}_{km}-\frac{1}{2}q_{km})^{2}
\right]\ ,\nonumber\\
&(10)&:\ g_{ij}R_{\alpha\beta\gamma\delta}R^{\alpha\beta\gamma\delta}
 =  \bar{g}_{ij}\left[ q_{km}q^{km}+\bar{R}_{k\ell mn}
\bar{R}^{k\ell mn}\right]\ .\nonumber
\ena
The corresponding odd parts of these tensors are:
\bea
O_{ij}^{(1)}&=&-\bar{\nabla}_{i}\bar{\nabla}_{j}
\bar{\nabla}_{n}a^{n}\ ,\nonumber\\
O_{ij}^{(2)}&=&\frac{1}{2}a_{k}\bar{\nabla}^{k}\bar{R}_{ij}
-\frac{1}{2}\bar{\nabla}^{2}\bar{\nabla}_{i}a_{j}
-\frac{1}{4}a_{i}a_{j}\bar{\nabla}_{k}a^{k}\ ,\nonumber\\
O_{ij}^{(3)}&=&-\frac{1}{2}\bar{R}_{injm}\bar{\nabla}^{n}a^{m}
+\frac{1}{8}a_{n}a^{n}\bar{\nabla}_{i}a_{j}
+\frac{1}{8}a_{i}a_{j}\bar{\nabla}_{n}a^{n}\ ,\nonumber\\
O_{ij}^{(4)}&=&\frac{1}{4}a_{k}a_{(i}\bar{\nabla}_{j)}a^{k}\ ,\nonumber\\
O_{ij}^{(5)}&=&-{1 \over 2}\bar{R}_{k(i}\bar{\nabla}_{j)}a^k+
\frac{1}{8}a_{k}a_{(i}\bar{\nabla}_{j)}a^{k}\ ,\label{odd2}\\
O_{ij}^{(6)}&=&-{1\over2}\bar{R}\bar{\nabla}_ia_j-\bar{R}_{ij}
\bar{\nabla}_na^n+{1\over4}a_ia_j\bar{\nabla}_na^n\nonumber\\
&&+{1\over4}a_na^n
\bar{\nabla}_ia_j\ ,\nonumber\\
O_{ij}^{(7)}&=&\bar{g}_{ij}\left[ \frac{1}{2}a^{k}\partial_{k}
(\bar{R}-\frac{1}{2}a_{m}a^{m})-\bar{\nabla}^{k}\partial_{k}
(\bar{\nabla}_{m}a^{m})\right]\ ,\nonumber\\
O_{ij}^{(8)}&=& \bar{g}_{ij}\left[-2(\bar{\nabla}^{k}a_{k})\bar{R}
+(\bar{\nabla}^{k}a_{k})a^{m}a_{m}\right]\ ,\nonumber\\
O_{ij}^{(9)}&=&\bar{g}_{ij}\left[ \frac{1}{4}(\bar{\nabla}^{k}a_{k})a^{m}a_{m}
-(\bar{\nabla}_{k}a_{m})\bar{R}^{km}\right.\nonumber\\
&&\; \; \; \; \;\left. +\frac{1}{4}
(\bar{\nabla}_{k}a_{m})a^{k}a^{m}
\right]\ ,\nonumber\\
O_{ij}^{(10)}&=& \bar{g}_{ij}
(\bar{\nabla}_{k}a_{m})a^{k}a^{m}\ .\nonumber
\ena
It is seen that the only odd term of the form (\ref{odd}) is $O_{ij}^{(4)}$ which originated from
$A_4R_{\m\a\b\g}R_{\n}^{\ \a\b\g}$. Moreover, it is possible to verify that a term like $O_{ij}^{(4)}$ cannot
be obtained as a linear combination of the remaining terms.
Hence, in  conclusion we have shown that with the requirement of covariance of duality under the RG,
the two-loop terms in the beta function must be:
\beq
\beta^{(2)}_{\mu\nu}=\lambda R_{\mu\alpha\beta\gamma}{R_\nu}^{\alpha\beta\gamma}\ .
\eeq
One should, however, keep in mind that the constant in front 
cannot be determined from this simple background -- the correct result is $\l=1/2$ \cite{tsey1,tsey2}.

We now know that the $(ij)$ component satisfies its consistency condition. What about the $(00)$ component? 
As it has been demonstrated in \cite{HO} the consistency condition for the $(00)$ component is also
exactly satisfied, and we can conclude that $[T,R]=0$ remains true at this two-loop order for the bosonic
sigma model. 

Furthermore, considering the bosonic sigma model on a {\sl flat} world sheet it is natural to ask whether
scale invariant models at this order are mapped to scale invariant models under duality? By rephrasing the
consistency conditions (\ref{cc2loop}) for the Weyl anomaly coefficients into conditions for the beta
functions, this statement is readily found to hold true also at \oaa \cite{HO}.   

\section{Supersymmetric Sigma Models}               

Having treated the bosonic sigma models at one- and two-loop it becomes natural to ask to what
extend $[T,R]=0$ holds true in the supersymmetric versions, which contrary to the
bosonic case describes consistent string theories. We start by describing the relevant
sigma models \cite{bus1,callan}.
\subsubsection*{The ${\cal N}=1,2$ Models}

There is a simple procedure to construct supersymmetric sigma models from bosonic ones. Using
complex coordinates on the worldsheet, the standard bosonic sigma model (involving only metric
and antisymmetric tensor) can be written as
\beq\label{N=0action}
S=\frac{1}{2\pi\a'}\int d^2z k_{\m\n}(X)\pa_{\bar{z}}X^{\m}\pa_{z}X^{\n}\ ,
\eeq
where $k_{\m\n}=g_{\m\n}+b_{\m\n}$. A (1,1)
supersymmetric model on the worldsheet is then constructed as follows. To the coordinates $(z,
\bar{z})$ we add two anticommuting complex coordinates $\theta$ and $\bar{\theta}$, with
\beq
\theta^2=\bar{\theta}^2=\{ \theta ,\bar{\theta}\}=0\ .
\eeq
Then define the superderivatives according to
\beq
D_{\theta}=\pa_{\theta}+\theta\pa_{z}, \; \; 
D_{\bar{\theta}}=\pa_{\bar{\theta}}+\bar{\theta}\pa_{\bar{z}}\ . 
\eeq
They are seen to satisfy
\beq
D_{\theta}^2=\pa_{z}, \; 
D_{\bar{\theta}}^2=\pa_{\bar{z}}, \;
\{ D_{\theta}, D_{\bar{\theta}}\}=0\ .
\eeq
It is well-known that a conformal transformation $z\rightarrow z'$ is defined by the property that 
$\pa_z=(\pa_{z'}/\pa_{z})\pa_{z'}$; likewise a superconformal
transformation $(z,\theta)\rightarrow
(z',\theta')$ obeys 
$D_{\theta}=(D_{\theta}\theta')D_{\theta'}$.

A superconformal invariant
action can now be written as a simple generalization of (\ref{N=0action}):
\beq\label{N=1action}
S=\frac{1}{2\pi\a'}\int d^2z d^2\theta  k_{\m\n}(X)D_{\bar{\theta}}X^{\m}D_{\theta}X^{\n}\ ,
\eeq
where $X^{\m}$ should be understood as a function of both bosonic and fermionic
coordinates:
$X^{\m}=X^{\m}(z,\bar{z},\theta,\bar{\theta})$. On the worldsheet, $X^{\m}$ is a scalar superfield (it has no
worldsheet indices) and after a Taylor expansion in $\theta$ we can write it as
\beq
X^{\m}(z,\bar{z},\theta,\bar{\theta})=X^{\m}+\theta\psi^{\m}+\bar{\theta}\tilde{\psi}^{\m}+\theta\bar{\theta}F^{\m}\
,
\eeq
where $X^{\m}$ is a scalar, $\psi^{\m}$ and $\tilde{\psi^{\m}}$ are spinors and $F^{\m}$ is an
auxiliary bosonic field. By using this Taylor expansion and performing the integral over the
fermionic coordinates (with $\int d^2\theta(\theta\bar{\theta})=1$)  one
obtains \cite{bus1}
\bea
S&=&\frac{1}{2\pi\a'}\int d^2z\left[ k_{\m\n}(X)\pa_{\bar{z}}X^{\m}\pa_{z}X^{\n}
-g_{\m\n}(X)(\psi^{\m}{\cal D}_{\bar{z}}\psi^{\n}+\tilde{\psi}^{\m}{\cal D}_{z}\tilde{\psi}^{\n})
\right.\nonumber\\
&&\hspace{10mm}\left.
+\frac{1}{2}R_{\mu\nu\lambda\rho}(X)\psi^{\mu}\psi^{\nu}\tilde{\psi}^{\lambda}\tilde{\psi}^{\rho}
\right]\ ,
\ena
where we have defined the expressions,
\bea
{\cal D}_{\bar{z}} \psi^{\n}&=& \pa_{\bar{z}}\psi^{\n} +(\Gamma^{\n}_{\l\r}(X)+\frac{1}{2}H^{\n}_{\
\l\r}(X))\pa_{\bar{z}}X^{\l}\psi^{\r}\ ,\nonumber\\
{\cal D}_{z} \tilde{\psi}^{\n}&=& \pa_{z}\tilde{\psi}^{\n} +(\Gamma^{\n}_{\l\r}(X)-\frac{1}{2}H^{\n}_{\
\l\r}(X))\pa_zX^{\l}\tilde{\psi}^{\r}\ .
\ena
Also, the auxiliary field $F$ has been integrated out. 

Clearly, the first term in this
action is just that of a standard bosonic sigma model. What about spacetime supersymmetry?
For general target manifold $M$ the action (\ref{N=1action}) describes
a ${\cal N}=1$ supersymmetric model \cite{callan}, as it is written in a manifestly
supersymmetric notation.

The existence
of higher (${\cal N} >1$) supersymmetry in the case of metric {\it and} torsion 
is discussed in \cite{gates}. We now turn now to the case of purely metric backgrounds,
that is $b_{\m\n}=0$.  

The condition for ${\cal N}=2$ spacetime supersymmetry has been derived
by Zumino
\cite{zumino}: the sigma model has ${\cal N}=2$ spacetime supersymmetry 
if and only if the target space $M$ is a K\"ahler
manifold (a complex manifold is K\"ahler if the metric is hermitian, that is
$g_{ij}=g_{\bar{i}\bar{j}}=0$, and it is locally of the form $g_{i\bar{j}}=\frac{\pa}{\pa
z^i}\frac{\pa}{\pa
\bar{z}^{\bar{j}}}K(z,
\bar{z})$, where $K(z,\bar{z})$ is a function called the K\"ahler
potential). The condition for ${\cal N}=4$
symmetry is that
$M$ is hyper-K\"ahler \cite{alv}; however for the rest of this section we will mainly concentrate
on the ${\cal N}=1, 2$ models. 

\subsubsection*{RG Flow}

Restricting to the case of torsionless backgrounds we only
have the metric tensor beta function to think about. Let us start with ${\cal N}=2$ models.
Such models have a metric beta function which is of the general form
\beq
\beta^{g}_{i\bar{j}}=a_1T^{(1)}_{i\bar{j}}+a_2T^{(2)}_{i\bar{j}}+a_3T^{(3)}_{i\bar{j}}+\ldots\ ,
\eeq
where $T^{(i)}$ is the $i$-loop counterterm. The form of the possible counterterms is -- as in
the bosonic case -- restricted by scaling arguments, but because of
${\cal N}=2$ supersymmetry there
is a further restriction on these terms. By Zumino's theorem \cite{zumino} the possible tensor
counterterms must be "K\"ahler" at any given order (since the unrenormalized metric on target space
is $g_{i\bar{j}}+T_{i\bar{j}}$). A K\"ahler tensor is a second rank tensor
$T_{IJ}$ with vanishing unmixed components, $T_{ij}=T_{\bar{i}\bar{j}}$ and mixed components
which are locally of the form $T_{i\bar{j}}=\pa_{i}\pa_{\bar{j}}S(z,\bar{z})$.

Using the same scaling arguments as in the
last section, 
at one-loop order the possible counterterms are 
\beq\label{1counter}
\beta^{g}_{i\bar{j}}=a_1R_{i\bar{j}}+b_{1}g_{i\bar{j}}R\ .
\eeq
However, the term proportional to $b_1$ is not a K\"ahler tensor, as it does not have vanishing
torsion, so $b_1$ must be identically zero. The Ricci tensor
$R_{i\bar{j}}=g^{l\bar{k}}R_{i\bar{j}\bar{k}l}$ is actually a K\"ahler tensor; it satisfies
\beq
R_{ij}=R_{\bar{i}\bar{j}}=0, \; \; R_{i\bar{j}}=-\pa_i\pa_{\bar{j}}\ln\det g\ ,
\eeq 
and qualifies therefore as a one-loop counterterm. It has been shown \cite{freed} that the
one-loop beta function is unchanged by the inclusion of fermions in the ${\cal N}=2$ model and is
therefore the same as for the bosonic sigma model,
\beq
\beta^{(1)}_{i\bar{j}}=R_{i\bar{j}}\ .
\eeq  

At two-loop order the possible counterterms are again restricted by the scaling arguments.
This will just give us the same ten counterterms as we had in the bosonic case.
We can repeat the analysis of two-loop beta functions that was carried out in the bosonic
case. Demanding our basic relation $[T,R]=0$, the only possible counterterm is
$R_{\m\gamma\sigma\lambda}R_{\n}^{\ \gamma\sigma\lambda}$, or on a complex K\"ahler manifold
\beq
\beta^{(2)}_{i\bar{j}}=a_2 R_{i\bar{k}l\bar{m}}R_{\bar{j}}^{\ \bar{k}l\bar{m}}\ . 
\eeq 
However, this is not a K\"ahler tensor and therefore we conclude that $a_2\equiv 0$. This result
is known from the literature \cite{freed}: all ${\cal N}=2$ sigma models have vanishing beta function at
two-loop and we therefore have essentially derived this result from the requirement of
consistency between RG flow and duality!

What about ${\cal N}=1$ models? At one-loop the ${\cal N}=1$ 
models have a beta function which is proportional
to the Ricci tensor as in the ${\cal N}=0, 2$ models \cite{freed,alv}. 
This however follows trivially from the
${\cal N}=2$ result: the only possible counterterms are as in (\ref{1counter}). Now imagine that
$b_1\neq 0$. Because of universality this must also be true when restricting the target space
to be K\"ahler. But K\"ahler geometry implies ${\cal N}=2$ supersymmetry
and therefore that $b_1=0$, which
is a contradiction. 

The same argument can be carried out for the two-loop term: at two-loop the only possible
counterterm is  $R_{\m\gamma\sigma\lambda}R_{\n}^{\ \gamma\sigma\lambda}$, but this is not a
K\"ahler tensor when restricting to K\"ahler target spaces and can consequently not appear as a
counterterm in the ${\cal N}=1$ models either.

In conclusion, using scaling arguments together with $[T,R]=0$ we have been able to prove
that both ${\cal N}=1,2$ supersymmetric sigma models have a vanishing
beta function at two-loop (note that we only considered the
torsionless case). 
It is natural to ask what happens for the ${\cal N}=4$
models. Such models are known to be ultraviolet finite, that is, their beta function is
identically zero \cite{sohnius}. Their target spaces are hyper-K\"ahler manifolds so it readily
follows from our analysis that the two-loop term must vanish, as a hyper-K\"ahler manifold is in
particular a K\"ahler manifold. That also the one-loop term must vanish seems to contradict
universality, but of course this is not the case: a hyper-K\"ahler manifold has vanishing Ricci
tensor so there can be no one-loop beta function either. 

Therefore, we might ask if we can "prove" that all higher loop counterterms must vanish 
for the ${\cal N}=4$ models starting with $[T,R]=0$? 

\subsubsection*{The Heterotic Sigma Models}

While the problem here is basically the same as before -- namely to see to what extend duality
can determine the one-loop beta functions -- there is a new ingredient in the appearance of a
target gauge field. Everything else being the same, we will concentrate on this related gauge
field beta function. The sigma model relevant for the heterotic string theories was described
in \cite{hullwitt}.

The gauge field in question, $A_{\mu}^{\ I}$, is in the adjoint of the gauge group $G$, i.e. 
$I=1,\ldots ,{\rm dim}G$,
which for the heterotic string is $G=Spin(32)/{\bf Z}_2$ or $E_8\times E_8$ \cite{gross}. 
The relevant sigma model action is \cite{hullwitt,alv9507}:
\bea
S &=&\frac{1}{4\pi\a'}\int d^2\s \left[
\bigl( g_{\mu\nu} + b_{\mu\nu} \bigr)
\partial_+X^{\mu}\partial_-X^{\nu} +
ig_{\mu\nu} \lambda^{\mu}
\bigl( \partial_-\lambda^{\nu} +
\bigl( \Gamma^{\nu}_{\rho\sigma}+{1 \over 2}{H^{\nu}}_{\rho\sigma}
\bigr)\partial_-X^{\rho}\lambda^{\sigma} \bigr) +\right.\nonumber \\
&&\left.+ i\psi^I \bigl(
\partial_+\psi^I + {{A_{\mu}}^I}_J\partial_+X^{\mu}\psi^J \bigr) +
{1 \over 2} {F_{\mu\nu}}_{IJ}\lambda^{\mu}\lambda^{\nu}\psi^I\psi^J\right] \ ,
\ena
where
\beq
H_{\mu\nu\rho}=\partial_{\mu}b_{\nu\rho}+\partial_{\nu}b_{\rho\mu}+
\partial_{\rho}b_{\mu\nu} \qquad {\rm and} \qquad
F_{\mu\nu}=\partial_\mu A_\nu - \partial_\nu A_\mu +[A_\mu,A_\nu]\ .
\eeq
Furthermore, the $\l^{\m}$ are left-handed Majorana-Weyl fermions and the $\psi^I$ are right-handed
Majorana-Weyl fermions.

We will denote by $\xi$ the Killing vector that generates the
Abelian isometry which enables duality transformations. Invariance of the action requires
a transformation of the gauge field: $\delta_\xi A_\mu\equiv{\pounds}_\xi
A_\mu={\cal D}_\mu\kappa$, where $\kappa$ is a target gauge parameter
\cite{hull9310,alv9510}. In adapted coordinates (where
$\xi^\mu\partial_\mu\equiv\partial_0$) we have
\cite{hull9310,alv9510}
\beq
{\cal D}_{\mu}\kappa \equiv \partial_\mu\kappa+[A_\mu,\kappa]=0\ .
\eeq
The duality transformation of the gauge field is 
\cite{alv9507,alv9510}, 
\beq
\tilde{A_0}_{IJ} = -{1 \over g_{00}}\mu_{IJ}\ , 
\eeq
and
\beq
\tilde{A_i}_{IJ} = {A_i}_{IJ}-{g_{0i}+b_{0i} \over g_{00}}\mu_{IJ}\ ,
\eeq
where we have defined $\mu_{IJ}\equiv(\kappa-\xi^\alpha
A_\alpha)_{IJ}=(\kappa-A_0)_{IJ}$ (the latter in adapted coordinates). The corresponding one-loop
duality transformations of the metric, antisymmetric tensor and dilaton fields
are as usual (and can in any
case be found in (\ref{dualb}) and (\ref{dilatonshift})). To study the relation between the
renormalization group flow and the duality we need as usual the beta functions (or better the Weyl
anomaly coefficients). For the heterotic sigma model, the Weyl anomaly coefficient is to one-loop
order
\cite{bell,tsey3}:
\beq\label{AWAC}
\bar{\beta}^{A}_\mu=\beta^A_\mu+\a'{F_\mu}^\lambda\partial_\lambda\phi
+{\cal O}(\alpha'^2)\ .     
\eeq
and the beta function is \cite{callan,grignani}:
\beq\label{betaA=}
\beta^{A}_{\mu} = {1\over2}\a'({\bf D}^{\lambda}F_{\lambda\mu}+
{1\over2}{H_{\mu}}^{\lambda\rho}F_{\lambda\rho})+
{\cal O}(\alpha'^2)\ ,      
\eeq
with ${\bf D}^{\l}$ the covariant derivative that includes both the gauge and metric
connections, that is ${\bf D}^{\l}F_{\l\m}=\nabla^{\l}F_{\l\m}+[A^{\l}, F_{\l\m}]$. The Weyl anomaly
coefficients for the remaining fields have been presented in (\ref{barg}), (\ref{barb})
and (\ref{barphi}). The consistency conditions follows from our basic equation
\beq
[T,R]=0\ ,
\eeq
or in this case
\beq
{\bar{\beta}}^{\tilde{A}}_{0} = {1 \over g_{00}}{\bar{\beta}}^A_0+
{1 \over g_{00}^2}(\kappa - A_0){\bar{\beta}}_{00}^g\ ,   
\eeq
\beq
{\bar{\beta}}^{\tilde{A}}_{i} = {\bar{\beta}}^A_i-
{1 \over g_{00}}\bigr(
(\kappa - A_0)({\bar{\beta}}_{0i}^g+{\bar{\beta}}_{0i}^b)-
(g_{0i}+b_{0i}){\bar{\beta}}^A_0 \bigl)+
{1 \over g_{00}^2}(g_{0i}+b_{0i})(\kappa-A_0)
{\bar{\beta}}_{00}^g\ .  
\eeq
Are these equations satisfied by and only by (\ref{betaA=})? This can of course be checked by
brute force. To simplify the discussion we will restrict to a diagonal metric background with no
torsion, i.e. 
\beq
g_{\mu\nu}=
\pmatrix{
a & 0 \cr
0 & \bar{g}_{ij} \cr
}\ ,      
\eeq
and we take $b_{\mu\nu}=0$. "Classical" duality transformations will then guarantee that there
is no torsion in the dual background either, see Eq. (\ref{dualb}). 

With this background the consistency conditions become simply
\beq\label{hetcc}
{\bar{\beta}}^{\tilde{A}}_{0} = {1 \over a}{\bar{\beta}}^A_0+
{1 \over a^2}(\kappa-A_0)\,{\bar{\beta}}_{00}^g\ ,        
\eeq
\beq
{\bar{\beta}}^{\tilde{A}}_{i} = {\bar{\beta}}^A_i\ . 
\eeq
Upon inserting the known beta functions, and doing the duality transformations,
there is, however, a small surprise since this is actually not the
result that one finds. Rather the result is \cite{OS}:
\beq\label{nhetcc}
{\bar{\beta}}^{\tilde{A}}_{0}={1 \over a}{\bar{\beta}}^A_0+
{1 \over a^2}(\kappa - A_0)(-{\bar{\beta}}_{00}^g)\ ,     
\eeq
\beq
{\bar{\beta}}^{\tilde{A}}_{i}={\bar{\beta}}^A_i\ .      
\eeq
For (\ref{hetcc}) to be consistent with (\ref{nhetcc}) -- or in order for duality to be a {\sl
quantum} symmetry -- we need to have:
\beq\label{kappa=A}
(\kappa - A_0)\,{\bar{\beta}}_{00}^g=0\ . 
\eeq
In the case of
the most general background what this is saying is that
$(\kappa - A_0)=0.$ What is the origin of such a condition? 

The answer to this question is that because of anomalies
the theory and its dual will only be equivalent when certain
conditions are met. 
Such anomalies come from chiral fermions which are part of the background fields. 

The simplest way to cancel the anomalies is to assume that the spin connection and gauge
connection match, i.e. $\omega=A$ \cite{gsw,alv9507,alv9510}. That this is a
consistent condition -- in the sense of duality -- follows from the fact that if $\omega=A$ then
also
$\tilde{\omega}=\tilde{A}$ in the dual theory \cite{alv9507,alv9510}. It then follows 
\cite{alv9507,alv9510} that $\mu=\Omega$, where we define
\beq
\Omega_{\mu\nu}\equiv{1\over2}(\nabla_\mu\xi_\nu-\nabla_\nu\xi_\mu)\ ,
\eeq
In adapted coordinates $\xi_\mu=g_{\mu0}$ and therefore -- since the connection is metric
compatible -- we have $\Omega=0$.  This readily answers our question
as to the origin of the condition (\ref{kappa=A}):
\beq
\mu_{IJ}=(\kappa-A_0)_{IJ}=0\ ,  
\eeq
because we demand that anomalies must cancel.
What this is telling us is that in order for the consistency condition,
as applied to the gauge
field, to be satisfied requires the cancellation of anomalies (the
logic might seem reversed here since in
order to have a well defined renormalization group 
flow the theory should be free of anomalies; however
the attitude we are taking here is that we will see how far the requirement of $[T,R]=0$ can
take us). 

So far we have shown that the consistency conditions (\ref{hetcc}) are satisfied by the Weyl anomaly
coefficients given in (\ref{AWAC}). But what we want is the other way around -- do the consistency
conditions determine the renormalization group flow?

The gauge beta function is a covariant tensor (a vector) on the space of background fields.
Together with the scaling arguments in \cite{sen} this determines that to one-loop order it must
be of the form
\beq\label{betaA}
\beta^{A}_{\mu} = c_1\,\a'{\bf D}^{\lambda}F_{\lambda\mu}+
c_2\,\a'{H_{\mu}}^{\lambda\rho}F_{\lambda\rho}\ ,    
\eeq
where $c_1$ and $c_2$ are constants that can be computed using one-loop Feynman diagrams as in 
\cite{callan,grignani}.
Here, these constants are determined by the beta function constraint:
\beq\label{Acons}
{\beta}^{\tilde{A}}_i={\beta}^A_i+{1\over2}\a'{F_i}^k\partial_k\ln a\ ,
\eeq
which follows directly from taking the dual of Eq. (\ref{AWAC}) and using that
$\bar{\b}^{\tilde{A}}_{i}=\bar{\b}^{A}_{i}$. 
For torsionless background, however, the last term in
(\ref{betaA}) is absent and we can therefore set
$c_2=0$. Insertion of (\ref{betaA}) in (\ref{Acons}) then gives
\beq
(c_1-{1\over2}){F_i}^k\partial_k\ln a=0\ ,       
\eeq
from which we obtain $c_1=\half$, which agrees with the result in Eq. (\ref{betaA=}).
We conclude that
we were able to uniquely determine the one-loop gauge field beta function -- though in the
particular case of vanishing torsion. Considering the case of torsion included \cite{OS}
shows that also the constant $c_2$ can be determined by (\ref{Acons}). The result is
$c_2=\frac{1}{4}$, which agrees with (\ref{betaA=}).

Thus, the consistency conditions are verified by, and only by, the correct
RG flows of the heterotic sigma model. In other words, classical target
space duality symmetry survives as a valid quantum symmetry of the heterotic
sigma model.

It is, however, not obvious how one should extend this analysis to higher ($\ell \geq 2$) loop
orders, since in this case the beta functions contain Lorentz and gauge variant terms
\cite{grignani}.

\section{Open Questions}

In this section we will address a few open questions which arise naturally in connection with the
situations analyzed in the preceding part of this chapter. In the last part of this section we will describe how the
relation $[T,R]=0$ has been derived at one-loop \cite{balog9806}.

\subsubsection*{Open String}

We have been analyzing the relation between renormalization group flow and duality for bosonic,
supersymmetric (with ${\cal N}=1,2$) and the heterotic sigma models (which have
${\cal N}=1$). It could be interesting to include the case of open (super)strings. 
In the open string case, 
the duality transformations are \cite{dorn1},
\beq
\tilde{A}_0=0 \qquad , \qquad \tilde{A}_i=A_i\ . 
\eeq
The consistency conditions associated with these transformations,
\beq
\bar{\beta}^{\tilde{A}}_0=0 \qquad {\rm and} \qquad 
\bar{\beta}^{\tilde{A}}_i=\bar{\beta}^A_i\ ,      
\eeq
are the same as  for the heterotic string with vanishing anomaly. 
Naively, scaling arguments would then dictate
that the only possible form  of the gauge field beta function is that of
Eq. (\ref{betaA}). 
However, this is not the whole story. The reason is that $T$-duality  -- as discussed further in the
next chapter -- interchanges the usual Neumann boundary condition $\pa_{n}X=0$ with the Dirichlet
condition $\pa_{\t}X=0$ in the dualized directions. This implies that the ends of the open string
are confined to move on D-branes and the position of this hyperplane $f(X^{\m})$ becomes a
dynamical field which should be included in the RG flow. For further discussion see \cite{dorn1}.

It could be interesting to include these boundary fields to study the RG flow of the open string
sigma model in the presence of D-branes. In some sense one would then have to require $[T,R]=0$
both in the bulk and on the boundary of the worldsheet surface. Maybe this could be important
for finding a non-Abelian version of the Born-Infeld action.   

\subsubsection*{Higher Loop Corrections}

Having shown that the two-loop beta functions must vanish for the
supersymmetric (${\cal N}=1,2$)
models, it is natural to enquire as to what happens for higher loop
orders. 
The three-loop correction to
the supersymmetric models is known to vanish \cite{grisaru2}, while the four-loop
correction is non-vanishing and very complicated indeed \cite{grisaru2}. To perform such an
analysis we need to know how the beta functions differ from the Weyl anomaly coefficients.
There is a difference at one-loop but no further terms at two-loop. However, there are
new terms appearing at three and higher loops, since in this case one cannot set $L_{\m}=W_{\m}=0$
\cite{tsey1}.

Another complication is that the number of counterterms restricted by scaling
arguments can be large -- the
maximal number of possible terms presumably grows exponentially with the number of loops. At
three-loop the counterterm must scale as
$T_{\m\n}(\Omega g)=\Omega^{-2}T_{\m\n}(g)$ and the tensors can be described  -- in the case
of no torsion -- as (i) tensors of the form
$\nabla\nabla\nabla\nabla R$ (2 terms), (ii) tensors like $R\nabla\nabla R$ (13 terms), (iii)
tensors like $\nabla R \nabla R$ (11 terms), (iv) tensors of the form $RRR$ (16 terms) and
finally (v) tensors which are of the form $g_{\m\n}\times ({\rm contractions}$) (16 terms).
These tensors (which have not yet been symmetrized) can be found in a paper by Fulling et al. 
\cite{fulling}. So just to verify that the three-loop beta functions
must vanish for ${\cal N}=1,2$
models seems very complicated from this point of view!

\bigskip 

We will end this chapter by mentioning, that
more recently Balog et al. \cite{balog9806} has presented a
derivation of (\ref{haagolsen}) to
one-loop order. Following \cite{balog9806}, let us write the 
worldsheet metric as
$h_{ab}=e^{\s(z)}\delta_{ab}$. Now let $g$ denote collectively the set of renormalized couplings
$(g, b,\f)$ and $g^0$ the bare couplings and let $Z^R[g;\s]$ denote the renormalized partition
function ($Z[g;\s]$ is the corresponding bare partition function). The starting point is the
Weyl anomaly coefficients being defined through the anomalous Ward identity:
\beq
\frac{\delta Z^R}{\delta \s(z)}=\langle T^{a}_{\ a}(z)\rangle
=\langle {\cal L}(\bar{\b})\rangle \ ,
\eeq 
(the meaning of the last expression is that the anomalous trace is given by the Lagrangian
${\cal L}$ with the Weyl anomaly coefficients $\bar{\b}(g)$ inserted instead of the background
couplings, see Eq. (\ref{wanomaly})) and the invariance of the bare partition function under
$T$-duality:
\beq
Z[g;\s]=Z[\tilde{g};\s]\ .
\eeq 
Using an infinitesimal Weyl transformation, $\s(z)\rightarrow \s(z)+\l(z)$, one derives
\beq
Z[g^0;\s+\l]=Z[g^0+\l \bar{B}(g^0); \s]\ ,
\eeq
where $\bar{B}(g^0)$ are the bare Weyl anomaly coefficients:
\beq
\bar{B}(g^0)=\frac{\delta g^0}{\delta g}\bar{\b}(g)\ .
\eeq
Using $T$-duality of the partition function one finds
\footnote{The duality transformation is denoted by $\tilde{g}=\G(g)$.}
\beq\label{wthend}
Z[g^0;\s+\l]=
Z[\tilde{g}^0+\l \bar{\delta}\G(g^0);\s]\ ,
\eeq
where
\beq
\l \bar{\delta}\G(g^0)= \G(g^0+\l \bar{B}(g^0))-\G(g^0)\ .
\eeq
Here we first performed a Weyl transformation, then a duality transformation. Instead
performing the duality transformation before the Weyl transformation one derives
\beq\label{dthenw}
Z[g^0;\s+\l]=Z[\tilde{g}^0+\l \bar{B}(\tilde{g}^0); \s]\ .
\eeq
Comparing (\ref{wthend}) with (\ref{dthenw}) then results in
\beq
\bar{\delta} \G(g^0)=\bar{B}(\tilde{g}^0)\ ,
\eeq
as $\l(z)$ was arbitrary. Formally this result is valid to all orders in perturbation theory,
but only to one-loop order is it true that bare and renormalized Weyl coefficients agree. Then
at one-loop one finally derives that
\beq
\bar{\b}(g)\frac{\delta \tilde{g}}{\delta g}= \bar{\b}(\tilde{g})\ .
\eeq
This is the same as having $[T,R]=0$. It would be interesting to generalize this result to two, or even
higher loop orders. A necessary condition for this relation to be true has also been derived in the mentioned
paper. Here it is shown -- at one-loop -- that if $[T,R]=0$ then given a renormalization of the original
theory, this will translate into a consistent definition of renormalized couplings of the dual theory,
meaning that the dual couplings are a finite function of the original couplings. Furthermore it is
demonstrated that $[T,R]=0$ at  one-loop for the $SU(2)$ WZW model and for certain models related by
Poisson-Lie $T$-duality. It would be interesting to study further the models with Poisson-Lie $T$-duality
since such models do not require an isometry of the target space for duality to work.


\chapter{String Duality}

In chapters 2 and 3 we have been considering some important dualities in
field theory, and to a lesser extent, in string theory. 
However, it seems almost impossible to discuss duality without saying
something about the astonishing results obtained only recently in string
theory; to mention a few: the understanding of the importance of so-called D-brane states and 
the connections between string theories and an eleven-dimensional "M-theory". 
Another important
motivation for discussing string duality is that the
$S$-duality in low energy ${\cal N}=2$ supersymmetric Yang-Mills theory can be "derived" from such
results. This chapter will therefore serve as a survey in which we
give an intentionally short introduction to some 
non-perturbative dualities in string theory. 

In the first section we review some basic facts about the known
perturbative
superstring theories in ten dimensions. We then turn to D-branes
which are a key element in establishing the
non-perturbative connection between the various string theories. In the
following section we review how all five of these theories can be connected
by considering their strong coupling limits. In the last section we
mention some important relations to field theory dualities in four
dimensions.

\section{Perturbative String Theory}

We start by describing some standard facts about superstring theory in
ten dimensions. Most important for the following discussion is the
massless R-R spectrum (which is connected to D-branes) and the massless NS-NS
spectrum (which plays a central role in the sigma model formulation of
string theory as described in Chapter 3). Useful introductions to
perturbative string theory can be found in
\cite{pol9411,ooguri9612,kirit9709}. 

It is known that there are five consistent perturbative superstring theories in
ten dimensions. Two of these theories, called the Type IIA and
Type IIB theories, have ${\cal N}=2$ spacetime supersymmetry. 
The three remaining theories have ${\cal N}=1$ spacetime 
supersymmetry, namely the
heterotic $SO(32)$ theory, the heterotic $E_8 \times E_8$ theory and the
Type I theory. The latter is a theory of open strings, while all the other
theories describe closed strings. 

The two Type II theories both have 32 supersymmetries in ten dimensions. 
In the Type IIA case, the supercharges are two Majorana-Weyl spinors of
opposite chirality with $Q_{\a}^1$ transforming under the ${\bf 16}$ of $SO(10)$ and 
$Q_{\a}^2$ the ${\bf 16'}$.
This notation refers to the fact that the 32-dimensional spinor representation
in ten dimensions can be decomposed as ${\bf 32}= {\bf 16}+ {\bf 16'}$. 
The NS-NS sector of this theory consists of a graviton $G_{\m\n}$, a
dilaton $\Phi$ and an antisymmetric tensor $B_{\m\n}$ (in form-language,
this is denoted $B_{2}$). The R-R sector consists of a one-form
$C_{1}$ and a three-form $C_{3}$. This theory is non-chiral and the supercharges have opposite
chirality.  

The Type IIB theory has two supercharges, or Majorana-Weyl
spinors, of the same chirality, with both $Q_{\a}^1$ and $Q_{\a}^2$ transforming under the 
${\bf 16}$ of
$SO(10)$.  
The NS-NS sector is the same as that of the Type IIA string: it contains a
graviton, a dilaton and an antisymmetric tensor. The R-R sector consists
of a zero-form $C_{0}$, a two-form $C_{2}$ and a four-form $C_{4}$.
The four-form has self-dual field strength the meaning of which is that
$F_{5}=dC_{4}=*F_{5}$. This theory is chiral.\footnote{In this chapter, $B_{p}$ will denote a
$p$-form coming from the NS-NS sector with field strength
$H_{p+1}=dB_{p}$; a R-R $p$-form is denoted by $C_{p}$ and its field
strength is $F_{p+1}=dC_{p}$.}

The Type I theory is obtained by gauging the worldsheet parity $\Omega$
(which interchanges left and right movers) of the IIB superstring. That
is, from the Type IIB theory one only keeps states with $\Omega=+1$. To
obtain a consistent theory one also adds open strings. The massless
bosonic spectrum then consists of a dilaton $\Phi$, a metric
$G_{\m\n}$ and from the R-R sector a two-form $C_{2}$. Also there are 496
gauge fields in the adjoint of $SO(32)$. The resulting theory has
${\cal N}=1$
spacetime supersymmetry. 

Finally, the two heterotic theories have 16 supersymmetries in ten
dimensions, that is ${\cal N}=1$ spacetime supersymmetry. The massless
bosonic fields of both theories are a metric 
$G_{\m\n}$, a scalar dilaton $\Phi$ and a
R-R two-form $C_{2}$. In addition one of the theories has 496 gauge fields
in the adjoint of $SO(32)$, the other theory in the adjoint of
$E_8\times E_8$.  

Perturbatively, all these superstring theories have of course a sigma model formulation as
in (\ref{maction}) with ten $X^{\m}$ embedding coordinates and various fermion and gauge field terms.

\section{D-Branes}
A central concept in the recent understanding of the various string
dualities is that of a D-brane (an excellent introduction can be found
in the paper by Polchinski \cite{pol9611}).

\begin{figure}[htb]
\begin{center}
\mbox{
\epsfysize=4cm
\epsffile{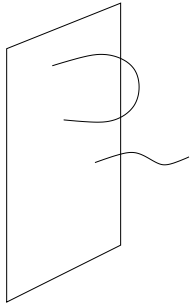}
}
\end{center}
\caption{Open strings attached to a D-brane.}
\label{dbrane}
\end{figure}

Loosely speaking a D-brane, of spatial dimension $p$, can be described as
a $p$-dimensional hypersurface on which open strings can terminate,
see fig \ref{dbrane}.
Topologically, the open string 
worldsheet is that of a ribbon and therefore,
we have to include boundary conditions of the worldsheet fields at the
end of the string. 

These boundary conditions can be determined by the variation of the
string action (\ref{maction}) with respect to $X^{\m}$:
\beq
\delta S_0=-\frac{1}{2\pi\alpha'}\int_{\Sigma}
d^{2}\sigma\delta X^{\mu}
\partial^2X_{\mu}+ \frac{1}{2\pi\alpha'}\int_{\partial\Sigma}
d^{2}\sigma\delta X^{\mu}\partial_nX_{\mu}\ ,
\eeq
with $\pa_n$ the derivative normal to the boundary.
The first term vanishes by the equations of motion and the second
term gives the two possible boundary conditions (for the open string we
take $0\leq\s\leq\pi$), namely the Neumann condition
\beq
\partial_nX^{\mu}|_{\s=0, \pi}=0\ ,
\eeq
or the Dirichlet condition
\beq
X^{\mu}|_{\s=0,\pi}={\rm constant}\ .
\eeq
The latter condition breaks translational invariance.
It is possible to disregard one of these conditions, say the Dirichlet
boundary condition; this condition then naturally appears after
using $T$-duality. Using (\ref{Tparity}) this is easily shown. Introducing worldsheet coordinates 
$\s^{\pm}=\t\pm\s$ and remembering  that in the directions in which
$T$-duality is performed we have $X^{m}\rightarrow X'^{m}=X_{L}(\s^+)-X_{R}(\s^-)$, we find
\bea
0&=&\pa_nX^m=\pa_+X^m-\pa_-X^m\nonumber \\
&=&\pa_+X'^m+\pa_-X'^m\nonumber \\
&=&\pa_{\t}X'^m\ .
\ena
This shows that Neumann boundary conditions have become Dirichlet
boundary conditions for the dual coordinates $X'^m$.

In the case of superstrings the
bosonic field $X^\mu$ necessarily appear together with fermionic fields
$\psi^\mu$. Therefore, i.e. because of worldsheet supersymmetry, the boundary
conditions for $X^{\m}$ should in principle be considered together with the
boundary conditions for
$\psi^{\m}$. However we will not consider fermions here.

By definition an open string ending on a D-brane obeys the Dirichlet
boundary condition in the direction transverse to the D-brane; 
more precisely, if we have
an open string joined to a
$p$ brane at
$x_{p+1}=\ldots =x_{25}=0$, the boundary condition for this string is
(for a bosonic string moving in 26 dimensions):
\bea\label{Dp}
(N)&& \partial_nX^m=0,\; \;  m=0,\ldots ,p\nonumber \\
(D)&& X^i= {\rm const}, \; \; i=p+1,\ldots ,25\ .
\ena
The Dirichlet condition implies that the ends of the
string are confined to move on the brane. This makes it possible to
have open strings stretched between two separated D-branes. 

A standard
way to introduce D-branes is by use of $T$-duality (actually, it was this
way that they were originally discovered, see \cite{dai89}). This is
because
$T$-duality interchanges Neumann and Dirichlet boundary conditions. For
example, in bosonic string theory, if we start with a theory of open
strings and $T$-dualize in  the $25-p$ directions, we then obtain a
D$p$-brane as in Eq.~(\ref{Dp}) (after taking the radius of the compact
dimensions to infinity).

This also implies that $T$-duality connects D-branes of different
dimensions. If $T$-duality is performed in a direction transverse to a
D$p$-brane then this brane is transformed into a D$(p+1)$-brane; if
done in a longitudinal direction the brane is turned into a D$(p-1)$-brane.

Where do such extended objects come from in string theory?
Here we should remember, that 
the spectrum of the different string theories
contains
$p+1$ forms $C_{p+1}$ coming from the R-R sector. The Type IIA
theory has forms with $p=0,2,4$ and the Type IIB with $p=-1,1,3$.
Consider now the integral
\beq
\mu_{p}\int_{V_{p+1}}C_{p+1}\ ,
\eeq   
where $V_{p+1}$ is some "surface" with $p$ spatial dimensions. By analogy
to electromagnetism, this is a natural
coupling of the
$p+1$ form where $\mu_p$ can be identified with the electric "charge". 
It is therefore tempting to conjecture that string theory contains
various higher-dimensional objects which couple electrically to the R-R
fields. Actually, in an important paper, 
Polchinski has identified such D-branes as BPS states which carry the R-R
charges \cite{pol9510a} (generally, the superstring $p$-brane configurations can be found
as solutions of the low energy effective supergravity theories, see \cite{duff}). 

There is a simple but important consequence of this identification:
in ten dimensions a $(p+2)$-form is Poincar\'e dual to a
$(8-p)$-form:
\beq
*(dC_{p+1})=*F_{p+2}=\tilde{F}_{8-p}=d\tilde{C}_{7-p}\ ,
\eeq
where $F_{p+2}$ is the field strength of $C_{p+1}$.
An "electrically" charged D$p$-brane is therefore dual to a
"magnetically" charged D$(6-p)$-brane in ten dimensions which is charged
under $\tilde{C}_{7-p}$. For example, a D0-brane of Type IIA theory is dual to
a D6-brane and a D3-brane of Type IIB theory is self-dual. As in the introduction, one
can -- using topological arguments alone -- derive a charge quantization
condition relating the electric and the magnetic charges
\cite{pol9611}:
\beq
\m_{6-p}\m_{p}=2\pi n\ , n\in {\bf Z}\ .
\eeq 
This relation turns out to be satisfied for $n=1$ \cite{pol9611}. 

The tension of a D-brane can be calculated by a vacuum loop diagram of
an open string with each end on the same kind of D-brane. For
a Dp-brane (of spatial dimension $p$) one finds \cite{pol9611}:
\beq\label{Dtension}
\t_p=\frac{2\pi}{g(4\pi^2\a')^{(p+1)/2}}\ ,
\eeq
where $g=e^{\Phi}$ is the string coupling; the tension is normalized
such that
$g=\t_{F}/\t_{D1}$, where $\t_F$ and $\t_{D1}$ are the string and D1-brane tensions respectively. 
We therefore see that a D-brane has the
special property that its mass goes like $1/g$, while a soliton
mass would vary like $1/g^2$. Anyhow, the result is clearly
non-perturbative. 

It is interesting to consider what happens when we have many D-branes. If
$N$ D-branes coincide, there will be new massless states since the mass
of the strings stretching between the branes is proportional to its
length. This gives a total of $N^2$ states (since the strings are
oriented) which exactly agrees with the dimension of the adjoint
representation of $U(N)$: in fact the gauge theory on the brane is a
$U(N)$ Yang-Mills theory \cite{witten9510}. For $N$ separated branes it would be
a $U(1)^N$ theory. 

The Lagrangian of the worldvolume theory is that of ${\cal N}=1$ ten-dimensional Yang-Mills theory,

\beq\label{tenYM}
S=\frac{1}{4g_{YM}^2}\int d^{10}x [F_{\mu\nu}F^{\mu\nu} + {\rm
fermions}]\ ,
\eeq
reduced to the ($p+1$)-dimensional worldvolume of the $p$-brane. The
Yang-Mills coupling on a D$p$-brane is:
\beq\label{gYM}
g_{YM}^2=\frac{1}{(2\pi\a')^2\t_{p}}
=2\pi g (4\pi^2\a')^{(p-3)/2}\ ,
\eeq
which can be derived from the fact that the gauge theory (\ref{tenYM})
is obtained 
by expanding a Born-Infeld action to leading order, 
see e.g. \cite{leigh89}.

Such D-branes are not only important in filling out duality
"multiplets", but has also played a central role in the problem of calculating
the entropy of black holes by counting microstates \cite{strom96,callan96}
and has offered a completely new approach to certain field theory problems
\cite{witten9703}. 

\section{String Dualities in $D=10$}

In this section we will consider the five superstring theories and their strong
coupling limits. This will demonstrate that they are (or at least seem to
be) all related as limits of a single theory. Maybe the
most important of these results is that one of the limits of this
theory is eleven-dimensional. A number of reviews on non-perturbative
string theory and $M$-theory can be found in Refs. 
\cite{townsend9612,kiritsis9708,sen9802}.

But first we consider the most basic duality in string theory, that is, the perturbative duality
relating the Type II theories: $T$-duality interchanges Type IIA and Type IIB
theories (so this is not really a duality in ten dimensions, but in nine
dimensions).
\subsubsection*{$T$-Duality of Type II Theories}

It is easy to see that the Type IIA and Type IIB theories are related by
$T$-duality \cite{dine89,dai89}. For example, start with the Type IIA in ten
dimensions and compactify the $X^9$ direction on a circle of radius $R$.
Taking the
$R\rightarrow 0$ limit is the same as taking the $R\rightarrow\infty$
limit in the dual coordinate (remember that $R'=\a'/R$ under duality)
with a right-moving coordinate which is reflected as in Eq. (\ref{Tparity}):
\beq
X'^9_{R}(\s^-)=-X^9_{R}(\s^-)\ .
\eeq
In addition, the right-moving 
worldsheet fermion $\psi^9_{R}(\s^-)$ must also be reflected as
\beq\label{refl}
\psi'^9_{R}(\s^-)=-\psi^9_{R}(\s^-)\ .
\eeq
This follows from the worldsheet supersymmetry of the Type II theories,
\beq
\delta X^{\m}=i\bar{\epsilon}\psi^{\m}\ , \; \; 
\delta \psi^{\m}=\gamma^a\pa_a X^{\m}\epsilon\ .
\eeq
with anticommuting parameter $\epsilon$ and $\g^a$ the Dirac matrices in two dimensions.
The transformation (\ref{refl}) changes the chirality of the right-moving
Ramond state and therefore the Type IIA theory is mapped to the Type IIB
theory and vice versa. While this establishes that the Type IIA and Type IIB are
perturbatively equivalent in 9 dimensions, one should check that
the duality maps the non-perturbative objects of Type IIA and Type IIB into
each other, which of course can be done \cite{witten9503}.
 
\subsubsection*{$SL(2,{\bf Z})$ Duality of Type IIB}

It has been conjectured that the Type IIB string theory is self-dual with a
duality group which is $SL(2,{\bf Z})$ \cite{witten9503,hull9410}.

The conjectured strong coupling limit of the Type IIB string theory can be motivated by looking
at its low energy action. The NS-NS fields of the Type IIB theory are a
graviton, a dilaton and an antisymmetric two-form. The R-R fields are a
zero-form, a two-form and a four-form with
selfdual field strength ($F_{5}=*F_{5}$).

It is not possible to write a covariant action for the selfdual
four-form -- the standard action $S=-\half \int F\wedge *F$ with $F=dC$
for
$F$ selfdual is $S=-\half \int F^2=-\half \int d(C\wedge F)$, i.e. the integrand is a total
derivative and therefore the action does not imply the equations of
motion, which are
$d(*F)=0$. 
The low energy IIB supergravity action does therefore not have a
term implying the $F_{5}$ equation of motion, but this equation must
be added as a constraint on the solutions, see \cite{schwarz83}.

Omitting fermions, the IIB supergravity action is a sum of three terms
\cite{howe} (with R-R fields and $H_3$ scaled by a factor of $1/\sqrt{2}\k_0$): 
\beq\label{SIIB}
S_{IIB}=S_{RR}+S_{NSNS}+S_{CS}\ ,
\eeq
with
\bea
S_{NSNS}&=& \frac{1}{2\k^2_{10}}\int d^{10}x \sqrt{G}e^{-2\Phi}
\left[ R+4\pa_{\m}\Phi \pa^{\m}\Phi \right]
-\half \int e^{-2\Phi}H_{3}\wedge *H_{3}\ ,\nonumber\\
S_{RR}&=& -\half\int\left[
F_{1}\wedge *F_{1}+F'_{3}\wedge *F'_{3}
+F'_{5}\wedge *F'_{5}\right]\ , \nonumber\\ 
S_{CS}&=&
-\sqrt{2}\k_{10}\int C_{4}\wedge H_{3}\wedge F_{3}\ , 
\ena
where the definitions are such that $F'_{3}=F_{3}-C_{0}\wedge H_{3}$
and $F'_{5}= F_{5}+ C_{2}\wedge H_{3}$ with $H_{p+1}=dB_{p}$ and
$F_{p+1}=dC_{p}$. This low-energy supergravity has an $SL(2,{\bf R})$
symmetry \cite{schwarz83}. 
This can be demonstrated by using instead the Einstein metric $\tilde{G}_{\m\n}$, 
in terms of which the IIB supergravity
action can be written as:
\bea\label{SIIBE}
S_{IIB}&=& \frac{1}{2\k^2_{10}}\int d^{10}x \sqrt{\tilde{G}}
\left[ \tilde{R}-\half\pa_{\m}\Phi \pa^{\m}\Phi \right]
-\half \int e^{-\Phi}H_{3}\wedge *H_{3}\nonumber\\
&& -\half\int\left[
e^{2\Phi}F_{1}\wedge *F_{1}+e^{\Phi}F'_{3}\wedge *F'_{3}
+F'_{5}\wedge *F'_{5}\right]\nonumber\\ 
&&-\sqrt{2}\k_{10}\int C_{4}\wedge H_{3}\wedge F_{3}\ . 
\ena
Choosing units where $2\k_{10}^2=1$ 
and collecting the axion and the dilaton into a complex coupling constant
according to:
\beq
\t=C_{0}+ie^{-\Phi}\ ,
\eeq
the action (\ref{SIIBE}) can be
written as
\bea
S_{IIB} &=&\int d^{10}x \sqrt{\tilde{G}}\left[
\tilde{R}-\frac{1}{2}\frac{\pa_{\m}\bar{\t}\pa^{\m}\t}{({\rm Im}\t)^2}\right]
\nonumber \\
&& - \half\int \left[ 
M_{ij}H^{i}_{3}\wedge *H^{j}_{3}+F'_{5}\wedge *F'_{5}
+\epsilon_{ij}C_{4}\wedge H_{3}^{i}\wedge H^{j}_{3}\right]\ .
\ena
Here we have defined the matrix
\beq\label{Mij}
M_{ij}=\frac{1}{{\rm Im}\t}\left( \begin{array}{cc}|\t|^2 & -{\rm Re}\t
\\ -{\rm Re}\t &1
\end{array}\right)\ ,
\eeq
and $H^{1}_{3}=dB_{2}\equiv dB_{2}^1$ (also $H^{2}_{3}=dC_{2}\equiv
dB_{2}^2$).  Now, the  $SL(2,{\bf R})$ symmetry generated by 
\beq\label{Lij}
\L=\left( \begin{array}{cc}a & b
\\ c &d
\end{array}\right)\ ,
\eeq
(i.e. $a$, $b$, $c$ and
$d$ are real with $ad-bc=1$) is:
\bea\label{SL2Rsym}
\t&\rightarrow& \frac{a\t+b}{c\t+d}\nonumber\ , \\
B_{2}^{i}&\rightarrow&\Lambda^{i}_{\ j}B_{2}^{j}\ ,
\ena
while keeping all other background fields invariant. 

Note that the $SL(2,{\bf R})$ mixes
$B_{2}$ with $C_{2}$, that is NS-NS with R-R fields. Since this is a
low energy symmetry the question is whether this symmetry survives in the
full quantum theory. Hull and Townsend conjectured \cite{hull9410} that the full Type IIB
theory is only invariant under
$SL(2,{\bf Z})$, because of the existence of solitonic objects
\footnote{More generally, let the symmetry group of the low energy supergravity in $d$
dimensions be denoted by $G({\bf R})$. Hull and Townsend have
conjectured \cite{hull9410} that the duality group of the full string
theory is an integer form of this group, $G({\bf Z})$. This group is
called $U$-duality.}. Let us see what this symmetry group corresponds to.
$SL(2,{\bf Z})$ can be generated by the two transformations
\beq
\t'=-1/\t\ ,
\eeq 
and
\beq
\t'=\t+1\ .
\eeq
Setting $C_{0}$ to zero, the
first of these transformations correspond to:
\bea\label{sl2z}
\Phi&\rightarrow&-\Phi \ , \; \; \; \;
G_{\m\n}\rightarrow e^{-\Phi}G_{\m\n}\ ,\nonumber
\\ B_{2}&\rightarrow&C_{2} \ , \; \; \; C_{2}\rightarrow -B_{2}\ ,\nonumber
\\ C_{4}&\rightarrow&C_{4}\ .
\ena
Since $g=\langle e^{\Phi}\rangle$ this is an 
$S$-duality that interchanges strong and
weak coupling. Also, it interchanges the NS-NS two-form that couples to
the fundamental string with the R-R two-form that couples to the D1-brane
of Type IIB string theory.

Actually, the relative tensions of the D-string and the F-string is
$\t_{F1}/\t_{D1}=g$.
Since this is a consequence of the BPS property of the D1-brane it must
be an exact expression and can therefore be extrapolated to strong
coupling without any corrections. We then see that at weak coupling the
D-string is very heavy compared to the F-string and effectively decouples
from the spectrum. What happens at strong coupling? Here the situation is
reversed, the D-string is light while the F-string is very heavy. 
So the natural assumption is that in  
the theory at strong coupling the D-string changes role with the
F-string (and vice versa) and the string coupling is
$1/g$.

Likewise, the $\t\rightarrow \t+1$ symmetry can be interpreted as a shift of the R-R
zero form:
\beq
C_{0}\rightarrow C_{0}+1\ .
\eeq
This keeps the field strength $F_{1}=dC_{0}$ invariant and is just a
symmetry of the perturbative string theory. 

While the D-string and F-string are interchanged under the $SL(2,{\bf
Z})$ duality it becomes natural to ask what happens with the other
extended objects? 
Under the transformation (\ref{sl2z}) the R-R four-form
is invariant so it keeps the D3-brane invariant also. 
The electrically charged D-string is dual to a magnetically charged
D5-brane. The 6-form $\tilde{C}_{6}$ that couples to this object can be
obtained by Poincar\'e duality: 
\beq
*dC_{2}=*F_{3}=\tilde{F}_{7}=d\tilde{C}_{6}\ .
\eeq
Using $SL(2,{\bf Z})$-duality this is transformed into a magnetic
source for the NS-NS two-form. 
This magnetic source is the so-called NS
5-brane. Its tension, which can be derived from the fact that one should
have  $\t_{D1}\t_{D5}=\t_{F1}\t_{NS5}$, is
\beq
\t_{NS5}=\frac{1}{2\pi^5g^2\a'^3}\ ,
\eeq
and therefore it is not a D-brane (which have a tension $\t_{D}\sim
1/g$), but a soliton which have a tension $\t_{NS}\sim 1/g^2$.

In conclusion, the conjecture is that the strong coupling limit of Type IIB
theory is again a Type IIB theory but with the D1 string playing the role of
the fundamental string and a string coupling $g'=1/g$.  
\subsubsection{Type IIA and $M$-Theory}

The most surprising of the string theory might be 
the Type IIA/$M$-theory duality: it is conjectured that the
strong coupling limit of the Type IIA theory is not a ten-dimensional string
theory, but rather an eleven-dimensional theory \cite{witten9503}. 

It has been known for a long time that eleven-dimensional
supergravity can be related to IIA supergravity by dimensional reduction.
To see how this works, we will only look at the bosonic parts of the
action. The eleven-dimensional supergravity has a three-form potential
$B_{3}$; this naturally couples to a 2-brane and is Poincar\'e
dual to a magnetically charged 5-brane. The bosonic part of
the eleven-dimensional supergravity is \cite{cremmer} 
(with $B_3$ scaled by a factor of $1/\sqrt{2}\k_{11}$):
\beq\label{S11}
S_{11}=\frac{1}{2\k_{11}^2}\int d^{11}x\sqrt{G}R
-\int \left[
\half H_{4}\wedge *H_{4}+\frac{\k_{11}}{3\sqrt{2}}B_{3}\wedge
H_{4}\wedge H_{4}
\right]\ .
\eeq
The dimensional reduction is performed by taking the eleventh
direction $x^{10}$ periodic,
\beq
x^{10} \sim x^{10}+2\pi R\ .
\eeq
The eleven-dimensional metric will then separate into 
$G_{\m\n}$, $G_{\m 10}$ and $G_{10 10}$. This is a metric, vector and
scalar from the ten-dimensional point of view. In ten-dimensional string
theory we only have one scalar, the dilaton $\Phi$, and it is therefore clear
that string coupling, which is determined by $\Phi$, and the radius of
the eleventh dimension must be related. It is convenient to define
$G_{10 10}=e^{2\g}$. Now, decompose the eleven-dimensional metric as
\beq\label{decom}
ds^2=G_{MN}dx^M dx^N = G_{\m\n}dx^{\m}dx^{\n}+
e^{2\g}(dx^{10}+A_{\m}dx^{\m})^2\ ,
\eeq 
and all background field are taken to be independent of the
compact direction $x^{10}$. This means that 
the momentum $p_{10}$ is zero.

The three-form potential $B_{3}$ will similarly result in a
three-form in ten dimensions, which we also call $B_{3}$, and a two-form
$B_{2}$ coming from $B_{\m\n 10}$. By dimensionally reducing in
this way we get for the three terms in the action (\ref{S11})
\footnote{With the metric in (\ref{decom}) the eleven-dimensional Ricci scalar becomes
$R^{(11)}=R^{(10)}-2e^{-\g}\nabla^2
e^{\g}-\frac{1}{4}e^{2\g}F_{\m\n}F^{\m\n}$, where $F=dA$. This can be
shown for example with the help of the formulas (\ref{riccitensor}) in Appendix A.}:
\bea\label{2a}
S_{1}&=&\frac{1}{2\k_{10}^2}\int d^{10}x\sqrt{G}e^{\g}R
-\half \int 
e^{3\g}H_{2}\wedge *H_{2}\ ,\nonumber\\
S_{2}&=&-\half \int\left[
e^{-\g}H_{3}\wedge
*H_{3}+e^{\g}H'_{4}\wedge *H'_{4}\right]\ ,\nonumber\\
S_{3}&=&-\frac{\k_{10}}{\sqrt{2}}\int B_{2}\wedge H_{4}\wedge H_{4}\ ,
\ena
where we have defined $H'_{4}=H_{4}+B_{1}\wedge H_{3}$ and 
$B_1=A_1/\sqrt{2}\k_{10}$; $B_{2,3}$ have been rescaled by a factor
$(2\pi R)^{-1/2}$.
The ten-dimensional gravitational coupling constant is related to the
eleven-dimensional by $\k_{10}^2=\k_{11}^2/2\pi R$. 
The bosonic fields of the ten-dimensional theory are a metric, a scalar,
two two-forms and a one-form. This coincides with the bosonic content of
the IIA theory. To see how this works in more detail rescale the metric
\beq
G_{\m\n}\rightarrow e^{-\g}G_{\m\n}\ ,
\eeq
and choose $\g=2\Phi/3$. Then the action consisting of the sum of the
three terms in (\ref{2a}) can be written as (note that we have already
identified it with the IIA supergravity): 
\beq
S_{IIA}=S_{NSNS}+S_{RR}+S_{CS}\ ,
\eeq
where
\bea
S_{NSNS}&=& \frac{1}{2\k^2_{10}}\int d^{10}x \sqrt{G}e^{-2\Phi}
\left[ R+4\pa_{\m}\Phi \pa^{\m}\Phi \right]
-\half \int e^{-2\Phi}H_{3}\wedge *H_{3}\ ,\nonumber\\
S_{RR}&=& -\half\int\left[
F_{2}\wedge *F_{2}+F'_{4}\wedge *F'_{4}
\right]\ , \nonumber\\ 
S_{CS}&=&
-\frac{\k_{10}}{\sqrt{2}}\int B_{2}\wedge F_{4}\wedge F_{4}\ , 
\ena
in terms of the new metric which we also denote by $G$ (to obtain the
first integral in the NS-NS part one can use the formula (\ref{rtilde}) from
Chapter 3). It follows that the IIA supergravity can be obtained as a
dimensional reduction of the eleven-dimensional supergravity. It seems
natural to ask what happens for the full Type IIA string theory? This has been
answered by Witten
\cite{witten9503} and Townsend \cite{town9501}. As in the Type IIB theory we
can learn important things by looking at the behaviour of the
non-perturbative objects. In the Type IIA theory the natural objects to look
at are the D0-branes. These are the objects that are electrically charged
under
$C_{1}$. 
The tension of such a D0-brane is by (\ref{Dtension}):
\beq
\t_{DO}=\frac{1}{g\a'^{1/2}}\ .
\eeq
This means that at weak coupling the D0-brane is very heavy and decouples
while at strong coupling it becomes very light. Furthermore, it has been
shown \cite{sethi,porrati} that for $n=2$ and $n$ prime there is a bound state of $n$ D0-branes at
threshold. Assuming this to be true for any positive integer $n$, we get a bound state with mass
\beq
n\t_{DO}=\frac{n}{g\a'^{1/2}}\ .
\eeq
This is a BPS state so this expression is exact and therefore can be used
even at strong coupling. At weak coupling, that is $g \rightarrow 0$,
these masses diverge and this is the reason why the states are not seen
in the elementary string spectrum. 
At strong coupling what happens is
that these states become very light and form a continuum.
Could a (perturbative) string theory have this kind of spectrum? Possibly
not, since there is no other known string theory that has Type IIA
supersymmetry in ten dimensions, but the spectrum could originate from a
Kaluza-Klein theory which has an infinite tower of excitations. Consequently, 
it is natural to assume that at strong coupling, spacetime is ${\bf R}^{10}\times {\bf S}^1$
and not ${\bf R}^{10}$. The radius of the eleventh dimension must then be
related to the string coupling according to: 
\beq\label{r10}
R=g\a'^{1/2}\ .
\eeq
This is the wanted relation between the radius of the compact dimension and the dilaton.
And a massless particle in eleven dimensions with one unit of
Kaluza-Klein momentum is interpreted as a D0-brane in the Type IIA theory.
Note that for this
interpretation to make sense, we must now include states with
$p_{10}\neq 0$ and not (as in the standard dimensional reduction) only
consider states with $p_{10}=0$. This also makes it clear that -- without
including the D0-branes in the description of Type IIA theory -- this extra
dimension is not seen at all in perturbation theory. 

It has been shown that the strong coupling limit of ten-dimensional IIA
supergravity is eleven-dimensional supergravity. The IIA supergravity is,
on the other hand, the low energy limit of Type IIA string theory. It is
natural then to ask: what is eleven-dimensional supergravity the
low energy limit of? The answer (even though we do not know it yet) has
tentatively been called $M$-theory, where $M$ could stand for  ``mystery'',
``mother'', ``membrane'' or ``matrix'' if you like.  So $M$-theory is a
hypothetical eleven-dimensional consistent quantum theory, whose low-energy
limit is eleven-dimensional supergravity.

What is $M$-theory? We will not try to answer this question, but at least
we can note some basic facts. $M$-theory has a metric and more
notably a three-form potential
$B_{3}$ which couples to an electrically charged 2-brane; this is
called the $M2$-brane. The three-form is Poincar\'e dual to a 6-form
potential $\tilde{B}_{6}$; the corresponding magnetically charged object
is a 5-brane and is called the $M5$-brane. Furthermore, the theory is
characterized by one length scale, the eleven-dimensional Planck
length $\ell_{11}$.

Now we have a Type IIA theory which naturally lives in ten dimensions and
$M$-theory which is eleven-dimensional. Compactifying $M$-theory on a circle
and taking this circle to be small we should obtain the Type IIA theory. This
means that all perturbative and non-perturbative objects in the Type IIA
theory should be obtainable from $M$-theory. Let us see how this works
out for a few interesting cases (see e.g. \cite{town9512} for further
discussion).

The D2-brane in Type IIA string theory is identified with an $M2$-brane which
is transverse to the compact dimension; it couples to $B_{3}$ which
originated form the three-form in $M$-theory. 

The fundamental IIA string is identified with an $M2$-brane which is
wrapped around the compact direction. Such an object is charged under
$B_{\m\n 10}$ which in the ten-dimensional theory was interpreted as
the NS-NS two-form and therefore couples to the fundamental string
consistent with the assumption. The tension of a wrapped $M2$-brane is
\beq
\t_{F1}=2\pi R_{10}\t_{M2}=\frac{1}{2\pi\a'}\ ,
\eeq
because of the above-mentioned identification and using (\ref{r10}). This
of course gives the correct result. 

The Type IIA theory fundamental string gives (like in the Type IIB theory) rise to
an NS 5-brane. This is interpreted as an $M5$-brane that is transverse to
the compact dimension. Therefore their tensions must be equal,
\beq
\t_{NS5}=\t_{M5}\ ,
\eeq
with
\beq
\t_{NS5}=\frac{1}{(2\pi)^5g^2\a'^3}\ .
\eeq
Note that
its tension goes like $1/g^2$ as it should for a soliton.

To conclude, while the Type IIB theory is
self-dual, meaning that its strongly coupled theory is again a Type IIB
theory but with other couplings, the Type IIA theory has an
eleven-dimensional theory as its strong coupling limit.
In detail $M$-theory compactified on a ${\bf S}^1$ is Type IIA string theory. Is
there a similar result for the Type IIB theory? In fact there is \cite{aspinwall}. The idea
is to compactify $M$-theory on a two-torus ${\bf T}^2$ with radii $R_{10}$
and
$R_{9}$. Keeping $R_{10}$ fixed and taking the limit $R_{9}\rightarrow
\infty$ gives as just stated Type IIA theory. Because of $T$-duality this theory
is $T$-dual to Type IIB theory compactified on a circle of radius $\a'/R_{9}$.
Then, in the limit of $R_{10}\rightarrow 0$, $R_9\rightarrow 0$ with
$R_{10}/R_9$ fixed one gets uncompactified Type IIB theory
\cite{townsend9612}. So the Type IIB theory can be understood as $M$-theory
compactified on a "vanishing" torus. In this picture the $SL(2,{\bf Z})$
duality group of the Type IIB theory comes from the
$SL(2,{\bf Z})$-transformations of the ${\bf T}^{2}$ \cite{schwarz9508}.  

From the Type II theories we now
turn to the heterotic and open string theories.
\subsubsection*{Type I/ $SO(32)$ Heterotic Duality}

The $SO(32)$ Type I open string theory is conjecture to be dual to the
heterotic $SO(32)$ theory \cite{witten9503,pol9510}. This again can be
motivated by looking at the respective low energy actions. The low energy
supergravity is in both cases ${\cal N}=1$ so it is maybe not that surprising
that a certain relation can interchange the two theories.

The bosonic part of the Type I low energy action is \cite{gsw}:
\beq\label{SI}
S_I=S_{1}+S_{2}\ ,
\eeq
where
\bea
S_{1}&=& \frac{1}{2\k^2_{10}}\int d^{10}x \sqrt{G}e^{-2\Phi}
\left[ R+4\pa_{\m}\Phi \pa^{\m}\Phi \right]
-\half \int e^{-2\Phi}F'_{3}\wedge *F'_{3}\ ,\nonumber\\
S_{2}&=& -\frac{1}{4}\int d^{10}x \sqrt{G}e^{-\Phi}
F^{a}_{\m\n}F^{a\m\n}\ .
\ena
The first term is readily constructed from the IIB supergravity action by
using that only the metric, the dilaton and the two-form $C_{2}$
survives the projection onto states with $\Omega=+1$.  The last term is
just the Yang-Mills action of the
$SO(32)$ gauge field with $F=dA_1+gA_{1}^2$, where $g$ is the
Yang-Mills coupling. The definitions are such that
\beq
F'_{3}=dC_{2}-\frac{\k_{10}}{\sqrt{2}}\omega_{3}\ ,
\eeq
and $\omega_{3}$ is the Chern-Simons form
\beq
\omega_{3}=A_{1}^a\wedge dA_{1}^a +\frac{2}{3}g
f^{abc}A_{1}^a\wedge A_{1}^b\wedge A_{1}^c\ .
\eeq
The heterotic string, on the other hand, has a low energy action which is
\beq\label{Shet}
S_{het}= \int d^{10}x \sqrt{G}e^{-2\Phi}
\left[ \frac{1}{2\k_{10}^2}R+\frac{2}{\k_{10}^2}\pa_{\m}\Phi \pa^{\m}\Phi
-\frac{1}{4}F^{a}_{\m\n}F^{a\m\n}
\right] -\half \int e^{-2\Phi}H'_{3}\wedge *H'_{3}\ ,
\eeq
where $H'_{3}$ is the same as the $F'_{3}$ of the Type
I string. The two actions (\ref{SI}) and (\ref{Shet}) can be mapped into
each other by the transformations
\bea\label{typeIhet}
G_{I\m\n}&\rightarrow&e^{-\Phi_h}G_{h\m\n}\nonumber\\
\Phi_I&\rightarrow&-\Phi_h\nonumber\\
F'_{3I}&\rightarrow&H'_{3h}\nonumber\\
A_{I\m}&\rightarrow&A_{h\m}\ .
\ena
It seems natural to conjecture that the strong coupling limit of
Type I string theory is heterotic $SO(32)$ theory, since the
dilaton transforms as $\Phi_I=-\Phi_h$ and the string couplings are
accordingly related as $g_{I}=1/g_{h}$.

The D1-brane of the open string theory is then identified with the
fundamental string in the heterotic theory \cite{pol9510}. The
tension of this D-string is 
\beq
\t_{D1}=\frac{1}{2\pi\a' g_I}\ .
\eeq 
The D-string must be dual to a magnetically charged D5-brane.
In the heterotic theory this D5-brane is identified with a solitonic NS
5-brane \cite{witten9511}.

So, the conjecture is that the strong coupling limit of $SO(32)$ heterotic
string theory is an open string theory -- and we only miss considering the
strongly coupled region of the
$E_8\times E_8$ heterotic theory.
\subsubsection*{$E_8\times E_8$ Heterotic Theory and $M$-theory on ${\bf S}^{1}/{\bf Z}_{2}$}

To determine the strong coupling behaviour of the
$E_8\times E_8$-theory \cite{horava9510} one uses the fact that this theory 
is connected to the $SO(32)$ heterotic theory by $T$-duality \cite{narain86}.

To use this duality we start by compactifying the $E_8\times E_8$ theory
on a circle of radius $R_9$; including a Wilson line in this direction
will break this group to $SO(16)\times SO(16)$. What is meant by this is that
we assume that the gauge field has a non-vanishing vacuum expectation value in
the compact direction. We therefore have that this is
$T$-dual to the
$SO(32)$ theory broken to
$SO(16)\times SO(16)$ \cite{narain86}. The relation between the couplings and
radii are then by (\ref{2tdual}):
\beq
R'_{9}=\frac{\a'}{R_9}\ , \; \; g'=\frac{g\sqrt{\a'}}{R_9}\ .
\eeq
The primed quantities are here referring to the $SO(32)$ theory. However,
from previous comments we know that this theory is related to the Type I
$SO(32)$ theory by
$S$-duality with
\beq
g_I=\frac{1}{g'}=\frac{R_9}{g\sqrt{\a'}}\ ,
\eeq
and (because of (\ref{typeIhet})):
\beq
R_{I}=\frac{R_{9}'}{g'^{1/2}}=
\frac{\a'^{3/4}}{g^{1/2}R_{9}^{1/2}}\ .
\eeq
The limit we are interested in is the decompactifying limit,
$R_9\rightarrow \infty$, with $g$ large. It is seen that the Type I
theory will be strongly coupled in this limit, but we need to relate the
original theory to a weakly coupled theory. A $T$-duality in the 9-direction
relates the Type I theory to another theory which is usually called the Type I' theory
and can be viewed as Type IIA theory compactified on an interval ${\bf
S}^1/{\bf Z}_2$ \cite{horava9510}. Since
$T$-duality interchanges Neumann with Dirichlet boundary conditions, the open
strings in this theory must have their ends stuck on D8-branes. It therefore
turns out \cite{townsend9612} that the compact dimension becomes an interval
of length $R_9''$ with 8 D8-branes at each end. Now, the $M$-theory interpretation
of this must be that the Type I' theory should be identified with $M$-theory
compactified on ${\bf S}^1\times {\bf S}^1/{\bf Z}_2$, with "radii" $R_{10}$
and $R_{9}''$. 
Also, the open strings of the Type I' theory are $M2$-branes
which are wrapped around the ${\bf S}^1$ and stretching between the D8-branes.  

In the end, this means that the strong coupling behaviour of the $E_8\times E_8$
is controlled by $M$-theory compactified on an interval. The group structure
$E_8\times E_8$ is here identified with the gauge fields living on the
two ends of the interval -- that is on the D8-branes.

\begin{figure}[htb]
\begin{center}
\mbox{
\epsfxsize=11cm
\epsffile{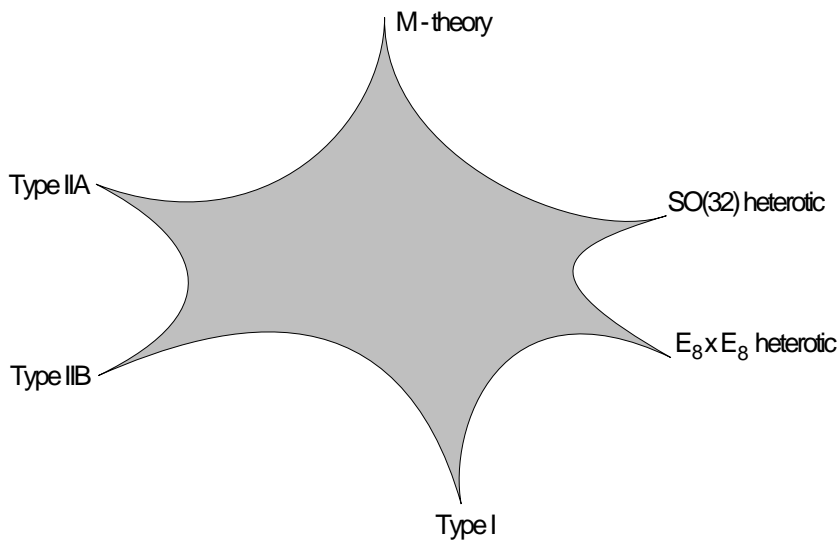}
}
\end{center}
\caption{The moduli space of superstring theory.}
\label{mspace}
\end{figure}

The upshot is that for every perturbative string theory in 
ten dimensions there is a candidate for its strong coupling limit.
This is usually illustrated as in fig. \ref{mspace}, where five of the  "cusps"
represent weakly coupled string theories and one is $M$-theory. 
There are
two things to say about this picture. First of all the picture -- as it
comes from duality -- is conjectural. There are no "proofs" of these
duality conjectures but only a number of consistency checks. However, so
far there seems to be no inconsistencies. Secondly, it is only certain
regions of the moduli space which is covered by known theories -- so
there is no fundamental definition of the theory behind all this (if it
is unique).

\subsubsection*{M as in Matrix?}

One attempt in the direction of giving a non-perturbative definition of
$M$-theory is the Matrix Theory conjecture by Banks, Fischler,
Shenker and Susskind \cite{banks9610} (and is reviewed
in \cite{bigatti,taylor9801}). It states that $M$-theory in the
so-called infinite momentum frame (IMF) is described by a supersymmetric
$U(N)$ Yang-Mills theory in $1+0$ dimensions. In a certain limit $N$ can
be identified with the number of D0-branes and should be taken to
infinity. 
The IMF is constructed as follows. Start with a state of energy
$E$ in $M$-theory and consider a boost in the eleventh direction such that
$p_{10}$ is large compared to any other scale in the theory. The energy
is in this limit accordingly:
\beq
E = \sqrt{p_{10}^2+p_{\perp}^2+m^2}=p_{10}+\frac{p_{\perp}^2+m^2}{2p_{10}}\ .
\eeq
Compactifying the theory -- in the $x_{10}$-direction -- on a circle of
radius $R$ relates it to the Type IIA theory with
\beq
E =\frac{N}{R}+\frac{R(p_{\perp}^2+m^2)}{2N}\ ,
\eeq
and the first term on the right hand side is identified with the mass
of $N$ D0-brane, as discussed earlier. The action describing the
interaction of such $N$ D0-branes is a $U(N)$ supersymmetric Yang-Mills
theory (\ref{tenYM}) reduced to the worldvolume of the D0-brane which
is one-dimensional. In this way one gets a $U(N)$ quantum mechanics with
an action like
\beq\label{m(atrix)}
S=\frac{1}{4g_{YM}^2}\int dt\ [2(D_{0}X^{i})^2+([X^i,X^j])^2 + {\rm
fermions}]\ .
\eeq
By dimensional reduction $F_{ij}=[X^i,X^j]$ and 
$F_{0j}=D_{0}X^j=\pa_0X^j+[A_0,X^j] $ with $X^i$ nine $N\times N$
matrices in the adjoint of $U(N)$. When the commutator-term in
(\ref{m(atrix)}) vanish, these matrices can be simultaneously
diagonalized and the diagonal elements are then identified with the
positions of the $N$ D0-branes.

To obtain eleven-dimensional $M$-theory, we must take the limit $R\rightarrow \infty$ and therefore also
$N\rightarrow \infty$.

This basically constitutes the definition of Matrix Theory. Subsequently,
Matrix Theory has been compared with "experiment". As an example, the scattering of
two gravitons has been computed to two-loop order (in the gauge coupling) in \cite{becker} 
and is found
to agree with calculations from supergravity. 
The same conclusion have recently been shown to hold also in the case of three-graviton scattering
\cite{fabri}.

It would of course be nice if this approach to $M$-theory could
elucidate anything interesting about four-dimensional theories. 
Compactification of Matrix Theory gives
gauge theories in higher dimensions (compactifying $d$ dimensions
results in
a $(d+1)$-dimensional gauge theory), but for $d>3$ they turn out not
to be renormalizable. So the connection to four-dimensional field theories
seems problematic. As another related problem we could mention that the large $N$ limit of  
Matrix Theory is not very well understood. 

Such problems make the Matrix Theory approach to $M$-theory less
appealing. It might, however, be possible that further study of the
relation between this approach and the Maldacena approach -- to be described below --
can lead to a better understanding of certain aspects of Matrix
Theory, e.g. of the large $N$ limit. 

\subsubsection*{M as in Maldacena?}

An -- a priori -- very different approach to $M$-theory is the
recent Maldacena conjecture \cite{malda} (see \cite{agmoo} for an
extensive review and references).
By studying D-brane scattering Maldacena formulated a series of conjectures that relates Type IIB theory
on Anti-de Sitter spaces to conformal field theories. As an example, with the geometry of $N$
D3-branes in Type IIB string theory, it is  conjectured that Type IIB string
theory on $AdS_5\times {\bf S}^5$ is equivalent to ${\cal N}=4$ supersymmetric $SU(N)$
Yang-Mills theory in $3+1$ dimensions! 
It is not obvious how a string theory in ten dimensions can be dual to a field theory in
four dimensions. But at least it is easy to see that the bosonic symmetries match: the conformal group in four
dimensions is $SO(2,4)$ and this is the same as the symmetry group of $AdS_5$ (the global bosonic symmetry groups are
$SO(2,4)\times SO(6)$).

Subsequently, this conjecture has
been made more precise \cite{gubser} and in general terms it states
that $M$-theory compactified on $AdS_{d+1}$ is dual to a conformal field
theory on the boundary $S^d$ of this space. 

If, for example, $\f$ is a massless scalar field in $AdS_{d+1}$ with boundary value $\f_{0}$, then it couples to an
operator ${\cal O}$ (of conformal dimension $d$) in the conformal field theory such that,
\beq
Z_{SUGRA}(\f|_{\pa AdS}=\f_0) = \langle e^{\int d^dx \f_0{\cal O}}\rangle_{CFT}\ .
\eeq 
On the left hand side the partition function is computed in $AdS$-space and is identified on the right hand side 
with a correlator in the boundary conformal field theory.

As mentioned in the introduction, it has also been shown that this
correspondence satisfies an interesting holographic bound \cite{susskind9805}: the bulk spacetime 
theory is described by a boundary field theory which has at the most one degree of freedom per Planck area. 

The $AdS$-space appears because in the so-called near-horizon limit
the geometry of for example $N$ D3-branes is \cite{malda}:
\beq
ds^2=\a'\left[
\frac{U^2}{g_{YM}N^{1/2}}dx_{\parallel}^2+
\frac{g_{YM}N^{1/2}}{U^2}(dU^2+U^2d\Omega_5^{2})
\right]\ ,
\eeq
in which the first two terms can be seen as a standard metric on
$AdS_5$ ($x_{\parallel}$ are coordinates on the worldvolume of the D3-brane, $U=r/\a'$ a radial coordinate, and
$d\O_5^{2}$ is a metric on the unit five-sphere). Here the Yang-Mills coupling is related to the string coupling
through
\beq
g=g_{YM}^2\ ,
\eeq
and is related to the radius of the ${\bf S}^5$ according to
\beq
\frac{R}{\sqrt{\a'}}=(4\pi g_{YM}^2N)^{1/4}\ .
\eeq  
For supergravity to be valid the radius $R$ should be large (so
that curvatures are small) compared to the string scale; that is, one
must have $gN\gg 1$. (In this sense the conjectured correspondence is a form of strong-weak coupling duality).
For fixed
$g$ one should therefore consider a limit
$N\rightarrow
\infty$, much as in Matrix Theory. But the interpretation of these limits
is of course very different in the two theories.

It should be noted that the Maldacena conjecture not only gives an
interesting approach to studying $M$-theory, but also to quantum field
theory. For example in certain limits on one side of the correspondence
one has supergravity compactified on $AdS$-space and on the other side a
quantum field theory. In this manner one can learn about strongly
coupled quantum field theory in e.g. four dimensions by studying
supergravity. But also theories with even less supersymmetry can be studied. 
For example, it has been shown \cite{witten9803} that four-dimensional ${\cal N}=0$ $SU(N)$ gauge theory has a dual
formulation in terms of $M$-theory on $M_7\times {\bf S}^4$, where $M_7$ is a certain seven-dimensional manifold.

This would mean that the
distinction between quantum field theories and string/$M$-theory might not be so fundamental after all.

\section{Some Consequences of String Duality in $D=4$}

We will end this chapter by mentioning two examples where the
conjectured duality relations between ten-dimensional string theories, 
and recent understanding of D-brane physics, have
been connected to four-dimensional physics. 
The main idea here being that 
if we believe that the
dualities relating the different string theories and $M$-theory in ten
dimensions are correct, their implications in lower dimensions can be
inferred by compactification.

First we look at the field theory Montonen-Olive duality (which was
mentioned in the introduction) in four dimensions.
\subsubsection*{Montonen-Olive Duality in $D=4$}

Starting with the conjectured $SL(2,{\bf Z})$ duality of Type IIB
theory, one can study its
consequences in lower dimensions. The
Montonen-Olive duality of $D=4$ ${\cal N}=4$ $U(n)$ gauge theory can for
example be derived from
the duality in ten dimensions \cite{town9512,tsey4}. 

The basic idea is as follows.
The Type IIB theory has D3-branes. The world volume theory of $n$ D3-branes in
Type IIB theory is a $D=4$ ${\cal N}=4$ $U(n)$ gauge theory, since reducing the
${\cal N}=1$ super Yang-Mills theory in ten dimensions to four dimensions gives
${\cal N}=4$
supersymmetry \cite{vecchia9803}. The gauge coupling is related to the string
coupling through (\ref{gYM}), or
\beq
g_{YM}^{2}=2\pi g\ ,
\eeq 
where $g$ is the string coupling and $g_{YM}$ the Yang-Mills gauge coupling. Now,
the $SL(2,{\bf Z})$ symmetry of the Type IIB theory keeps the D3-brane invariant but
changes the string coupling as $g\rightarrow 1/g$, the last because of
$S$-duality corresponding to the transformation $\t\rightarrow-1/\t$. Then
\beq
g_{YM}^{2}\rightarrow 4\pi^2/g_{YM}^{2}\ ,
\eeq
which is the transformation of the Yang-Mills coupling under
Montonen-Olive duality. Similarly, one can derive the shift of the vacuum
angle
$\theta
\rightarrow
\theta +2\pi$ as coming from the shift of the axion (that is the R-R
field $C_0$) in the Type IIB theory.

Thus, if $SL(2,{\bf Z})$-duality of the IIB string is correct it follows that
the
${\cal N}=4$ supersymmetric Yang-Mills theory in four dimensions has a
Montonen-Olive duality. Turning the argument around, if we think that
Montonen-Olive duality is correct something must be right about the
conjectured self-duality of the Type IIB theory.

This argument, however, is not very strong since there could be many
different ways to ``derive'' Montonen-Olive duality in four dimensions
from string theory (see e.g. \cite{vafa9707} for another realization of 
Montonen-Olive duality).

\subsubsection*{Seiberg-Witten Duality in $D=4$}

Seiberg-Witten duality has been identified as a consequence of duality of
Type II theories in \cite{vafa9604}. Introductions to the string theory
construction of the ${\cal N}=2$ gauge theories can be found in
\cite{lerche,mayr9807}.

In turning to the string theory realization of Seiberg-Witten theory, the first question that comes to mind is the
following.
In the Seiberg-Witten solution, the prepotential ${\cal F}$ (in Eq. (\ref{loweffsw})) is determined by the period
integrals of a certain meromorphic one-form $\l$ (\ref{swdiff}) on the basic cycles of the torus, or Seiberg-Witten
curve,
$\Sigma$ (\ref{torus}). Is there a concrete physical meaning of this curve? 

In one construction \cite{witten9703} of the ${\cal N}=2$ theory the
answer is simply that in Type IIA theory the worldvolume of a
certain five-brane is $\Sigma\times {\bf R}^4$ and that the ${\cal N}=2$ effective field theory comes from the low
energy Lagrangian of this five-brane theory. 

In order to construct a four-dimensional theory one can compactify Type IIA theory on a six-dimensional
Calabi-Yau
manifold $X_3$, which should be such that we get a theory with ${\cal N}=2$ supersymmetry in four dimensions. 
For example, compactifying Type IIA on ${\bf T}^6$ would give a theory of ${\cal N}=8$ supersymmetry.

More
concretely, $X_3$ is taken to be locally of the form ${\bf P}^1\times K3$. Here ${\bf P}^1$ is complex
one-dimensional projective space (topologically ${\bf S}^2$) and $K3$ is a compact complex K\"ahler manifold of
real dimension four with vanishing first Chern class and $h^{1,0}=0$ \cite{aspinwall96}.
However, to preserve the Calabi-Yau condition, $X_3$ cannot be globally a product manifold. 
One says that $K3$ is fibered over ${\bf P}^1$, where ${\bf P}^1$ is the base geometry. 

In order to identify the four-dimensional ${\cal N}=2$ theory, one can start with the compactification of Type
IIA on a $K3$ manifold $X_2$. The R-R three-form $C_3$ in the Type IIA theory gives rise to the vector multiplet
$A^a$ as $C_3\rightarrow A^a\wedge \omega^a$, where $\omega^a$ is a basis of $H^2(X_2)$ \cite{vafa9707}.   
Charged fields are obtained by wrapping $D2$-branes on two-cycles
$S_2^{a}$ dual to the two-forms $\omega_a$. 
These become identified with the $W^{\pm}$ massive gauge multiplets
that have  masses
proportional to the volume of the two-cycles $S_2$ (wrapping four-branes on four-cycles gives the corresponding
magnetic duals). 

To decouple gravity, that is to obtain a quantum field theory, these cycles must be vanishing. This is
because in the decoupling limit one should take $\a'\rightarrow 0$,
while keeping the $W^{\pm}$ boson masses $\sim \a'^{-1/2}$ finite 
-- it therefore turns out \cite{vafa9707} that the local geometry becomes
singular. Generally (that is for general gauge groups) the local geometry
is that of an ALE (Asymptotically Locally Euclidean) space with
so-called ADE Type singularities (see e.g. \cite{aspinwall96} for further discussion). 

So the classical $SU(2)$ Yang-Mills theory is constructed by compactifying Type IIA string theory on ${\bf P}^1\times K3$
with local singular geometry.

However, in the Seiberg-Witten solution, the prepotential ${\cal F}$ has also an infinite series of instanton corrections
\cite{lerche}. How does one determine these in the string theory construction? One possibility is to use mirror symmetry
\cite{greene97} which will relate Type IIA on $X_{3}$ to Type IIB theory compactified on its mirror
$\tilde{X_3}$ and with K\"ahler structure
and complex structure interchanged.

The Calabi-Yau three-manifold can be characterized by 
K\"ahler structure and complex structure moduli.
If $S_2^{a}$ is a basis of $H_2(X_3)$ and $J$ is the K\"ahler form
\footnote{In local complex coordinates the K\"ahler form is $J=ig_{i\bar{j}}dz^i\wedge d\bar{z}^{\bar{j}}$.}, then
$t_a=\int_{S_2^{a}}J$ parametrizes the K\"ahler moduli space.  The complex structure moduli space is parametrized by
complex numbers $z_i$; here, if
$S_3^{i}$ is a basis of
$H_3(X_3)$ and $\Omega$ is the unique three-form on $X_3$, then $z_i=\int_{S_3^{i}}\Omega$.  

In the Type IIA theory the scalars of the vector multiplets (containing gauge fields and gauginos) are determined by the
K\"ahler structure and those of the  hyper 
multiplets (containing matter fields) by the complex structure. In the Type IIB
theory the situation is reversed: the vector multiplet scalars are determined by the
complex structure and the hyper multiplets by the K\"ahler structure. 

The instanton corrections are so-called worldsheet instanton corrections -- in the IIA theory they can be understood as
coming from the wrapping of the string worldsheet on the base manifold ${\bf P}^1$. Since
there is no neutral coupling between hypermultiplets and vector multiplets in the ${\cal N}=2$ theory only vector
multiplets are important for determining these corrections. 

So, in the Type IIB theory -- where things are reversed -- there are no worldsheet instanton corrections to the vector
multiplet. This implies \cite{vafa9707} that 
the classical period integrals $\int \Omega$ in $\tilde{X}_3$ describe the exact
vector multiplet moduli space in the ${\cal N}=2$ theory. The general solution to the ${\cal N}=2$ theory is,
therefore, given in terms of period integrals $\int_{S_3^i}\Omega$ on a local Calabi-Yau geometry in $\tilde{X}_3$.  

The period integrals of the meromorphic one-form $\l$ on the Seiberg-Witten curve $\Sigma$ can then be obtained by
'integrating out' the two extra dimensions on a certain two-cycle $S_2$. In this way, one can get an explicit
representation of the Seiberg-Witten curve (\ref{torus}) as determined by the local geometry of the Calabi-Yau
manifold
$\tilde{X}_3$ \cite{vafa9604}. 

In this representation, one can also compute the stable BPS spectrum in a rather simple way --
something which usually is quite hard from the quantum field theory point of view. 

Finally, let us mention, that there is a third representation of the ${\cal N}=2$ theory. This is obtained by
using a $T$-duality, which maps Type IIB in the neighbourhood of an $A_1$ singularity of ALE space to Type IIA
on a symmetric five-brane \cite{mayr9807}. 
The worldvolume of this five-brane is $\Sigma\times {\bf R}^4$. Compactifying the resulting theory
on $\Sigma$ gives a four-dimensional ${\cal N}=2$ theory which is identified with the low energy effective theory
of the Seiberg-Witten solution \cite{witten9703}.

Having sketched how the Seiberg-Witten duality (which played an
important role in the topological field theories studied in our first
chapter) in four dimensions follows naturally from string duality in ten
dimensions, we believe this is a natural place to end.


\chapter{Discussion}

We have studied aspects of duality in various situations: in topological field theory
(as the Seiberg-Witten duality), in field theory (as the $T$-duality of sigma models) and in string
theory (where duality seems to connect all five known theories in ten dimensions). 

In testing dualities the topological field theory approach seems to be an important one. Not only
because topological field theories are very simple but also because one might
hope that the results that follow from using the Seiberg-Witten invariants can be proven in an exact
way.  

In this thesis we have shown that by dimensionally reducing the  topological field theories corresponding to the
two (dual) approaches to Donaldson theory, one obtains theories which by construction are topological, and
we have derived the corresponding "monopole" equations in three and
two dimensions \cite{ols}. 
While the
three-dimensional versions have been studied quite extensively in the literature, there is still a gap to be filled
in the two-dimensional case.  Here one studies what are called the Hitchin equations on a Riemann surface. The
connection from duality in moduli space could in principle be used to carry out an analysis
from the point of view of the Seiberg-Witten theory. 
To this end, the
vanishing theorems we derived in two dimensions should be important. 

Even in the original four-dimensional effective ${\cal
N}=2$ theory do such two-dimensional topological field theories appear to play a role. In
two-dimensional Landau-Ginzburg models, the three-point correlators can be shown to obey 
the so-called WDVV (Witten-Dijkgraaf-Verlinde-Verlinde) equations. As discussed in \cite{marsha},
the same equations have been shown to hold for certain derivatives of the prepotential, which seem
to indicate that the low-energy effective theory contains a certain two-dimensional topological
structure. 

It is important to establish whether the two-dimensional topological field theory
obtained by dimensionally reducing the monopole action is in any way connected to the WDVV equations.

Also, since everything
is rather simple in two dimensions, one might speculate that duality can be derived more
directly. Here the connection to string theory is also important; it is quite likely that  
string theory will provide a new understanding of the
relations between invariants of various moduli spaces. Hence, it would be interesting to
study the moduli problems from the point of view of non-perturbative string theory. This does not
seem an impossible task. It is well known that topological quantum field theories (at least
those associated with ${\cal N}=4$ supersymmetry) can be obtained from
D-branes in string theory \cite{vafa9511}.    

One might also speculate about getting interesting topological field theory dualities from the
Maldacena conjecture \cite{malda}. According to this conjecture, a conformal field theory in $d$
dimensions is described by Type IIB string theory compactified on ($d+1$)-dimensional Anti-de Sitter
space (on the boundary of which the field theory resides). Twisting the conformal field theory, one
should obtain a $d$-dimensional topological field theory which is dual to a certain string theory in
$d+1$ dimensions! Could this be important for studying the topology of four-manifolds?
\footnote{As a variant of the AdS/CFT correspondence, Gopakumar and Vafa have suggested recently \cite{gopaku} 
that $SU(N)$ Chern Simons theory on ${\bf S}^3$ is dual to a certain topological string theory (described by
the A-model topological sigma model mentioned in Chapter 2).}

We then studied some restrictions imposed by $T$-duality in two-dimensional sigma models. 
The key result here is the relation, put forward by the author and P. Haagensen, 
that at any order in sigma model perturbation theory the RG flow (as denoted by the
operator $R$) commute with
$T$-duality:
$[T,R]=0$ \cite{HO}. This has been verified to leading order in $\a'$
-- for bosonic, supersymmetric and heterotic sigma models
\cite{HO,haag,HOS}. 
The validity has also been
demonstrated to second order in $\a'$ for bosonic models with a purely metric background \cite{HO}. It
seems natural to generalize this latter result to the case of torsionful backgrounds even though 
there here is the further complication of scheme dependence. 

Also, starting with $[T,R]=0$, we have been able to essentially
compute the one-loop beta functions in these models, and the two-loop
functions in the case of bosonic and supersymmetric models with purely
metric backgrounds \cite{HOS}.

Recently Balog et al. \cite{balog9806} have presented a derivation of
$[T,R]=0$ at one-loop order, which explains (at least to one-loop
order) why we have found the consistency conditions studied
in Chapter 3 being exactly satisfied for the Weyl anomaly coefficients. 
This work naturally calls for a generalization to higher-loop orders. In this way one should be
able to address the question of corrections to duality at two-loop order including the
antisymmetric tensor background \cite{kalop9705}.

It has also turned out that a requirement such as
$[T,R]=0$, with $T$ some duality symmetry, not only provides constraints on RG flow in sigma models
but also in models with
$S$-duality and statistical systems \cite{damg9609,ritz9710}. But in these cases it needs to be
identified what principles are needed to ask for commutativity between RG flow and dualities in
the first place.
In this context, we also propose to study further such relations between duality and
RG flow for the open string sigma models, here with the ingredient of
D-branes added.   

Finally, we have tried to give an outline of the present status of
string theory duality in ten dimensions.
Here it has turned out that all ten-dimensional theories are connected in a manner such that the
strong coupling limit of any of these five theories has a corresponding naturally weakly coupled
theory (many similar results have been obtained in lower dimensions as is apparent from
\cite{witten9503} and its citations). Most surprising is the Type IIA theory 
where one gets an eleven-dimensional
theory, called $M$-theory (note that all other theories can be obtained as certain limits of
$M$-theory). While we do not know the correct way to formulate $M$-theory or what the relevant
degrees of freedom are, we do know that in superstring theory D-branes play a central
role in describing the dynamics at strong coupling.
Hence, they seem to be important in finding the right way to formulate string/$M$-theory. As such they have
played an important role both in the formulation of the Matrix Theory conjecture and in the more
recent Maldacena conjecture. And in the latter case there is even the added bonus that one can make certain
predictions about quantum field theories at strong coupling. 
 
Let us conclude by saying that duality still has not fully answered the more fundamental questions
such as: "what is field theory?",  "what is string theory?" or "what is $M$-theory?". But at least we
seem to be on the way.

\appendix
\chapter{Kaluza-Klein Reduction}

We write a generic background
metric $g_{\m\n}$ as in (\ref{metric}) , and the components of the
antisymmetric
background tensor $b_{\mu\nu}$ as $b_{0i}\equiv w_i$ and $b_{ij}$. In this
notation, barred quantities refer to the metric $\bar{g}_{ij}$.
\begin{itemize}
\item {\it Inverse metric}:  $g^{00}\!=\!1/a+v_iv^i,\
g^{0i}\!=\!-v^i,\ g^{ij}\!=\!\bar{g}^{ij}$.  On decomposed tensors,
indices $i,j,\ldots$ are raised and lowered with the metric
$\bar{g}_{ij}$ and its inverse. With the metric decomposition
(\ref{metric}) we also have 
$\det g=a\det\bar{g}$.

\item {\it Connection coefficients}:
\bea
\G^0_{00}&=&{a\over2}v^ia_i\ ,\ \G^0_{i0}={a\over2}\left[ {a_i\over a}+
v^ja_jv_i+v^jF_{ji}\right]\ ,\nonumber\\
\G^i_{00}&=&-{a\over2}a^i\ ,\ \G^i_{0j}=-{a\over2}\left[ F^i_{\
j}+a^iv_j\right]\ ,\nonumber\\
\G^0_{ij}&=&-\bar{\G}^k_{ij}v_k+\half (\pa_iv_j+\pa_jv_i+a_iv_j+a_jv_i)-
{a\over2}v^k\left[ v_jF_{ik}+v_iF_{jk}-a_kv_iv_j\right]\ ,\nonumber\\
\G^i_{jk}&=&\bar{\G}^i_{jk}+{a\over2}\left[ v_jF_k^{\ i}+v_kF_j^{\ i}
-a^iv_jv_k\right]\ ,
\ena
where $a_i\!=\!\pa_i\ln a\ ,\ F_{ij}\!=\!\pa_iv_j-\pa_jv_i$.

\item {\it Ricci tensor}:
\bea\label{riccitensor}
R_{00}&=&-{a\over2}\left[ \bar{\nabla}_ia^i+\half a_ia^i-{a\over2}F_{ij}
F^{ij}\right]\ ,\nonumber\\
R_{0i}&=&v_iR_{00}+{3a\over4}a^jF_{ij}+{a\over2}\bar{\nabla}^jF_{ij}\ ,\nonumber\\
R_{ij}&=&\bar{R}_{ij}+v_iR_{0j}+v_jR_{0i}-v_iv_jR_{00}-\half
\bar{\nabla}_ia_j
-{1\over4}a_ia_j-{a\over2}F_{ik}F_j^{\ k}\ .
\ena

\item{\it Riemann tensor}:
\bea
R_{i0k0}&=& -{a\over2}\left({1\over2}a_{i}a_{k}+\bar{\nabla}_{i}a_{k}+
{a\over2}F_{i}^{\ l}F_{lk}\right)\ ,\nonumber\\
R_{ijk0}&=& v_{j}R_{i0k0}-v_{i}R_{j0k0}-{a\over2}\bar{\nabla}_{k}F_{ij}
-{a\over2}\left( a_{k}F_{ij}+{1\over2}a_{j}F_{ik}-{1\over2}a_{i}F_{jk}
\right)\ ,\nonumber\\
R_{ijkm}&=& \bar{R}_{ijkm}+R_{ijk0}v_{m}+R_{jim0}v_{k}+R_{mkj0}v_{i}
+R_{kmi0}v_{j}\ ,\nonumber\\
&&\ -R_{m0j0}v_{i}v_{k}+R_{k0j0}v_{i}v_{m}-R_{k0i0}v_{j}v_{m}
+R_{m0i0}v_{j}v_{k}\ ,\nonumber\\
&&\ -{a\over4}\left(F_{im}F_{kj}+F_{ki}F_{mj}+2F_{ji}F_{mk}
\right). 
\ena

\item {\it Torsion}:
\bea
H_{0ij}&=&-\pa_iw_j+\pa_jw_i\equiv -G_{ij}\ ,\nonumber\\
H_{ijk}&=&\pa_ib_{jk}+\pa_jb_{ki}+\pa_kb_{ij}\ ,
\ena 
and all other
components vanish. For the one-loop beta function the following
quantities are needed:
\bea
H_{0\m\n}H_0^{\ \m\n}&=&G_{ij}G^{ij}\ ,\nonumber\\
H_{0\m\n}H_i^{\ \m\n}&=&-2G_{ij}G^{jk}v_k-H_{ijk}G^{jk}\ ,\nonumber\\
H_{i\m\n}H_j^{\ \m\n}&=&2\left({1\over a}+v_mv^m\right) G_i^{\ k}G_{jk}
-2v^kv^mG_{ik}G_{jm}+2H_{km(i}G_{j)}^{\ k}v^m\nonumber\\ 
&&+H_{ikm}H_j^{\ km}\ ,
\ena
and
\bea
\nabla_{\mu} H^{\mu}_{\ 0i}&=&\bar{\nabla}^jG_{ji}-aG_{ij}F^{jk}v_k+\half G_{ij}a^j
-{a\over2}F^{jk}\left( H_{ijk}+v_iG_{jk}\right)\ ,\nonumber\\ 
\nabla_\m H^\m_{\ ij}&=&\bar{\nabla}^k\left( H_{kij}+v_kG_{ij}\right) -\half \left[ G_i^{\
k}\bar{\nabla}_{(k}\ v_{j)} -G_j^{\ k}\bar{\nabla}_{(k}\ v_{i)}
\right]-{a\over2}v_{[i}H_{j]km}F^{km}\nonumber\\ 
&&+v_{[i}G_{j]k}\left(
a^k-aF^{km}v_m\right)+\half a^kH_{kij}+\half v_ma^mG_{ij}-\half
F_{[i}^{\ k}G_{j]k}\ ,
\ena
where $[ij]=ij-ji$ and $(ij)=ij+ji$.
                              
\item {\it Dilaton terms}:
\bea
\nabla_0\pa_0\f&=&{a\over2}a^i\pa_i\f\ ,\nonumber\\
\nabla_0\pa_i\f&=&{a\over2}\left(F^j_{\ i}+a^jv_i\right) \pa_j\f\ ,\nonumber\\
\nabla_i\pa_j\f&=&\bar{\nabla}_i\pa_j\f -{a\over2}\left( v_iF_j^{\ k}+
v_jF_i^{\ k}-a^kv_iv_j\right)\pa_k\f\ .
\ena

\item {\it Tangent space geometrical tensors}:

When referred to the tangent space, the Ricci tensor becomes
\bea
R_{\hat{0}\hat{0}}&=&-{1\over2}
\left[ \bar{\nabla}_ia^i+\half a_ia^i-{a\over2}F_{ij}
F^{ij}\right]\ ,\nonumber\\
R_{\hat{0}\a}&=&\bar{e}^{\ i}_{\a}
\left[ {3\sqrt{a}\over4}a^jF_{ij}+{\sqrt{a}\over2}\bar{\nabla}^jF_{ij}
\right]\ ,\nonumber\\
R_{\a\b}&=&\bar{e}^{\ i}_{\a}\bar{e}^{\ j}_{\b}
\left[\bar{R}_{ij}-\half \bar{\nabla}_ia_j
-{1\over4}a_ia_j-{a\over2}F_{ik}F_j^{\ k}\right]\ ,
\ena
where $\bar{e}^{\ i}_{\a}$ is the inverse vielbein for the metric
$\bar{g}_{ij}$. Likewise, the Riemann tensor is
\bea
R_{\a \hat{0} \b \hat{0}}&=& -{1\over2}\bar{e}^{\ i}_{\a}\bar{e}^{\
j}_{\b}
\left({1\over2}a_{i}a_{j}+\bar{\nabla}_{i}a_{j}
+{a\over2}F_{i}^{\ s}F_{sj}\right)\ ,\nonumber\\
R_{\a \b \g \hat{0}}&=& -{\sqrt{a}\over2}\bar{e}^{\ i}_{\a}\bar{e}^{\
j}_{\b}
\bar{e}^{\ k}_{\g}\left( a_{k}F_{ij}+{1\over2}a_{j}F_{ik}
-{1\over2}a_{i}F_{jk}+\bar{\nabla}_{k}F_{ij}\right)
\ ,\nonumber\\
R_{\a \b \g \d}&=&
\bar{e}^{\ i}_{\a}\bar{e}^{\ j}_{\b}\bar{e}^{\ k}_{\g}
\bar{e}^{\ m}_{\d}\left( \bar{R}_{ijkm}-{a\over4}
( F_{im}F_{kj}+F_{ki}F_{mj}+2F_{ji}F_{mk})\right) . 
\ena
\end{itemize}                                                      
\chapter{Two-Loop Tensor Structures}
The possible tensor structures which can appear at two-loop are
symmetric tensors and must scale as $\Omega^{-1}$ under
global scalings of the background metric. A moment of thought reveals
that the structure of the terms can be only of the three following
kinds: $\nabla_{\cd}\nabla_{\cd}R_{\cd\cd\cd\cd}$,
$R_{\cd\cd\cd\cd}R_{\cd\cd\cd\cd}$ or $g_{\mu\nu}\times$(traces of the
previous tensors). Here the $\cd$ indicates a $\mu$, $\nu$ index or a
contracted index; $R_{\cd\cd\cd\cd}$ is the Riemann tensor. The
resulting tensors are:

\begin{itemize}
\item{Tensors of the form
    $\nabla_{\cd}\nabla_{\cd}R_{\cd\cd\cd\cd}$:}
\beq
\na_{\m}\na_{\m}R, \na^2R_{\m\n},\ 
\na_{(\m}\na_{\a}R^{\a\ \ \ \b}_{\ \b\n)},\
\na_{\a}\na_{(\m}R^{\a \ \ \ \b}_{\ \b\n)},\
\na_{\a}\na_{\b}R^{\a\ \ \b}_{\ (\m \ \n )}
\eeq

\item{Tensors of the form $R_{\cd\cd\cd\cd}R_{\cd\cd\cd\cd}$:}
\bea
&&R_{\m\a\n\b}R^{\a\b},\ R_{\m\a\b\g}R_{\n}^{\ \a\b\g}, \ 
R_{\m\g\a\b}R_{\n}^{\ \a\b\g}, \ R_{\m\a\b\g}R_{\n}^{\ \g\a\b},
\nonumber\\
&&R_{\m\g\a\b}R_{\n}^{\ \g\a\b}, \ R_{\m\a}R_{\n}^{\ \a}, \ R_{\m\n}R
\ena
\item{Tensors of the form $g_{\mu\nu}\times$(traces):}
\bea
&&g_{\m\n}\na^2R, \ g_{\m\n}\na_{\a}\na_{\b}R^{\a\b}, \ 
g_{\m\n}R^2, \ g_{\m\n}R_{\a\b}R^{\a\b}, \ 
g_{\m\n}R_{\a\b\g\d}R^{\a\b\g\d}\nonumber\\
&&g_{\m\n}R_{\a\b\g\d}R^{\a\g\d\b}
\ena
\end{itemize}
Here we have reduced from a larger set by using the known symmetries
of the Riemann tensor.
As described in the main text only 10 of these 18 tensors are linearly
independent. To show linear dependence between different tensors one uses
the first and second Bianchi identities and the expression for $[\na,
\na]$ acting on tensors. The first Bianchi identity determines for
example that
\beq
R_{\m\a\b\g}R_{\n}^{\ \a\b\g}=
-R_{\m\a\b\g}R_{\n}^{\ \g\a\b}
-R_{\m\g\a\b}R_{\n}^{\ \a\b\g},
\eeq
while the second Bianchi identity implies that
\beq
g_{\m\n}\na_{\a}\na_{\b}R^{\a\b}=\half g_{\m\n}\na^2R\ .
\eeq

\newcommand{\NP}[1]{Nucl.\ Phys.\ {\bf #1}}
\newcommand{\NPPS}[1]{Nucl.\ Phys.\ Proc.\ Suppl.\ {\bf #1}}
\newcommand{\PL}[1]{Phys.\ Lett.\ {\bf #1}}
\newcommand{\CMP}[1]{Comm.\ Math.\ Phys.\ {\bf #1}}
\newcommand{\PR}[1]{Phys.\ Rev.\ {\bf #1}}
\newcommand{\PRL}[1]{Phys.\ Rev.\ Lett.\ {\bf #1}}
\newcommand{\PTP}[1]{Prog.\ Theor.\ Phys.\ {\bf #1}}
\newcommand{\PTPS}[1]{Prog.\ Theor.\ Phys.\ Suppl.\ {\bf #1}}
\newcommand{\MPL}[1]{Mod.\ Phys.\ Lett.\ {\bf #1}}
\newcommand{\IJMP}[1]{Int.\ Jour.\ Mod.\ Phys.\ {\bf #1}}
\newcommand{\JP}[1]{Jour.\ Phys.\ {\bf #1}}
\newcommand{\JMP}[1]{Jour.\ Math.\ Phys.\ {\bf #1}}
\newcommand{\AP}[1]{Ann.\ Phys.\ {\bf #1}}
\newcommand{\CQG}[1]{Class.\ Quant.\ Grav.\ {\bf #1}}
\newcommand{\PREP}[1]{Phys.\ Rept.\ {\bf #1}}

\end{document}